%% file: mass-loading.tex
\documentclass[fleqn,usenatbib]{mnras}

\usepackage{newtxtext,newtxmath}

\usepackage[T1]{fontenc}
\usepackage{ae,aecompl}

\usepackage{graphicx,xspace}	
\usepackage{subfigure}
\usepackage{amsmath}	
\usepackage{array}      
\usepackage{pdflscape}
\usepackage{rotating}
\usepackage{longtable}
\usepackage{verbatim}
\usepackage{pifont}
\usepackage{booktabs}
\usepackage{bigints}
\usepackage{soul}

\usepackage[dvipsnames]{xcolor}

\newcommand{\gr}{$\gamma$-ray\xspace}
\newcommand{\grs}{$\gamma$-rays\xspace}
\newcommand{\g}{$\gamma$\xspace}
\newcommand{\pg}{p$\gamma$\xspace}
\newcommand{\hadjet}{\texttt{HadJet}\xspace}
\newcommand{\bhjet}{\texttt{BHJet}\xspace}
\newcommand{\bh}{BHXB\xspace}
\newcommand{\bhs}{BHXBs\xspace}

\defcitealias{chatterjee2019accelerating}{CLTM19}
\newcommand{\cltm}{{\citetalias{chatterjee2019accelerating}}\xspace}

\title[Composition and mass-loading in hadronic jets]{Exploring the role of composition and mass-loading on the properties of hadronic jets}

\author[D. Kantzas et al.]{
D.~Kantzas,$^{ 1,2}$\thanks{E-mail: kantzas@lapth.cnrs.fr}\thanks{Currently at: LAPTh, CNRS, USMB, F-74940 Annecy, France}
S.~Markoff$,^{ 1,2}$ 
M.~Lucchini,$^{ 3}$
C.~Ceccobello$^{ 4}$
and K.~Chatterjee$^{ 5}$
\\
$^{1}$Anton Pannekoek Institute for Astronomy (API), University of Amsterdam, Science Park 904, 1098 XH Amsterdam, the Netherlands\\
$^{2}$GRavitation AstroParticle Physics Amsterdam (GRAPPA), University of Amsterdam, Science Park 904, 1098 XH Amsterdam, the Netherlands\\
$^3$MIT Kavli Institute for Astrophysics and Space Research, Massachusetts Institute of Technology, Cambridge, MA 02139, the USA\\
$^4$Department of Space, Earth and Environment, Chalmers University of Technology, Onsala Space Observatory, 439 92 Onsala, Sweden\\
$^5$Black Hole Initiative at Harvard University, 20 Garden Street, Cambridge, MA 02138, the USA\\
}

\date{Accepted XXX. Received YYY; in original form ZZZ}

\pubyear{2022}

\begin{document}
\label{firstpage}
\pagerange{\pageref{firstpage}--\pageref{lastpage}}
\maketitle

\begin{abstract}
Astrophysical jets are relativistic outflows that remain collimated for remarkably many orders of magnitude. Despite decades of research, the origin of cosmic rays (CRs) remains unclear, but jets launched by both supermassive black holes in the centre of galaxies and stellar-mass black holes harboured in X-ray binaries (BHXBs) are among the candidate sources for CR acceleration. When CRs accelerate in astrophysical jets, they initiate particle cascades that form \grs and neutrinos. In the so-called hadronic scenario, the population of accelerated CRs requires a significant amount of energy to properly explain the spectral constraints similarly to a purely leptonic scenario. The amount of energy required often exceeds the Eddington limit, or even the total energy available within the jets. The exact energy source for the accelerated protons is unclear, but due to energy conservation along the jets, 
it is believed to come from the jet itself via transfer of energy from the magnetic fields, or kinetic energy from the outflow. To address this hadronic energy issue and to self-consistently evolve the energy flux along the flows, we explore a novel treatment for including hadronic content, in which instabilities along the jet/wind border play a critical role. We discuss the impact of the different jet composition on the jet dynamics for a pair dominated and an electron-proton jet, and consequently the emitted spectrum, accounting for both leptonic and hadronic processes. Finally, we discuss the implications of this mass-loading scenario to address the proton energy issue. 

\end{abstract}

\begin{keywords}
acceleration of particles -- stars: jets -- galaxies: jets
\end{keywords}




\input{Sections/Introduction}


\input{Sections/Steady_jet}

\input{Sections/Steady_jet_results}


\input{Sections/Mass_loaded_jets}

\input{Sections/Mass_loaded_jets_results}

\input{Sections/Discussion}

\section{Summary and conclusions}\label{sec: summary}
Relativistic jets are efficient CR accelerators, but we still do not fully understand the particle acceleration mechanism. To fully interpret the jet kinematics, and how they relate to particle acceleration, we need to better understand how to link the observed spectra emitted by jetted sources over more than ten orders of magnitude in photon frequency to the jet physical properties.  Currently uncertainties about the composition as well as a lack of conserved dynamical models have contributed to a degeneracy between leptonic and lepto-hadronic models.

To break this degeneracy, we have developed a new multi-zone approach that links the jet composition to the jet dynamics. The total energy flux along the jet is conserved, while magnetic energy can be dissipated into both kinetic energy and gas enthalpy via particle acceleration.  This new approach makes clear the key role that the specific enthalpy $h$ can have on the evolution and exchange of energy along the jet.  In particular the enthalpy should be explicitly taken into account in models where:  i) electrons accelerate to large average energies, ii) protons accelerate in the jets as well, and/or iii) when the jet is pair-dominated, as suggested for numerous Galactic and extragalactic jets launched by black holes. 

When protons are accelerated into a non-thermal power law,
the energy requirement often exceeds the total energy that can be provided by the jet and/or the accretion energy onto the black hole, potentially violating energy conservation. We have developed a new model \hadjet based on our earlier lepto-hadronic work, that now conserves energy and includes a prescription for proton entrainment. Such a mass loading may in fact inhibit proton acceleration. By allowing the jets to entrain protons over a range of distance, as seen to occur in GRMHD simulations via eddies forming at the jet/accretion disc interface (\cltm), we demonstrate a new method to avoid the ``hadronic power'' problem in a more self-consistent approach. In a future work, we plan to further explore the impact of mass loading on the multiwavelength emission of both \bh jets and AGN jets.

\section*{Acknowledgements}

We would like to thank the anonymous reviewer for the thorough commenting that significantly improved the manuscript. 
DK and SM are grateful for support by the Netherlands Organisation for Scientific Research (NWO) VICI grant (no. 639.043.513).

\section*{Data availability}
No new data were generated or analysed in support of this research.



\bibliographystyle{mnras}
\bibliography{mass-loading} 


\appendix

\section{Specific enthalpy for a hard power law of accelerated particles}\label{app: h specific enthalpy for p=1.7}

In Fig.~\ref{app:fig: specific enthalpy for p=1.7 in appendix} we plot the evolution of $h$ for different jet composition. See Section~\ref{sec: specific enthalpy} for a detailed description of the subplots. In this figure, we show the evolution of $h$ assuming that the particles accelerate in a harder power law with an index of $p=1.7$ compared to $p=2.2$ we discuss in the main text. 

In the top subplots, we notice a similar behaviour to Fig.~\ref{fig: specific enthalpy} but $h$ goes to larger values for the case of a pair-dominated jet ($h\sim \Gamma_e \langle \varepsilon_{\rm e}\rangle $) according to equation~\ref{eq: h specific enthalpy versus eta}).

In the case where protons accelerate as well, $h$ can attain values as large as $\sim 2000$ for a particle acceleration with $\varepsilon_{\rm e, min}=\varepsilon_{\rm p, min}=10$ as we show in the lowermost subplots. This value is significantly larger than the expected values of \g of the bulk flow and in combination with the case where $\sigma$ takes large values to lead to hard power laws of particles \citep[][]{sironi2015relativistic,Sironi2020,Ball2018}, we see that the equation $\mu = \gamma (\sigma +h +1)$ (equation~\ref{eq: mu}) would not be a good approximation for the bulk Lorentz factor anymore \citep[][]{McKinney2006,Komissarov2007magnetic,Komissarov2009ultrarelativistic,Beskin_2010}.

\begin{figure*}
    \centering
 	\subfigure[Purely leptonic acceleration with $\varepsilon_{\rm e,min}=1.5$.]{                       
 	    \includegraphics[width=1.\columnwidth]{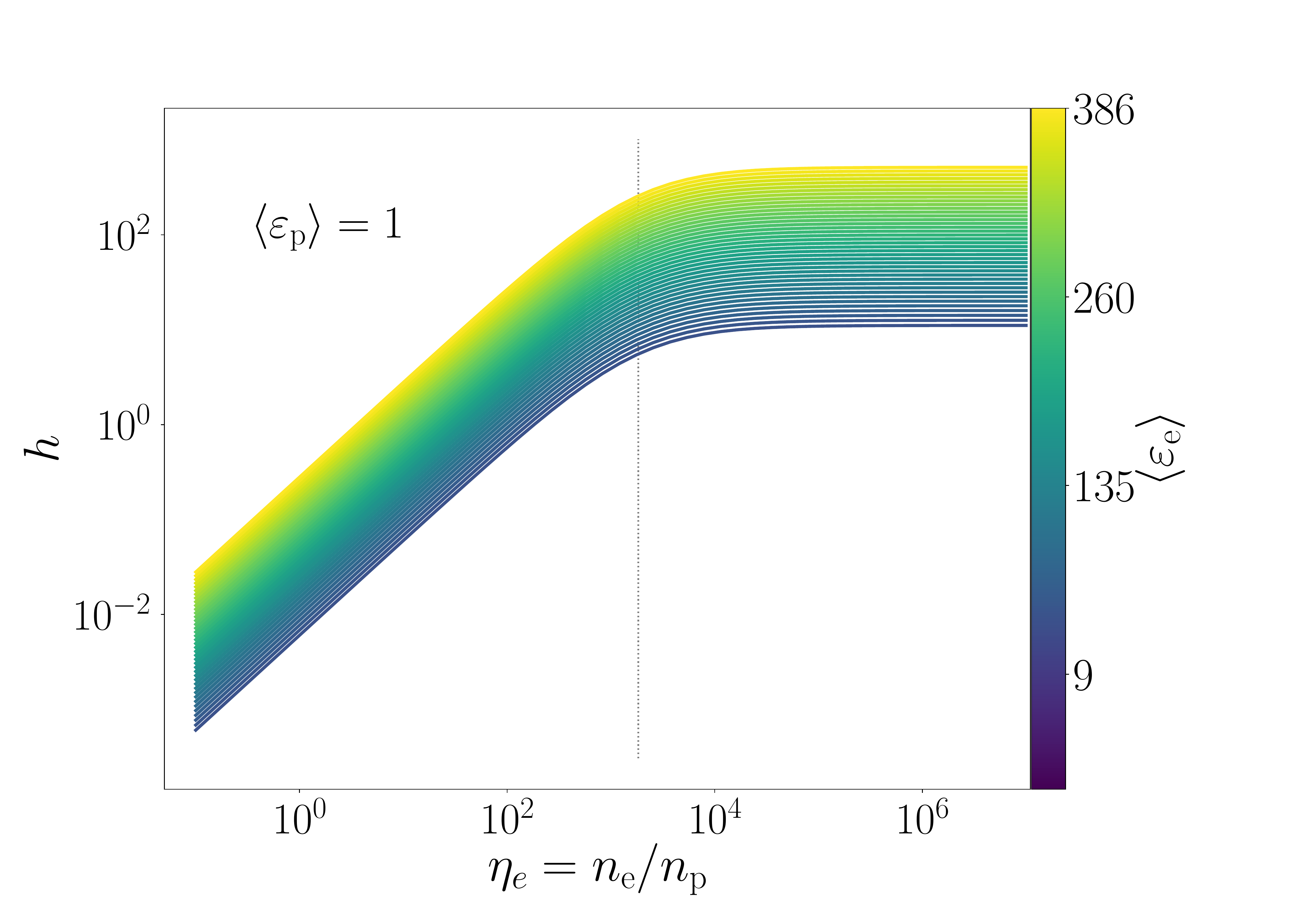}
    }
    \subfigure[Purely leptonic acceleration with $\varepsilon_{\rm e,min}=10$.]{
        \includegraphics[width=1.\columnwidth]{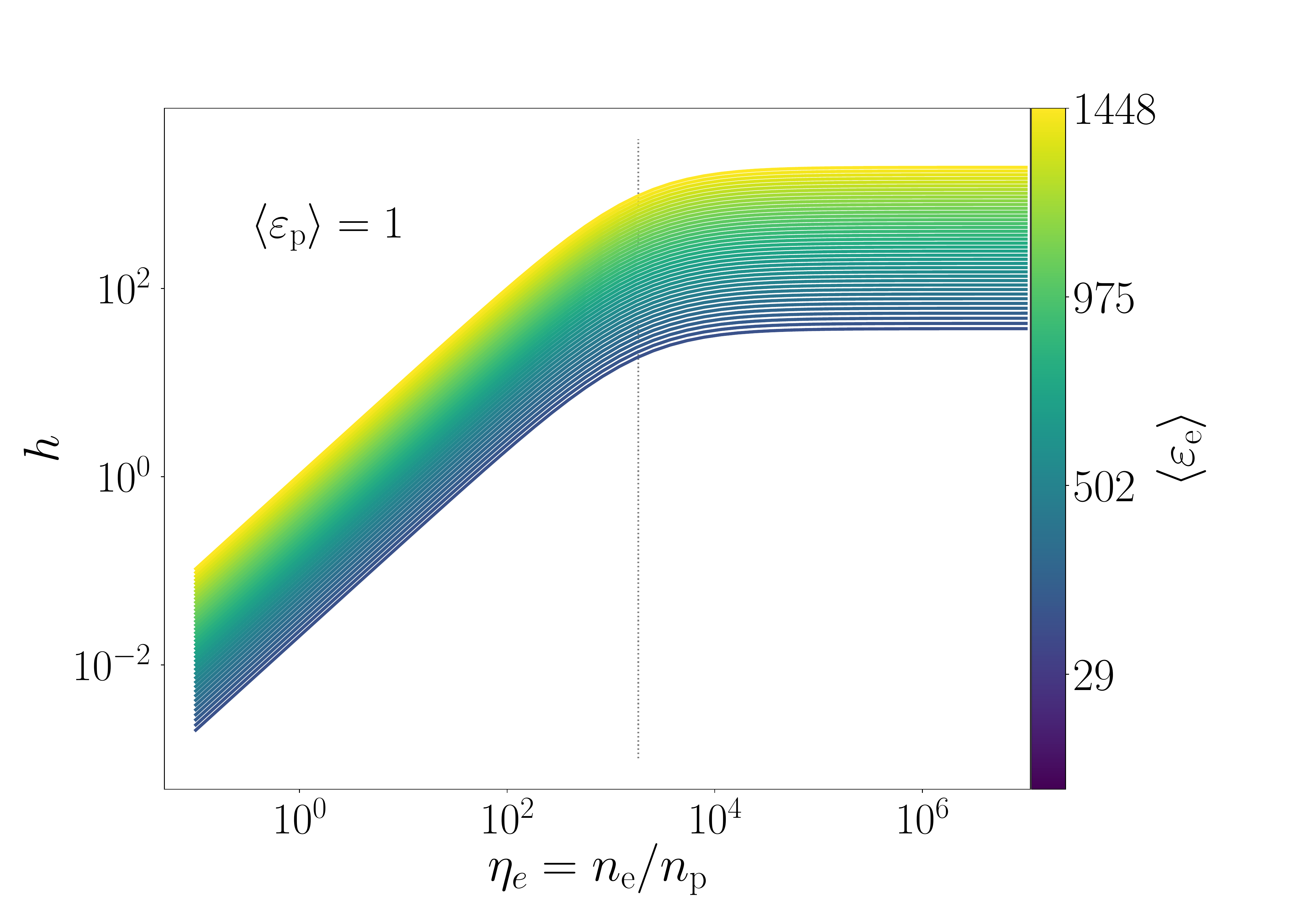}
    } 
 	\subfigure[Leptohadronic acceleration with $\varepsilon_{\rm e,min}=1.5$.]{
        \includegraphics[width=1.\columnwidth]{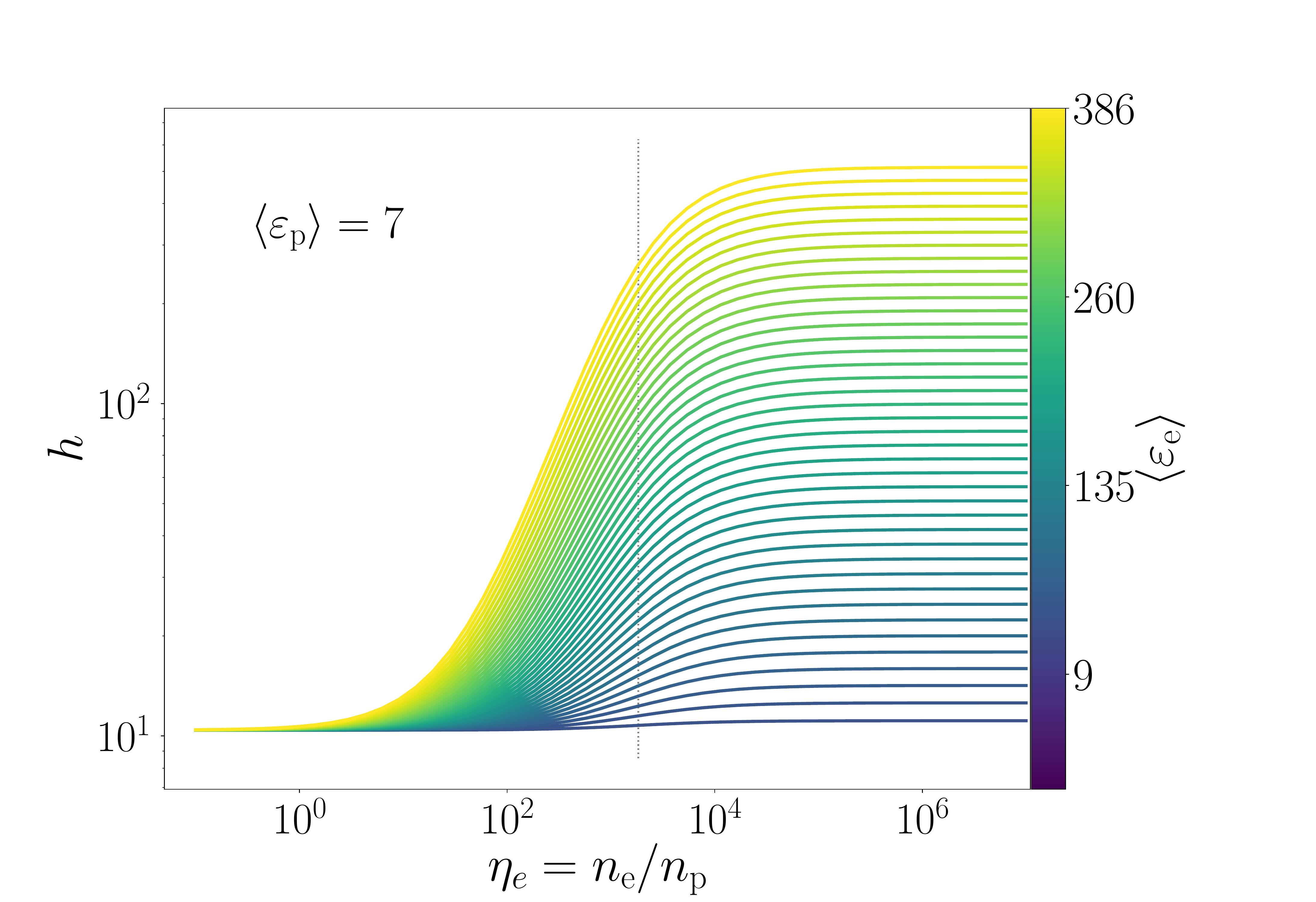}
    }
 	\subfigure[Leptohadronic acceleration with $\varepsilon_{\rm e,min}=10$.]{
        \includegraphics[width=1.\columnwidth]{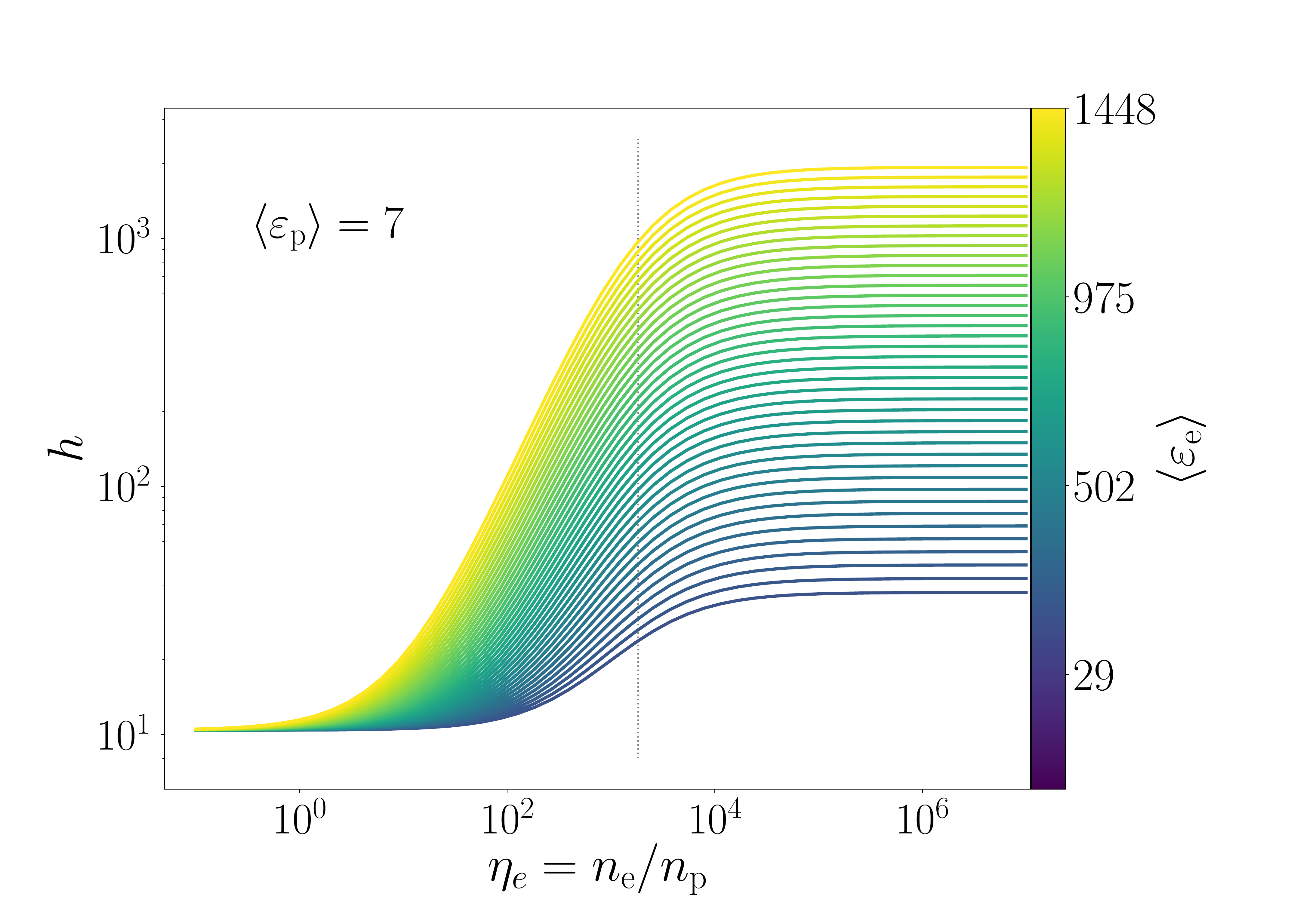}
    }
	\subfigure[More efficient hadronic, and leptonic acceleration with $\varepsilon_{\rm e,min}=1.5$.]{
	    \includegraphics[width=1.\columnwidth]{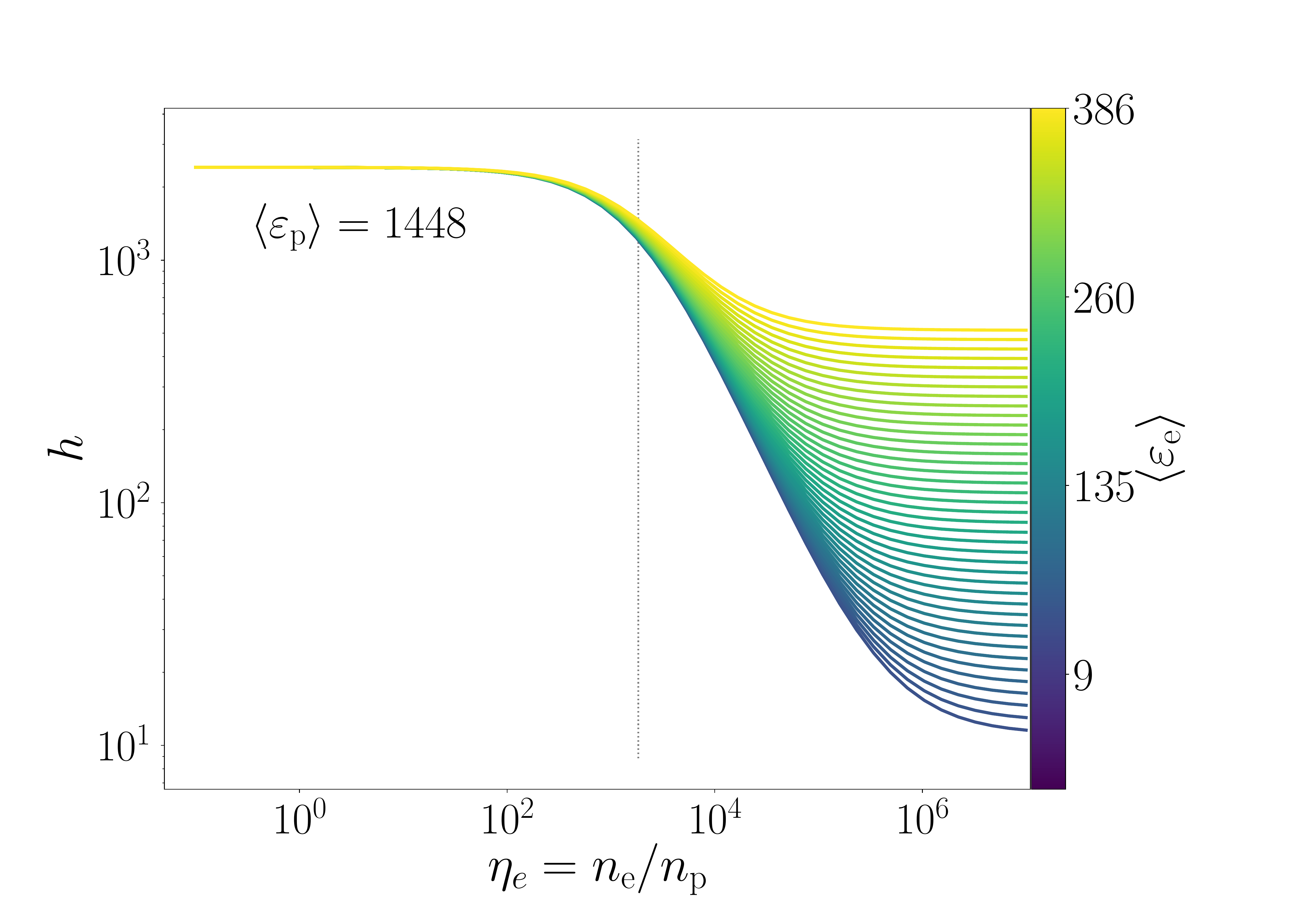}
    }
 	\subfigure[More efficient hadronic, and leptonic acceleration with $\varepsilon_{\rm e,min}=10$.]{
        \includegraphics[width=1.\columnwidth]{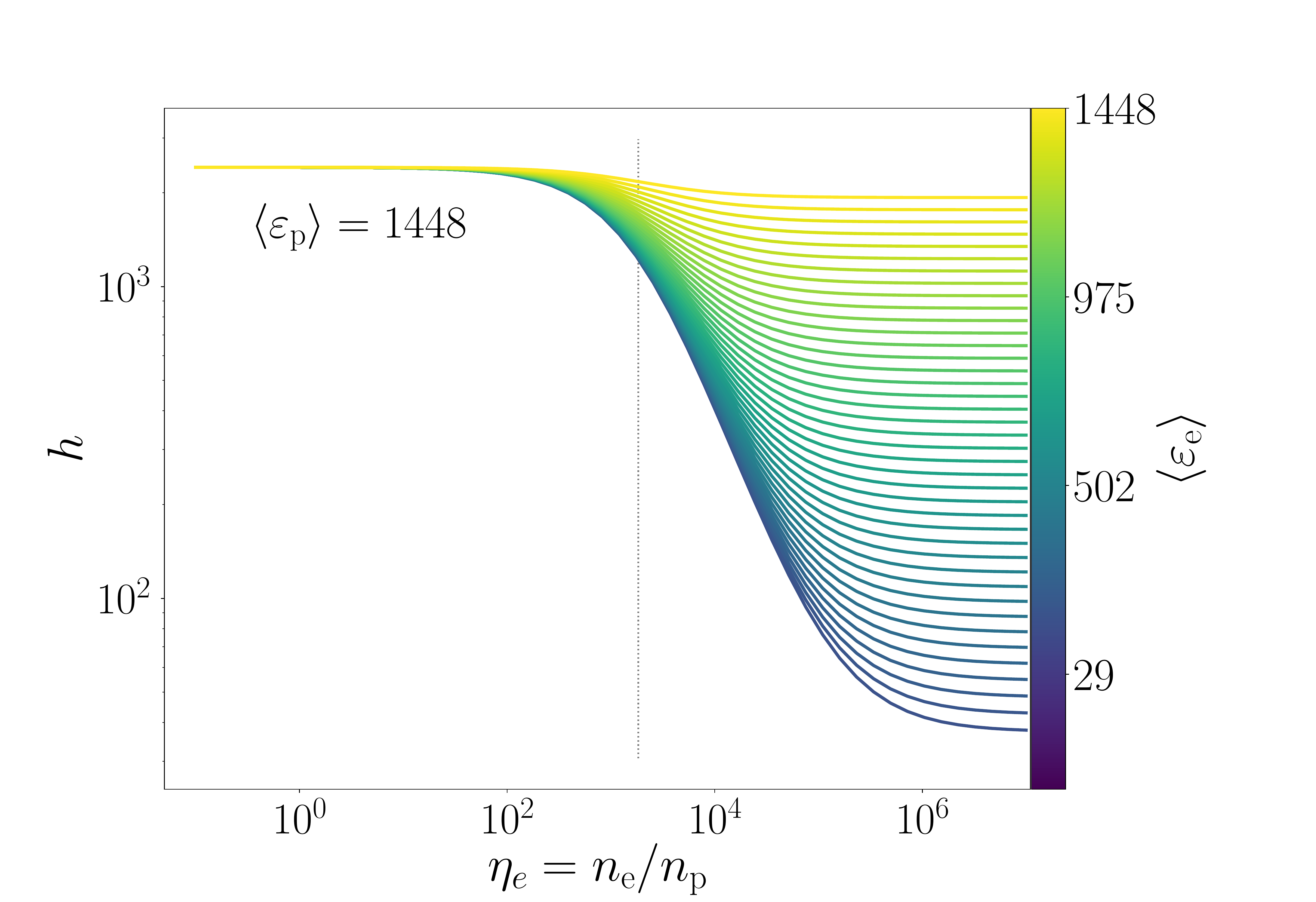}
    }
    \caption{The jet specific enthalpy $h$ as a function of the jet content $\eta_e=n_{\rm e}/n_{\rm p}$. In all plots, we assume $p=1.7$ to derive the average particle Lorentz factors from equation~(\ref{eq: average Lorentz factor of particles}). The color-map corresponds to the average Lorentz factor of pairs with lighter colors to indicate larger values. In the left column, we set the minimum Lorentz factor of pairs to be $\varepsilon_{\rm e,min}=1.5$, and on the right column, we use $\varepsilon_{\rm e, min}=10$. In the top raw, we assume that only leptonic acceleration takes place, in the middle raw we assume hadronic acceleration as well with $\varepsilon_{\rm p, min} = 1$ and $\varepsilon_{\rm p, max} = 100$, and in the bottom raw, we assume $\varepsilon_{\rm p, min} = 10$ and $\varepsilon_{\rm p, max} = 10^7$. 
    }
    \label{app:fig: specific enthalpy for p=1.7 in appendix}
\end{figure*}

\section{All energy components plots}\label{app: mu all plots}
\begin{figure*}
    \centering
    \subfigure[$\eta_e = 10$]{
    \includegraphics[width=1.0\columnwidth]{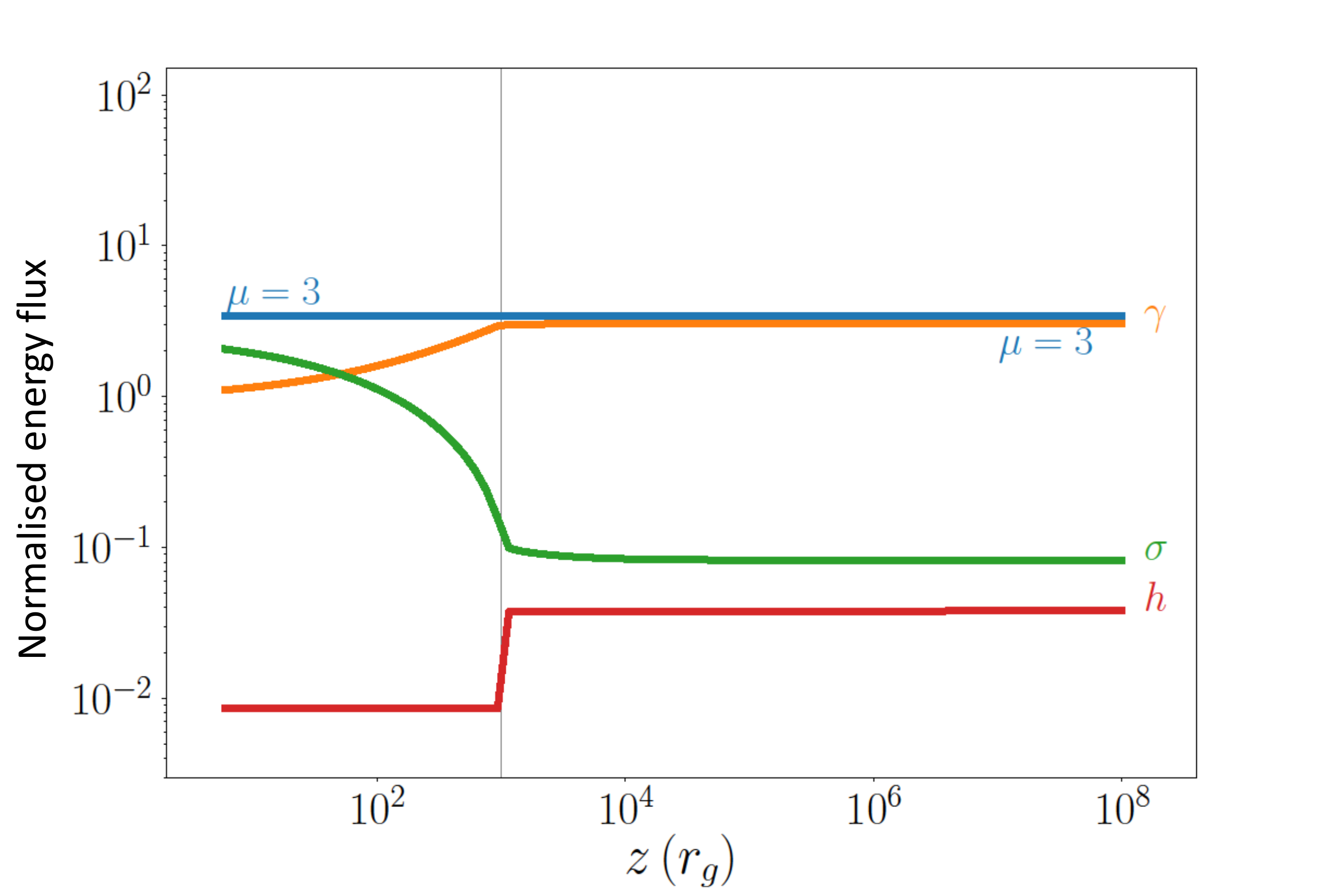}
    }
    \subfigure[$\eta_e = 100$]{
        \includegraphics[width=1.0\columnwidth]{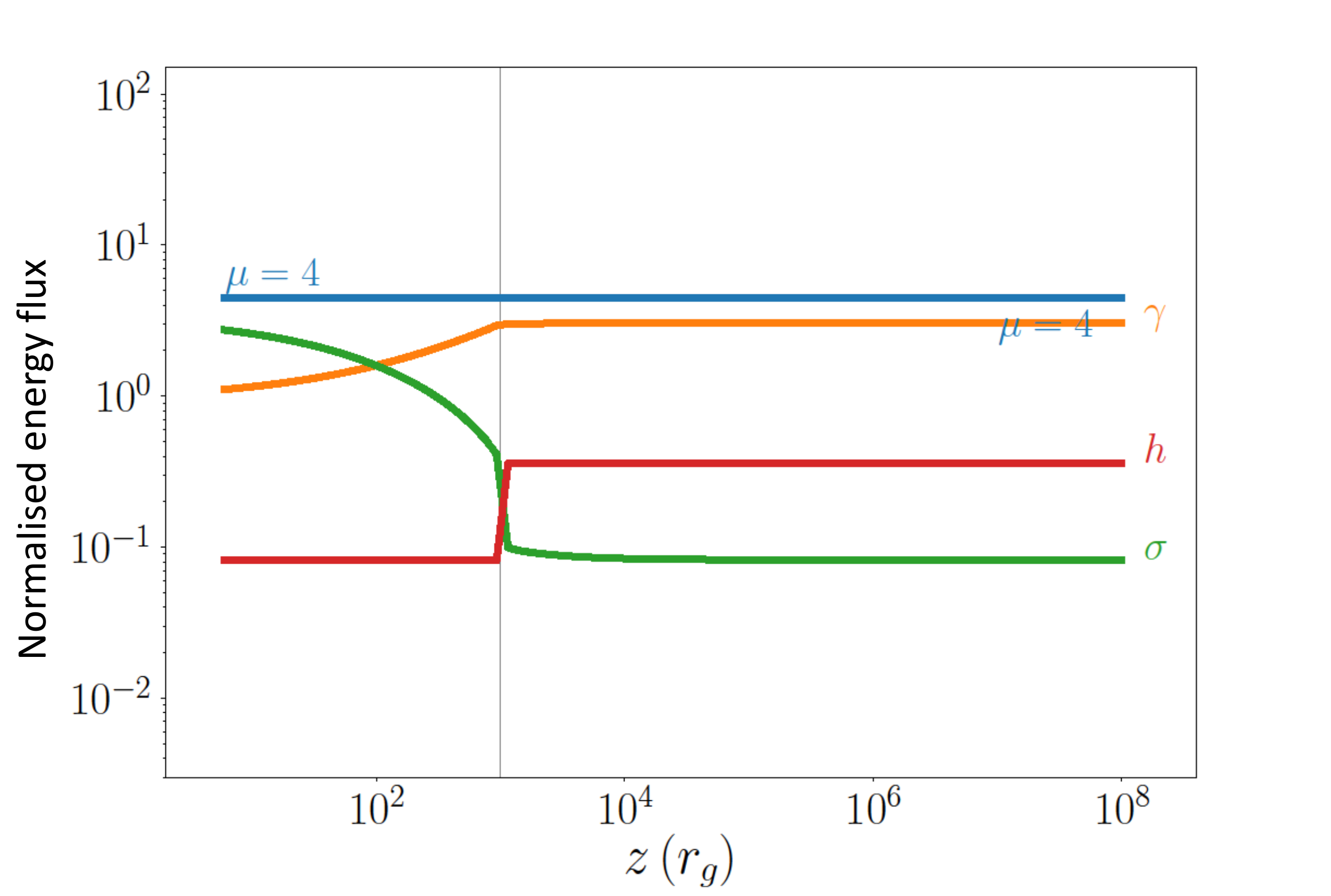}
    }
    \subfigure[$\eta_e = 1000$]{
        \includegraphics[width=1.0\columnwidth]{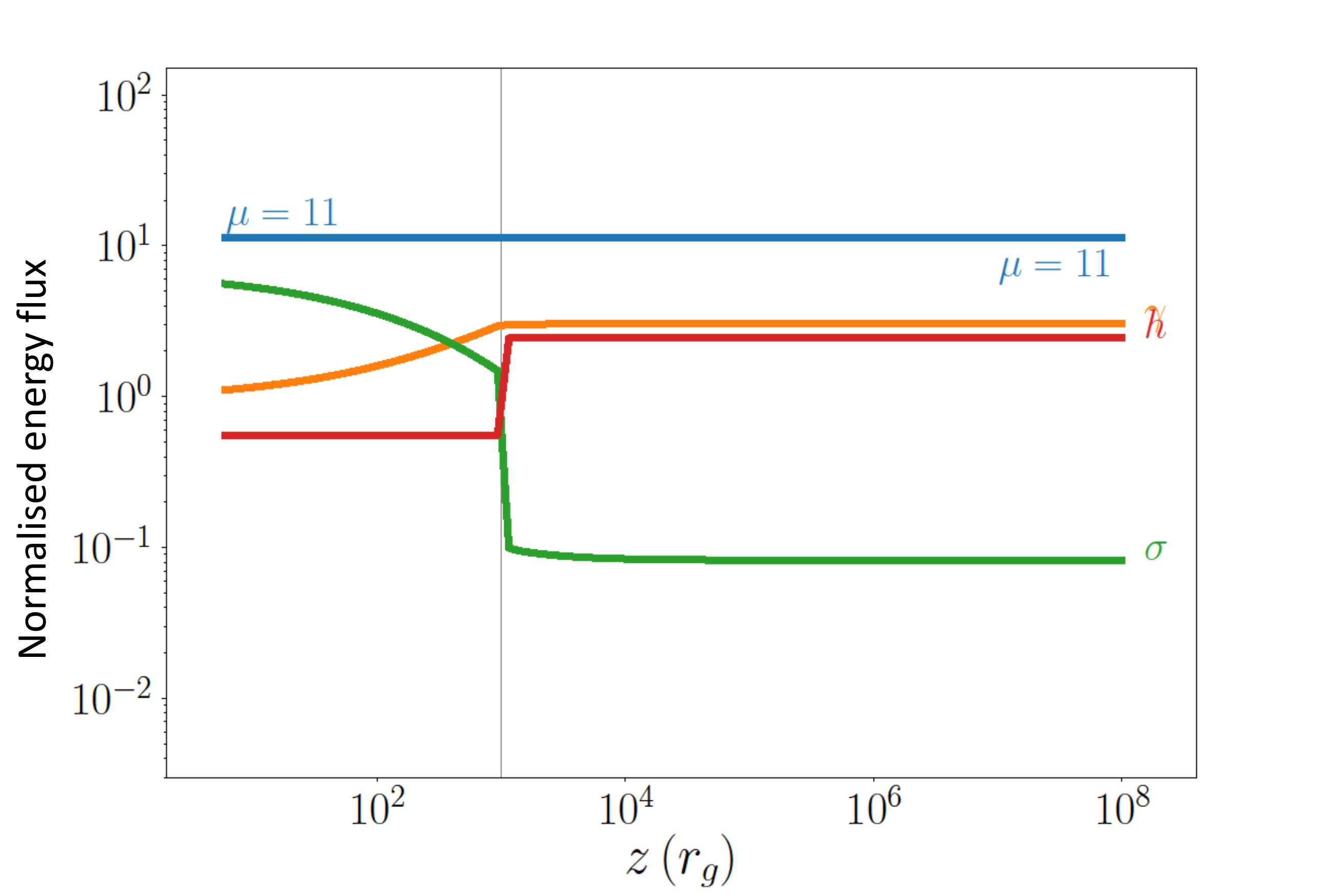}
    }
    \subfigure[$\eta_e = 10000$]{
        \includegraphics[width=1.0\columnwidth]{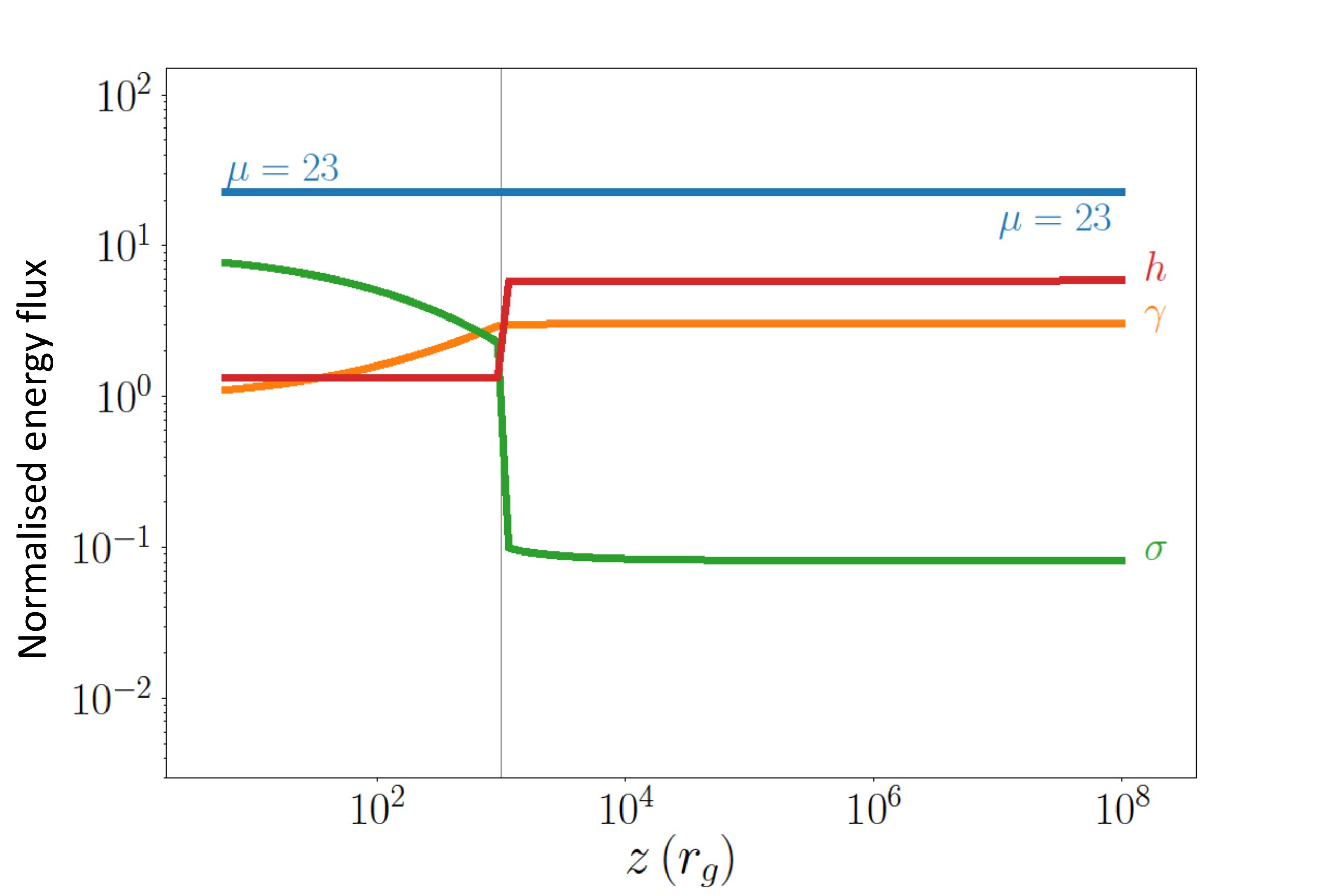}
    }
    \caption{The evolution of the different energy components \g, $\sigma$ and $h$ as indicated in each subplot, and the total $\mu$ based on equation~(\ref{eq: mu}). All subplots are for an average electron Lorentz factor of $\langle \varepsilon_{\rm e}\rangle =6$ and the jet content is shown in each subplot.}
    \label{app:fig: mu for leptonic geav=6 all plots appendix}
\end{figure*}

\begin{figure*}
    \centering
    \subfigure[$\eta_e = 10$]{
        \includegraphics[width=1.0\columnwidth]{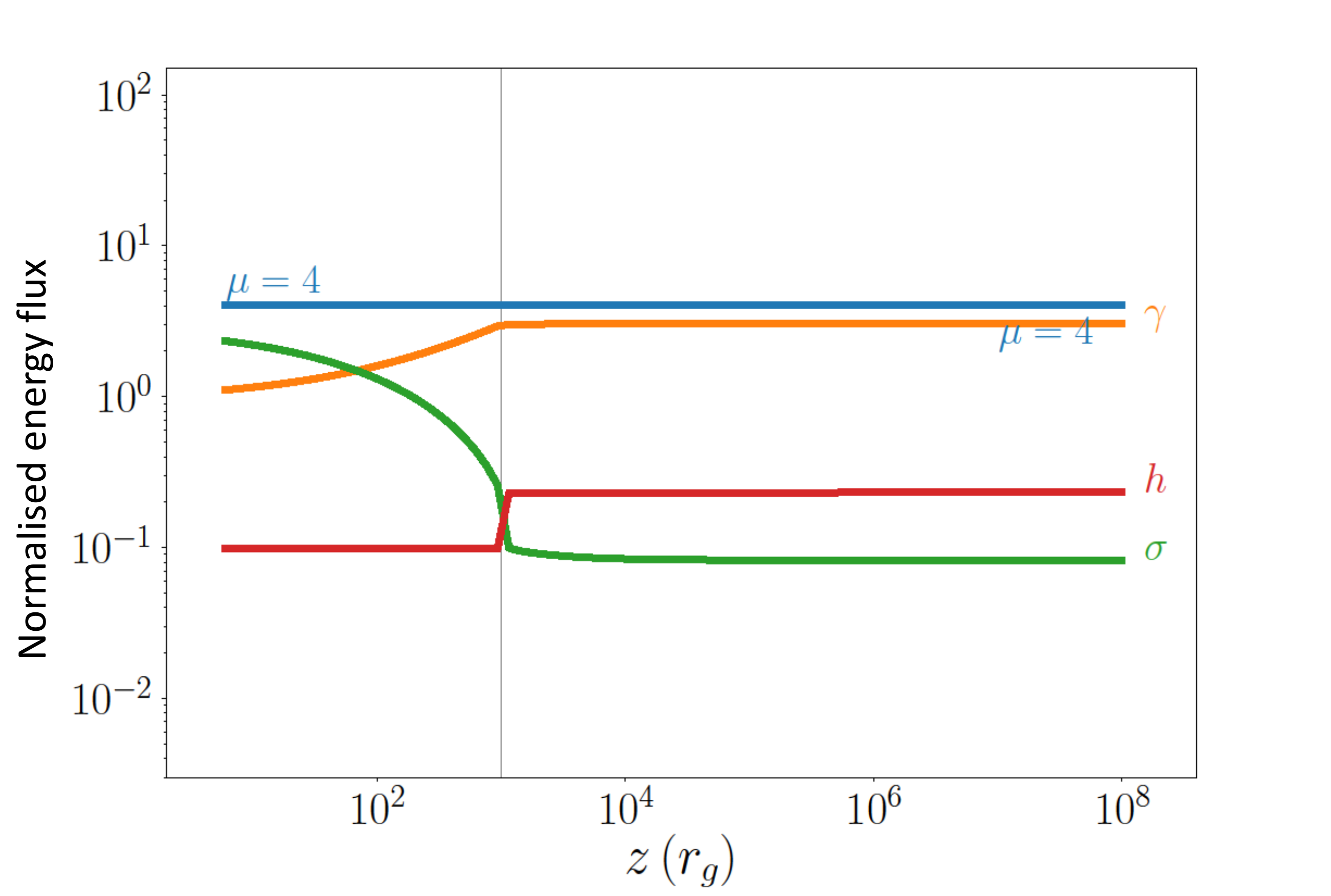}
    }
    \subfigure[$\eta_e = 100$]{
        \includegraphics[width=1.0\columnwidth]{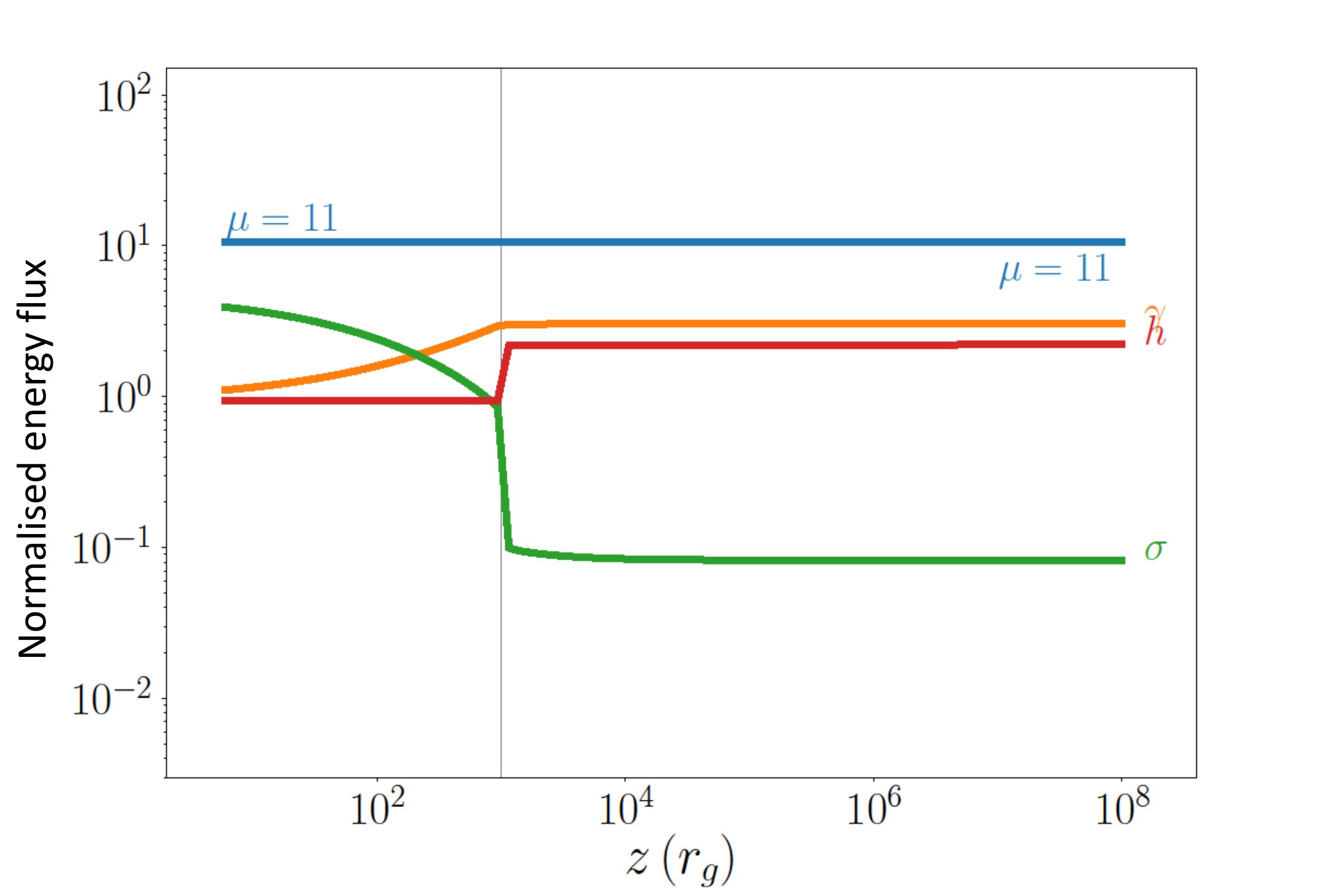}
    }
    \subfigure[$\eta_e = 1000$]{
        \includegraphics[width=1.0\columnwidth]{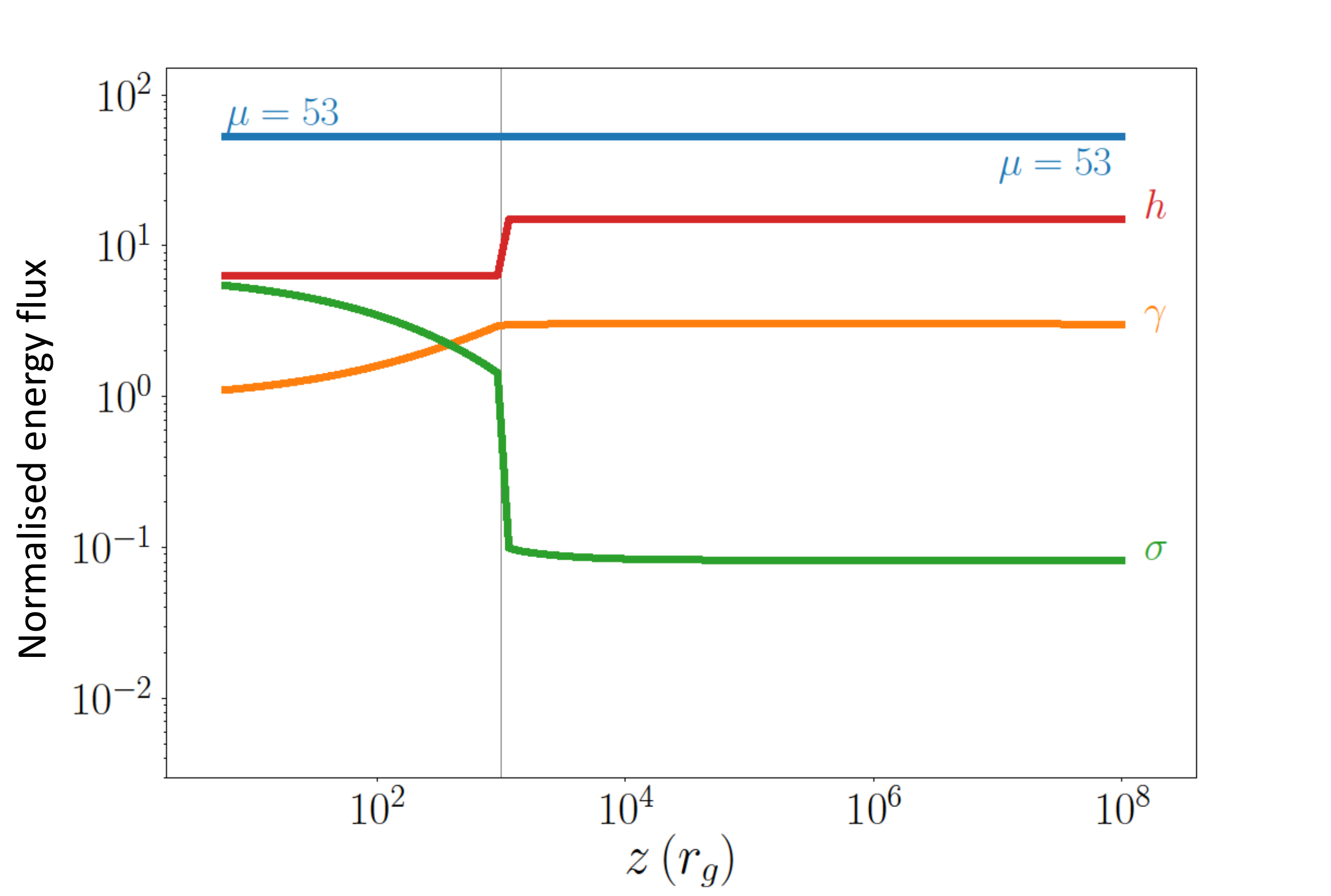}
    }
    \subfigure[$\eta_e = 10000$]{
        \includegraphics[width=1.0\columnwidth]{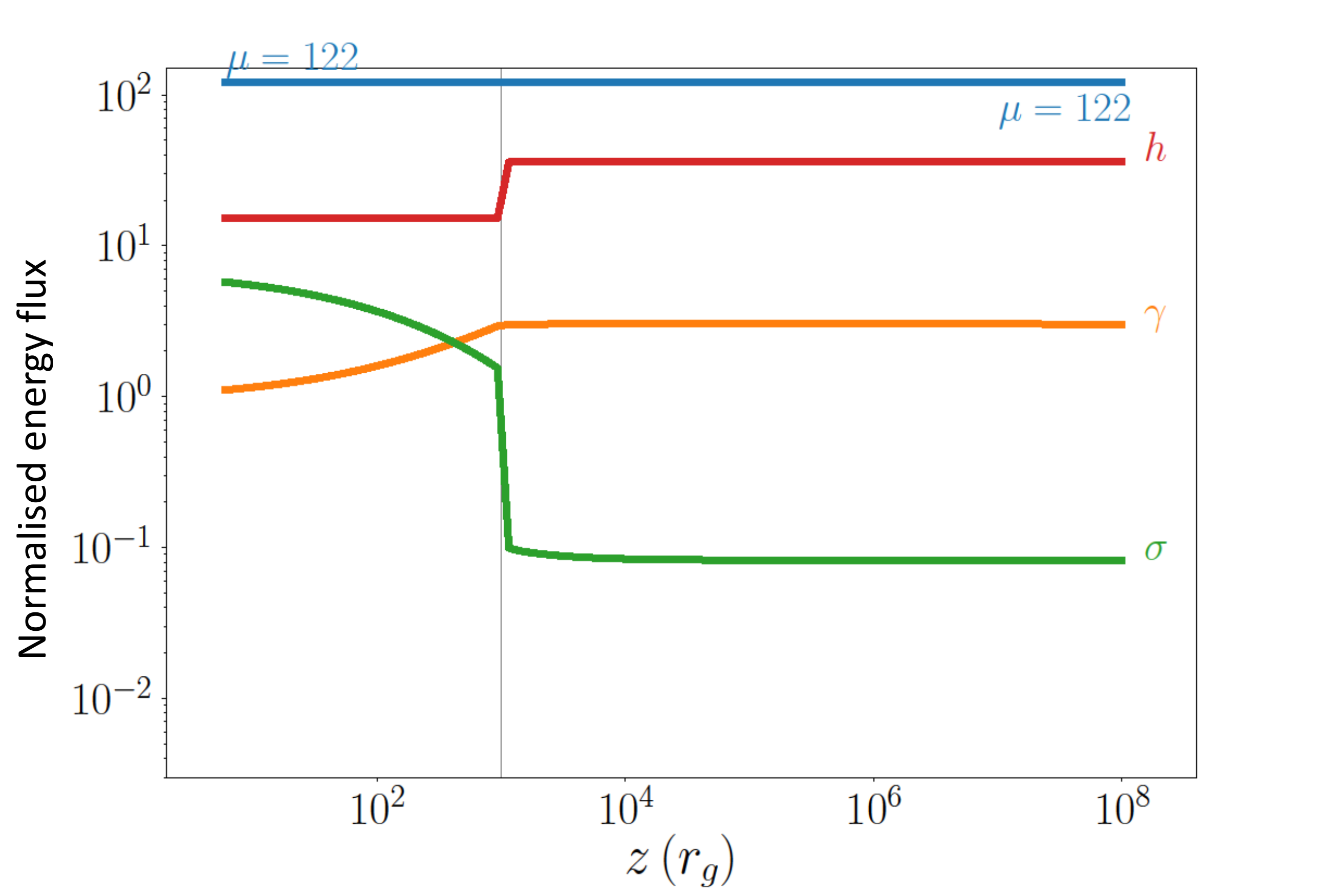}
    }
    \caption{Same as Fig.~\ref{app:fig: mu for leptonic geav=6 all plots appendix} but for $\langle \varepsilon_{\rm e}\rangle =32$.}
    \label{app:fig: mu for leptonic geav=32 all plots appendix}
\end{figure*}

\begin{figure*}
    \centering
    \subfigure[$\eta_e = 10$]{
        \includegraphics[width=1.0\columnwidth]{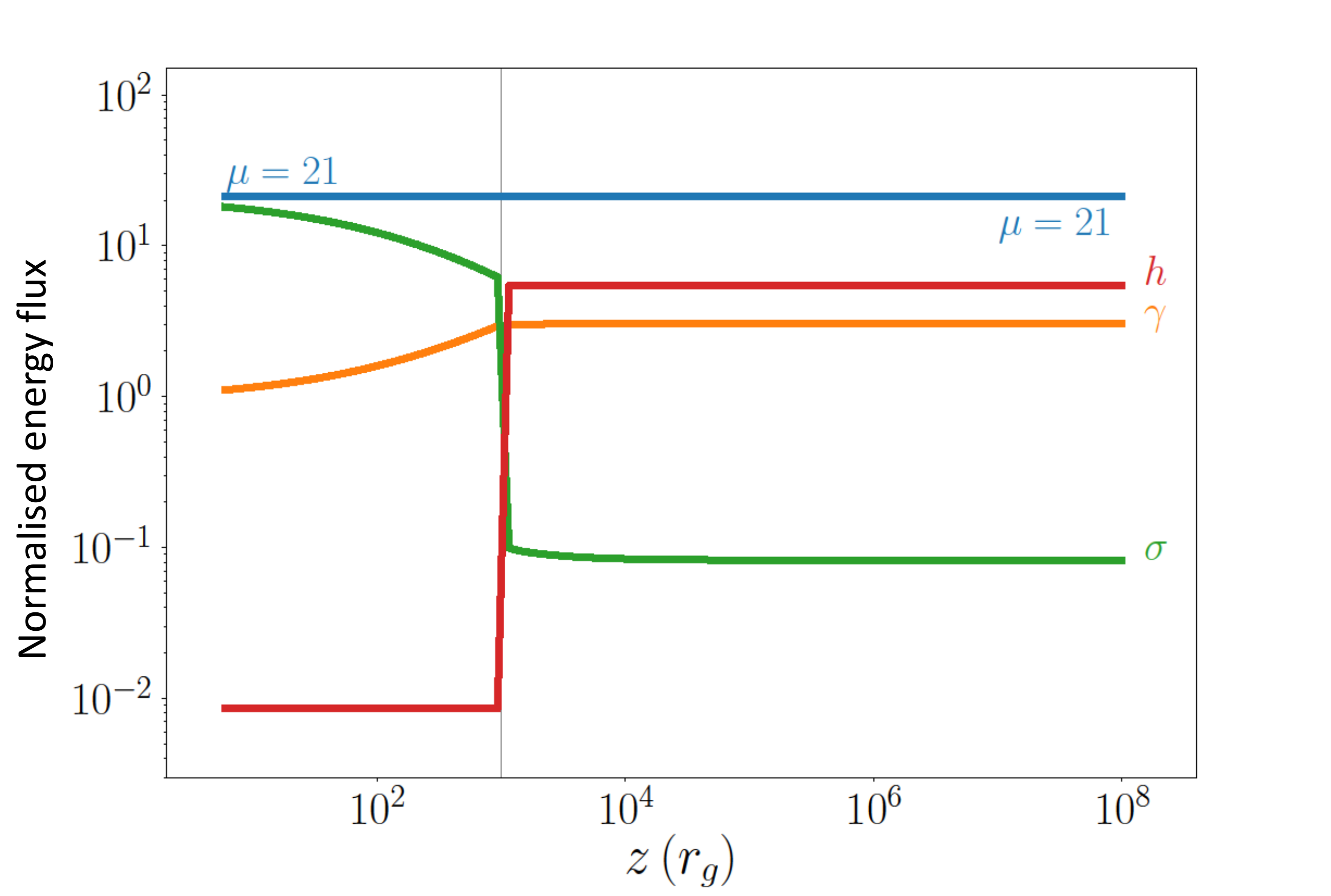}
    }
    \subfigure[$\eta_e = 100$]{
        \includegraphics[width=1.0\columnwidth]{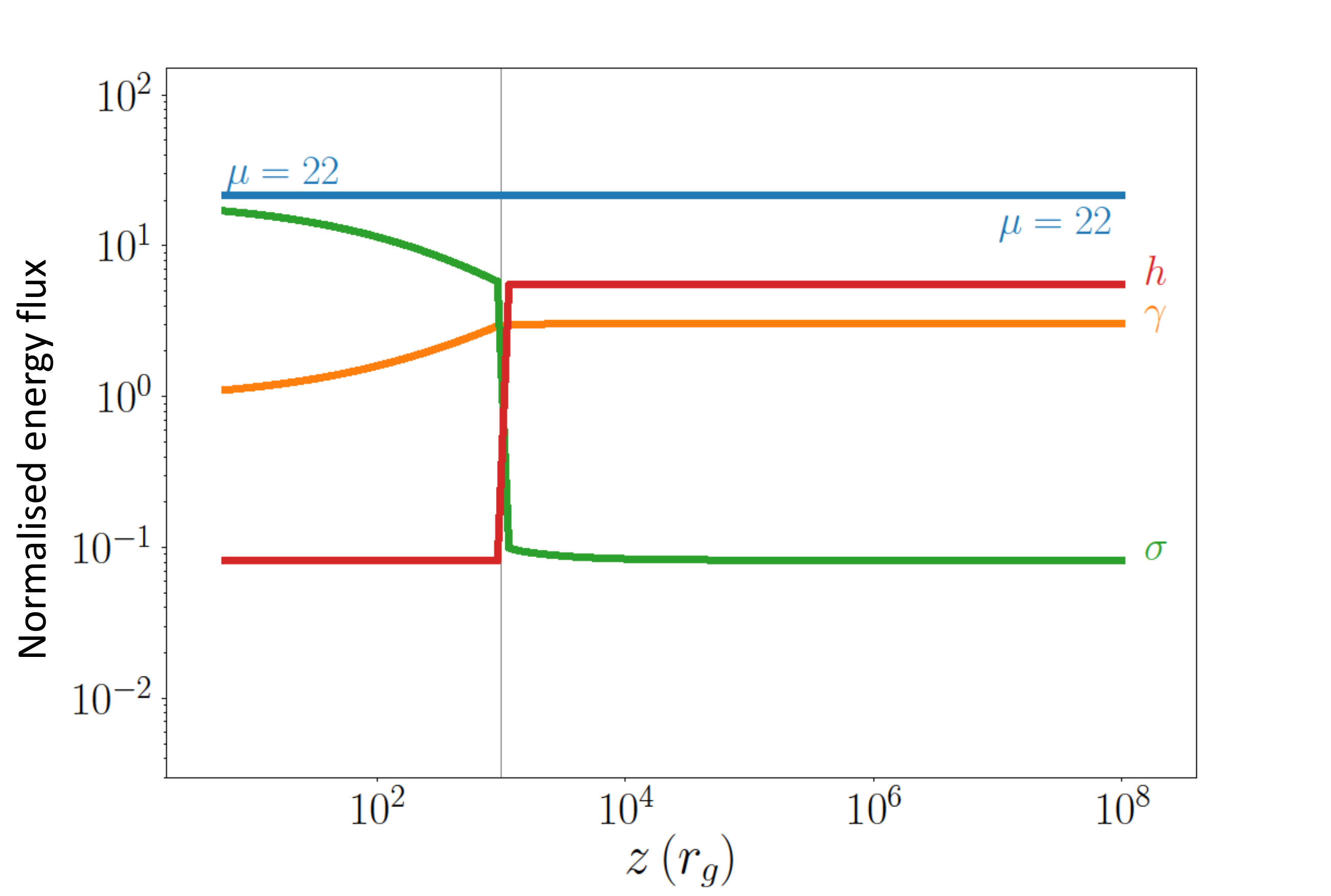}
    }
    \subfigure[$\eta_e = 1000$]{
        \includegraphics[width=1.0\columnwidth]{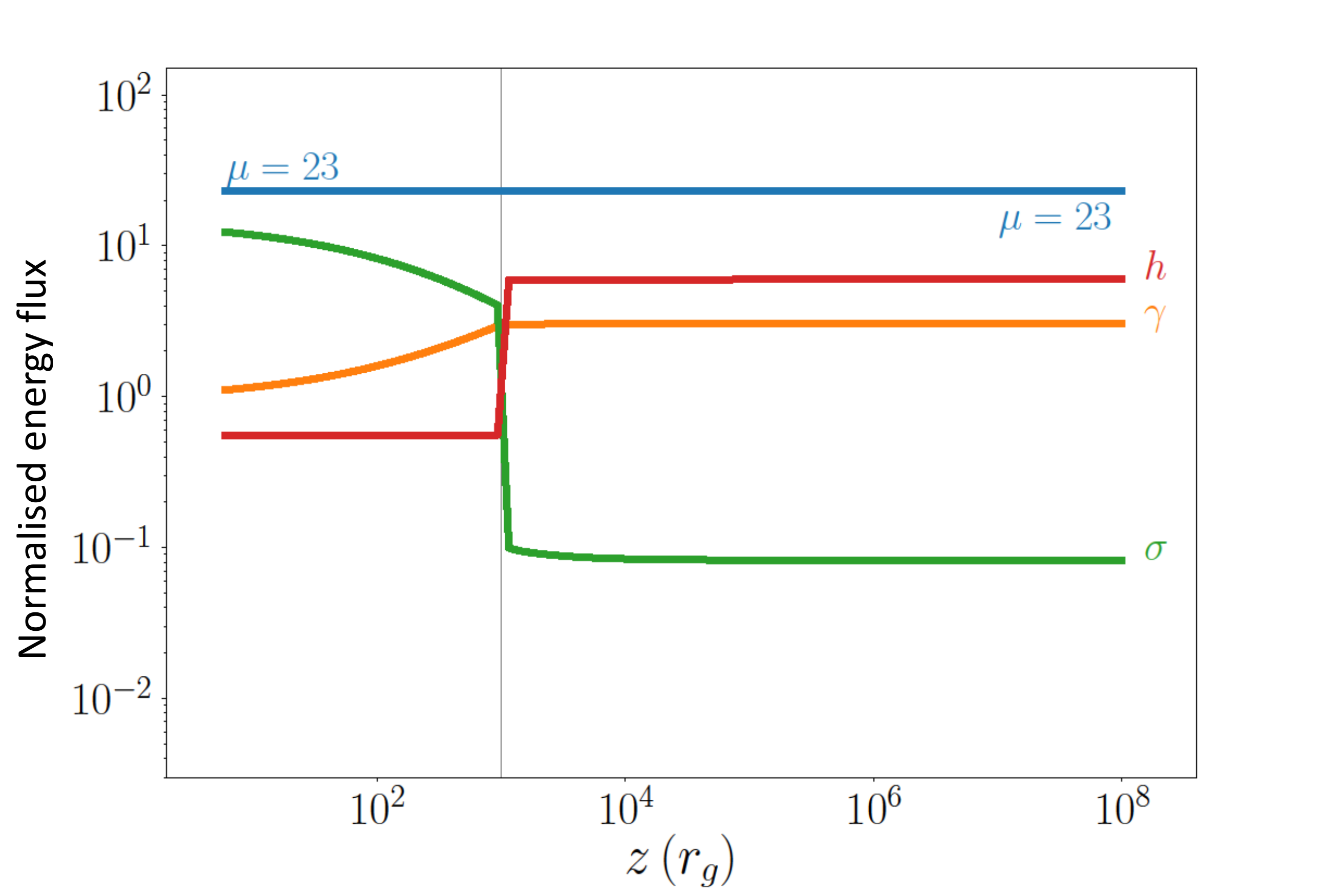}
    }
    \subfigure[$\eta_e = 10000$]{
        \includegraphics[width=1.0\columnwidth]{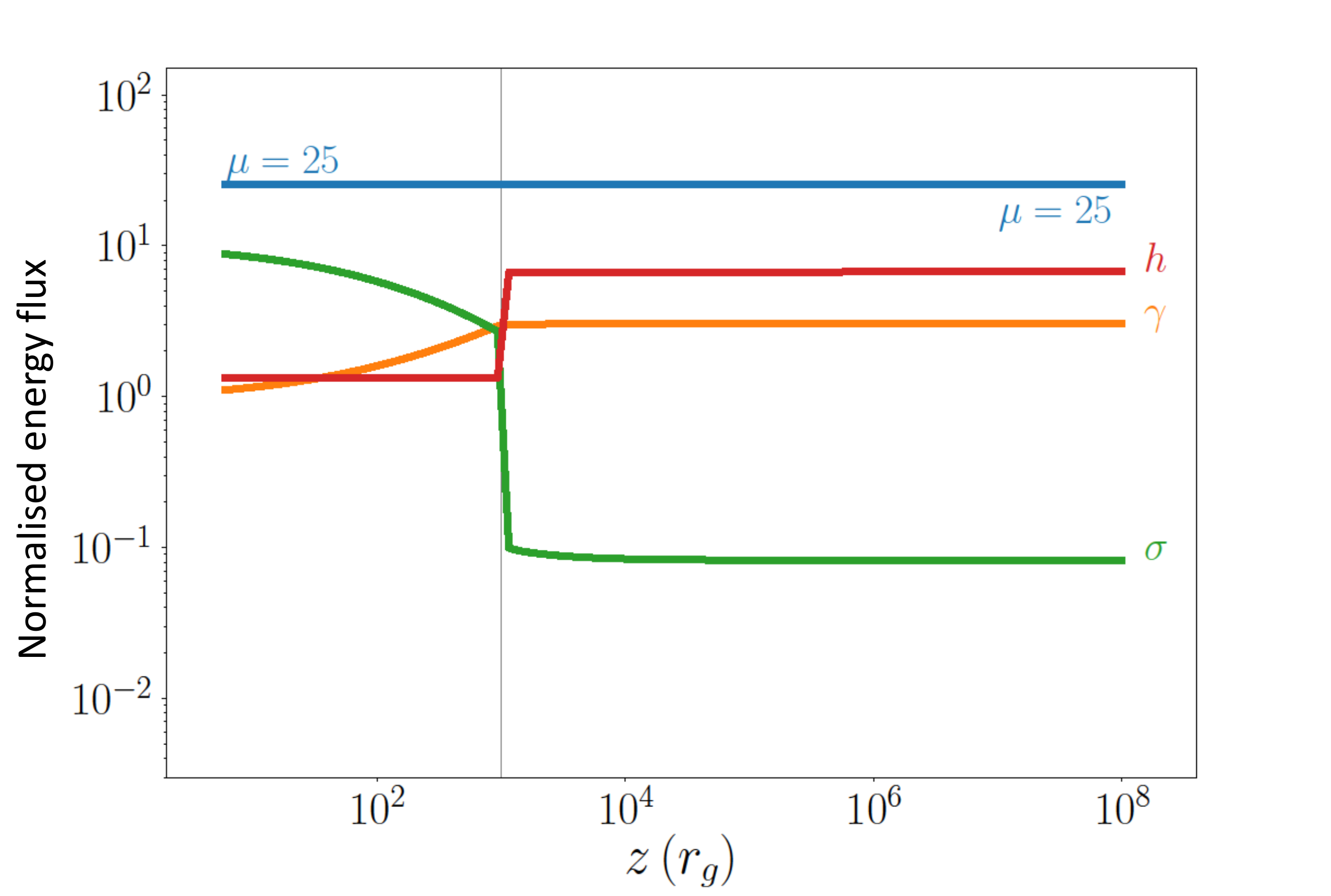}
    }
    \caption{Same as Fig.~\ref{app:fig: mu for leptonic geav=6 all plots appendix} but for $\langle \varepsilon_{\rm e}\rangle =6$. We further account for hadronic acceleration with $\langle \varepsilon_{\rm p}\rangle =4$.}
    \label{app:fig: mu for hadronic geav=6 and gpav=4 all plots appendix}
\end{figure*}

\begin{figure*}
    \centering
    \subfigure[$\eta_e = 10$]{
        \includegraphics[width=1.0\columnwidth]{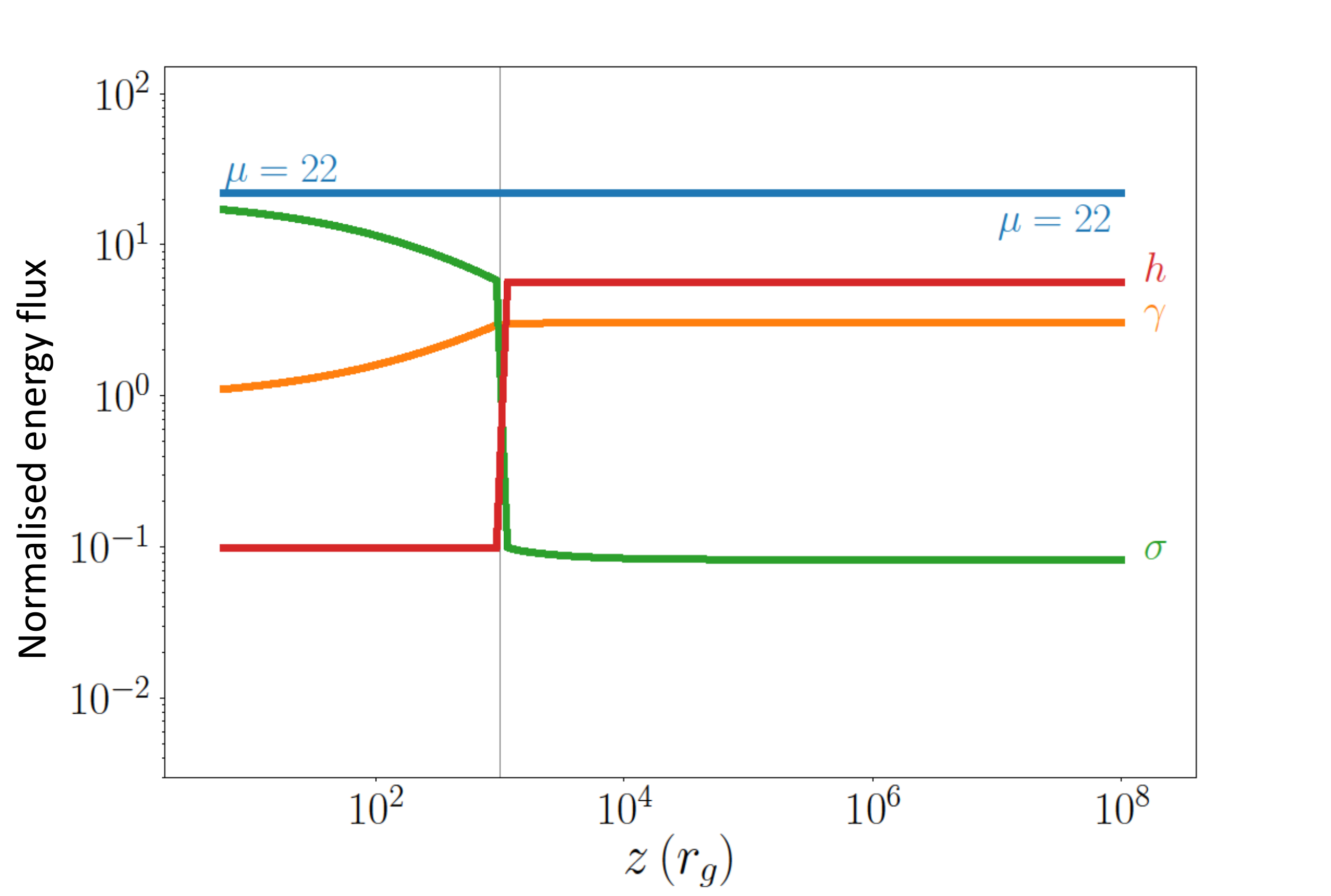}
    }
    \subfigure[$\eta_e = 100$]{
        \includegraphics[width=1.0\columnwidth]{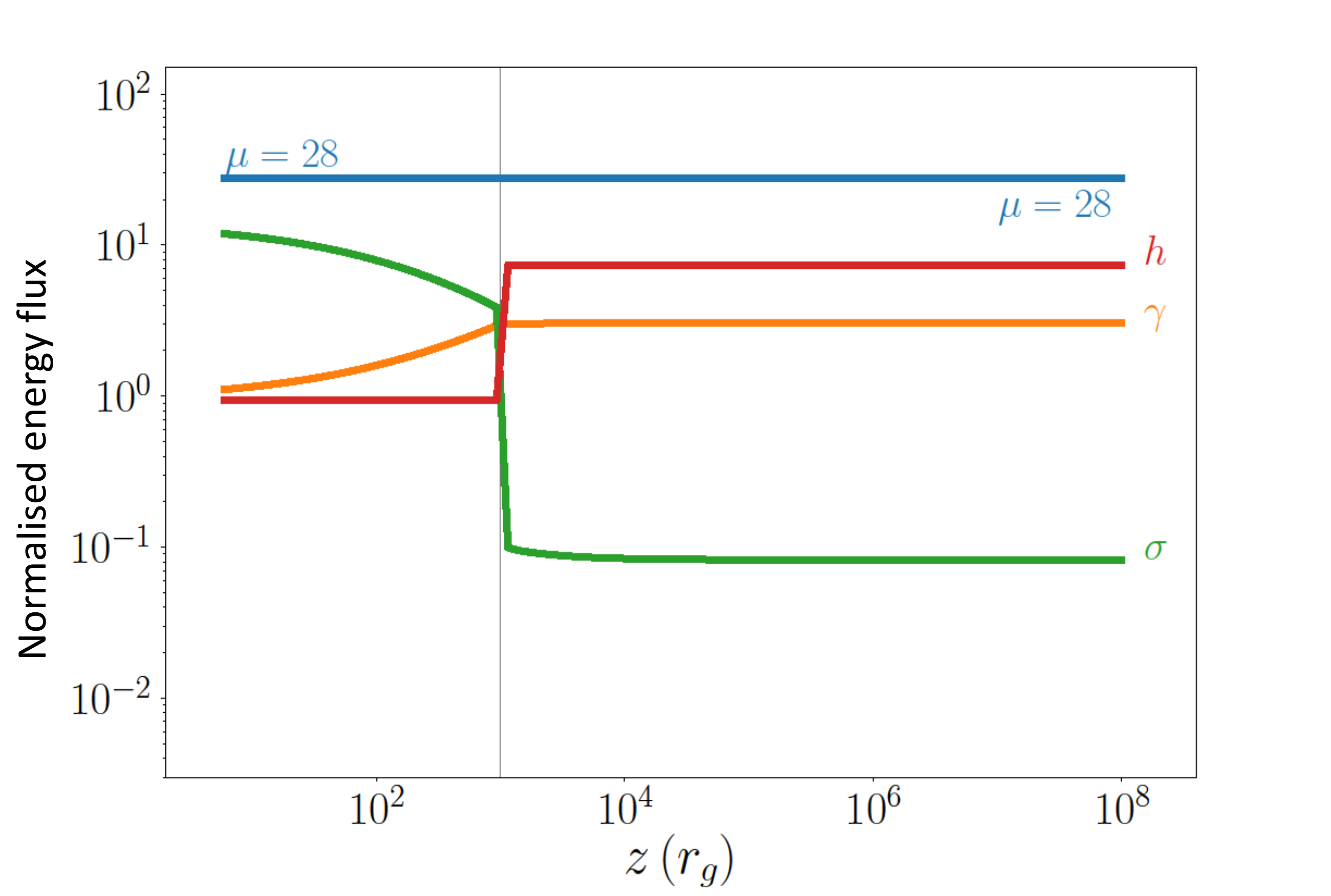}
    }
    \subfigure[$\eta_e = 1000$]{
        \includegraphics[width=1.0\columnwidth]{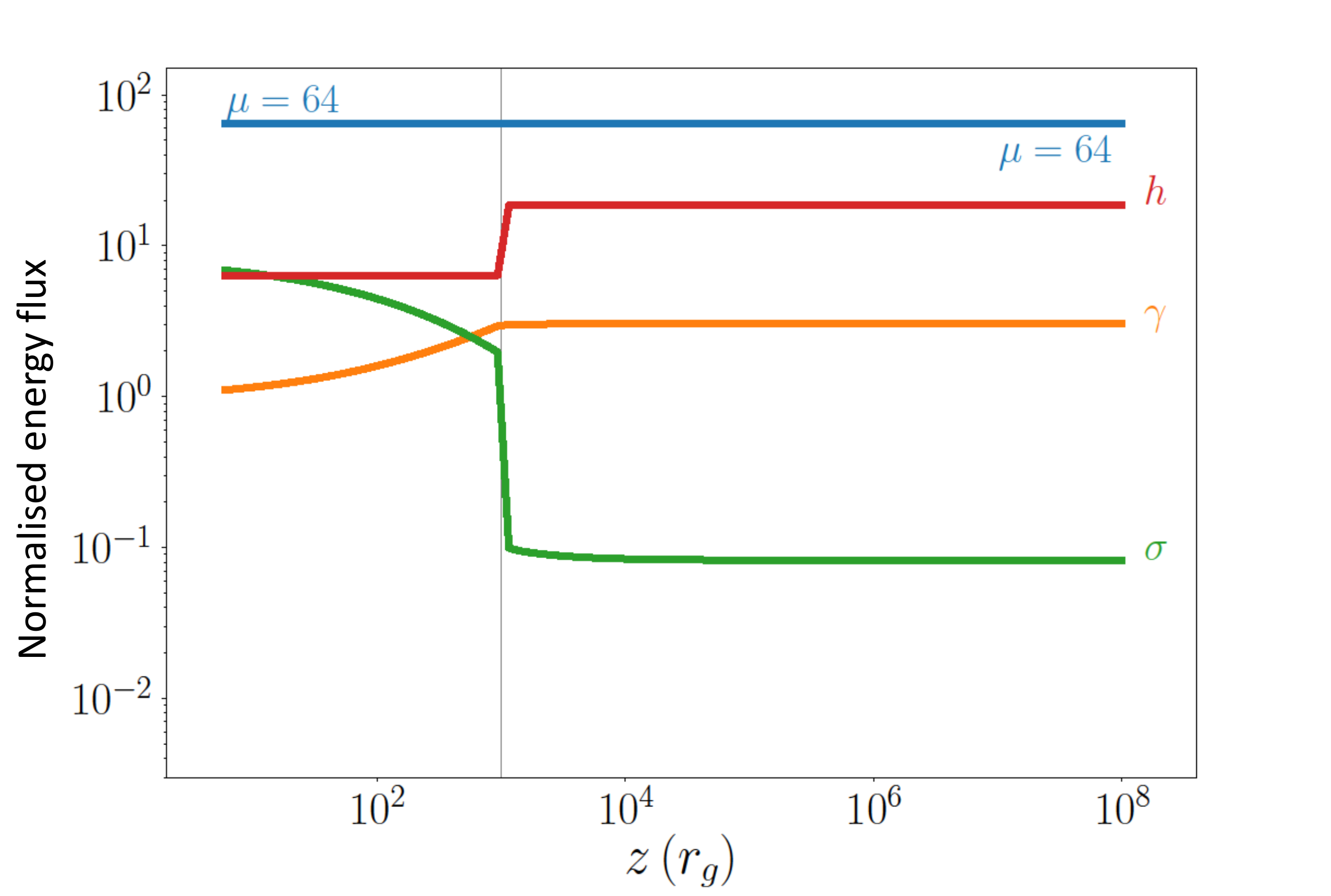}
    }
    \subfigure[$\eta_e = 10000$]{
        \includegraphics[width=1.0\columnwidth]{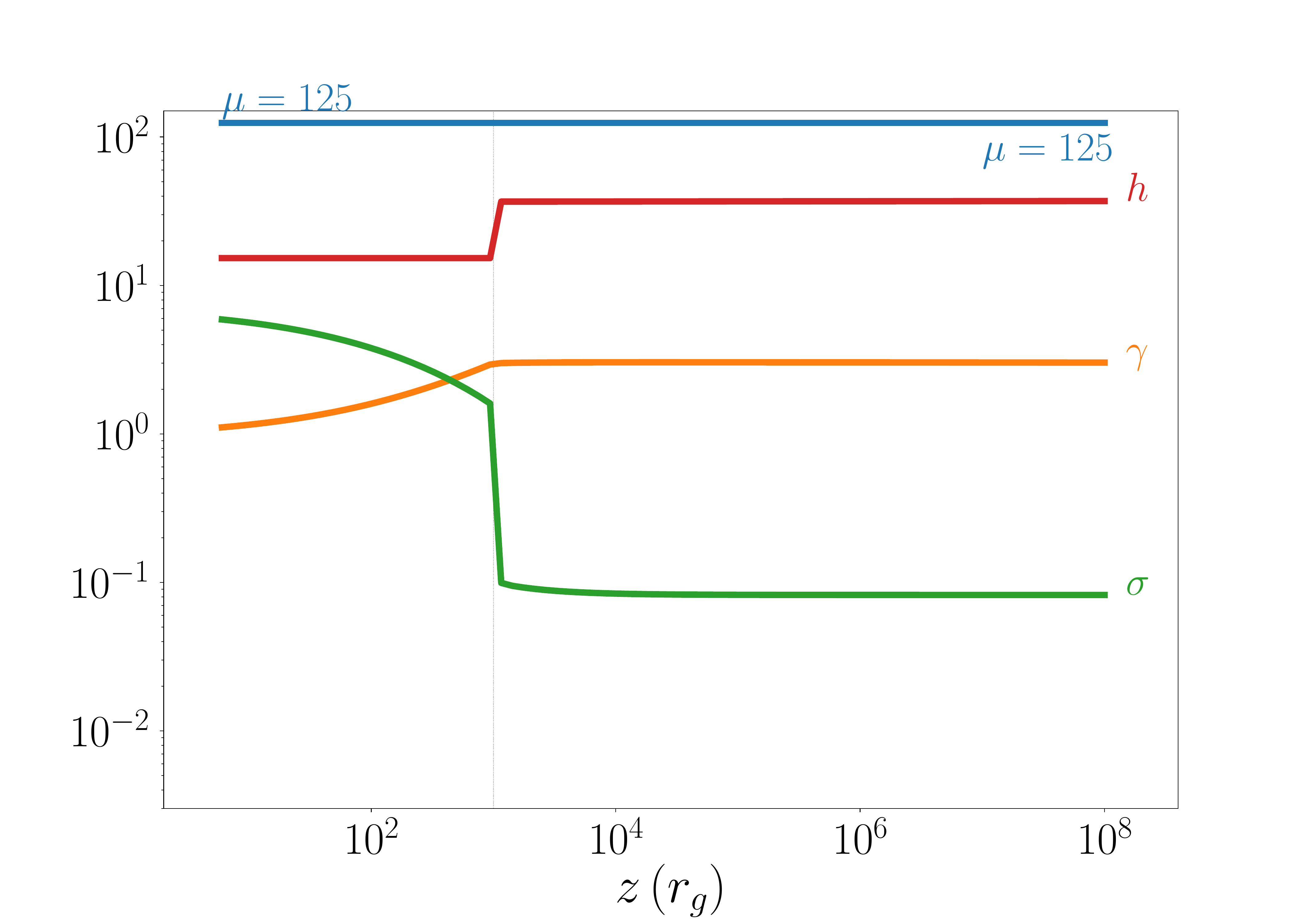}
    }
    \caption{Same as Fig.~\ref{app:fig: mu for hadronic geav=6 and gpav=4 all plots appendix} but for $\langle \varepsilon_{\rm e}\rangle =32$ and $\langle \varepsilon_{\rm p}\rangle =4$.}
    \label{app:fig: mu for hadronic geav=32 and gpav=4 all plots appendix}
\end{figure*}

In Figures~\ref{app:fig: mu for leptonic geav=6 all plots appendix}-\ref{app:fig: mu for hadronic geav=32 and gpav=4 all plots appendix}, we show the evolution of $\mu$ along the jet for different values of h. In particular, for all subplots of Figures~\ref{app:fig: mu for leptonic geav=6 all plots appendix}-\ref{app:fig: mu for hadronic geav=32 and gpav=4 all plots appendix}, we assume that the accelerated particles follow a power law with an index of $p$=2.2. The outflow launches at a distance of 6\,$r_g$ from the black hole and the particle acceleration initiates at $10^3\,r_g$. While the jets accelerate at some maximum Lorentz factor $\varepsilon_{\rm acc}=3$, we assume that the magnetisation at this region has dropped to $\sigma_{\rm acc}=0.1$. For every subplot, we assume $\eta_e=10$ (top left), $\eta_e=100$ (top right), $\eta_e=10^3$ (bottom left) and $\eta_e =10^4$ (bottom right) constant along the outflow. 

In Fig.~\ref{app:fig: mu for leptonic geav=6 all plots appendix}, we plot the jet evolution assuming only leptonic acceleration with an average Lorentz factor of $\langle \varepsilon_{\rm e} \rangle = 6$. in agreement with Fig.~\ref{fig: specific enthalpy}, we see that while the pair content increases in the jets, the specific enthalpy increases accordingly, and hence the total $\mu$ increases. In the cases of $\eta_e=10^3$ and $10^4$, in particular, we see that the specific enthalpy h has values comparable or even larger than the bulk Lorentz factor of the jet flow. 

In Fig.~\ref{app:fig: mu for leptonic geav=32 all plots appendix}, we plot a purely leptonic acceleration similar to Fig.~\ref{app:fig: mu for leptonic geav=6 all plots appendix} but assuming $\langle \varepsilon_{\rm e} \rangle = 32$ instead. The pair-dominated jets where $\eta_3=10^3$ and $10^4$ (bottom subplots), indicate that an efficient acceleration mechanism would lead to high values of h, which for the case of $\eta_e=10^4$ the overall value of $\mu$ is of the order of 100, a much higher value than commonly found in the literature. 

In Figures~\ref{app:fig: mu for hadronic geav=6 and gpav=4 all plots appendix} and \ref{app:fig: mu for hadronic geav=32 and gpav=4 all plots appendix}, we further account for hadronic acceleration with $\langle \varepsilon_{\rm p} \rangle = 4$. In the cases where the jets are pair-dominated, to obtain the specific enthalpy $h$ calculated at the particle acceleration region, we require a jet base that is Poynting flux dominated with a magnetisation of the order of 50-100.

\section{Artificial mass loss}\label{app: constrain h}
\begin{figure*}
    \centering
	\begin{minipage}{\columnwidth}
    \includegraphics[width=1.1\columnwidth]{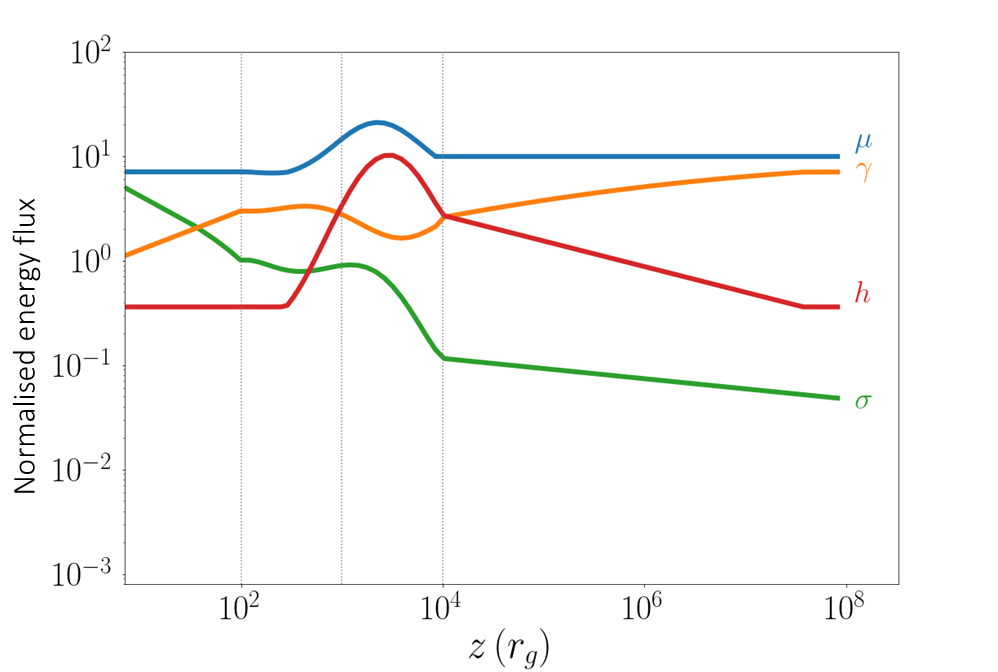}
    \end{minipage}
    \begin{minipage}{\columnwidth}
    \includegraphics[width=1.1\columnwidth]{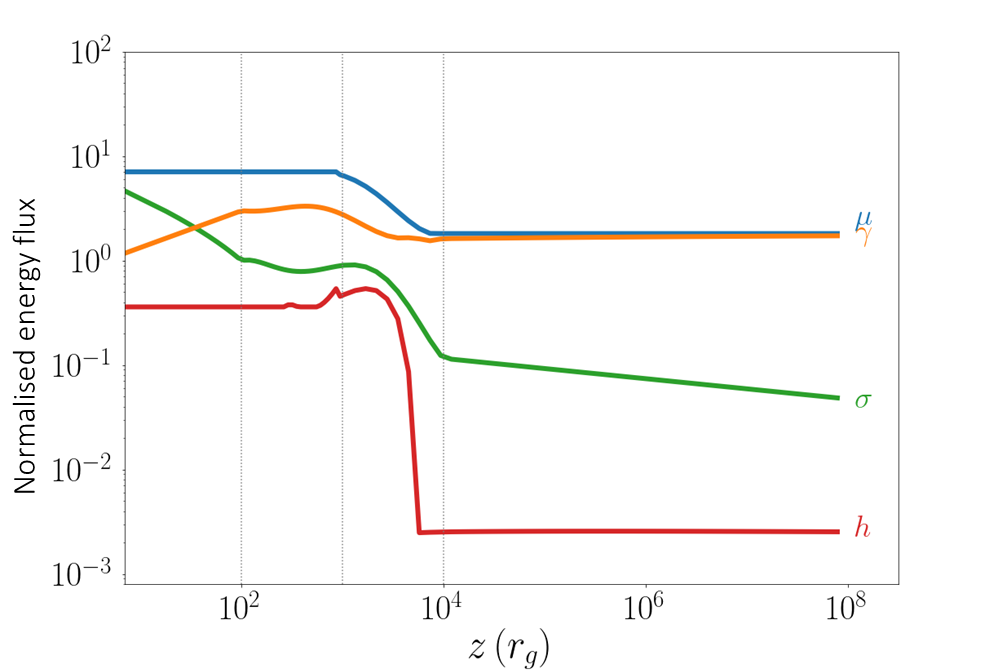}
    \end{minipage}
    \caption{The jet energy components similar to Fig.~\ref{fig: mu for mass loading, shifted and renormalised} but for the case of a jet base with $\eta_e = 10^5$. Following the description we discuss in Section~\ref{sec: mass loaded jets}, the particular profile of $h$ leads to an artificial increase of $\mu$ that would mean mass loss instead, which is unphysical (\textit{left}). Using the profile of $\mu$ from equation~\ref{app:eq: polynomial for mu}, we constrain $h$ to follow the mass-loading scenario (\textit{right}). The initial magnetisation is $\sigma_0=5$ and the Lorentz factor at the dissipation region is $\gamma_{\rm acc}=3$. 
} 
    \label{app:fig: constrain false increase for eta>>1000}
\end{figure*}

\begin{figure*}
    \centering
	\begin{minipage}{\columnwidth}
    \includegraphics[width=1.1\columnwidth]{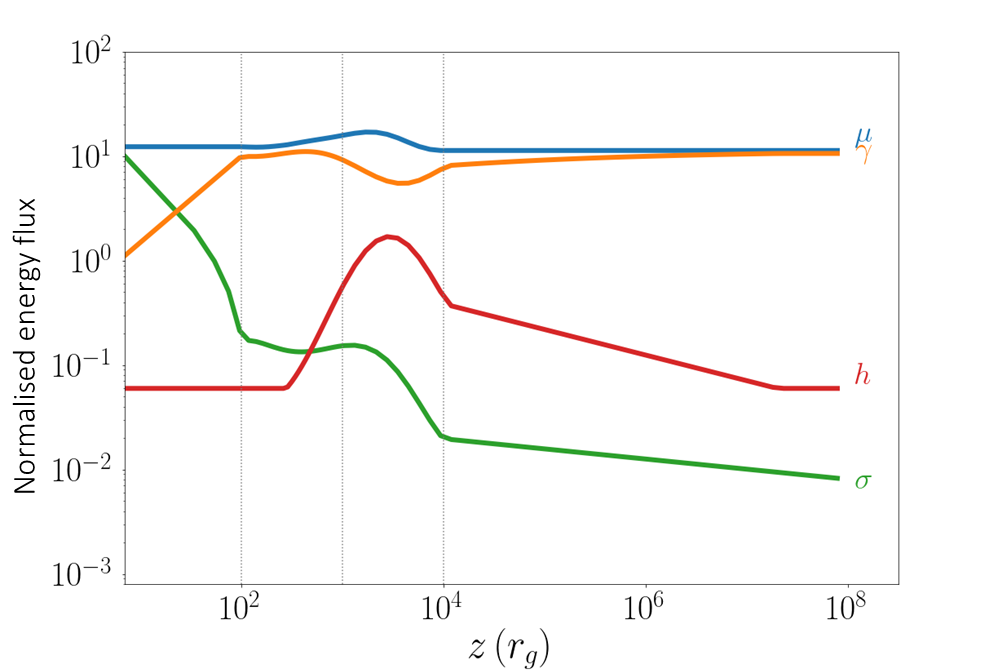}
    \end{minipage}
    \begin{minipage}{\columnwidth}
    \includegraphics[width=1.1\columnwidth]{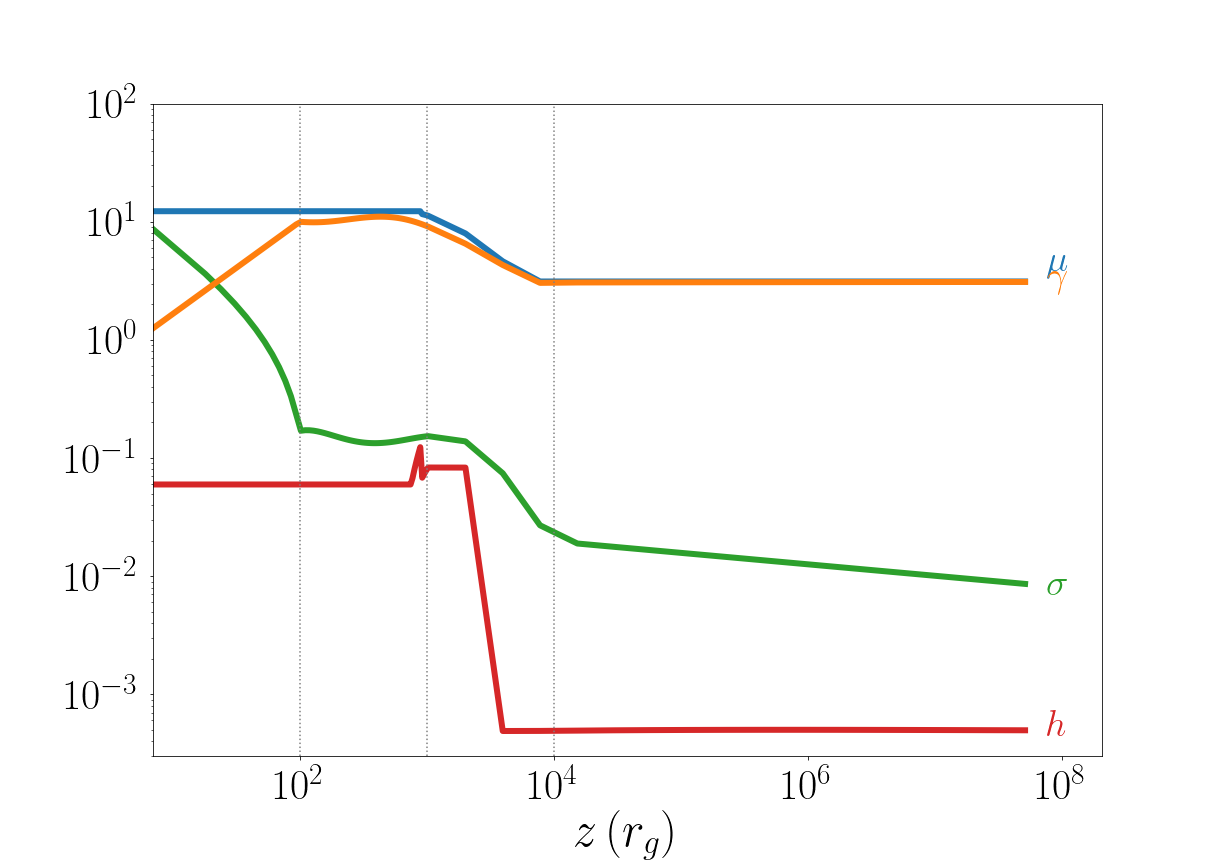}
    \end{minipage}
    \caption{Similar to Fig.~\ref{app:fig: constrain false increase for eta>>1000} but for the case of $\gamma_{\rm acc}=10$, $\sigma_0=10$ and $\eta_e=10^4$. 
}
    \label{app:fig: constrain false increase for gacc 10}
\end{figure*}

In Fig.~\ref{app:fig:  constrain false increase for eta>>1000} we show how the increase of the specific enthalpy $h$ in the mass-loading region may lead to a unphysical increase of $\mu$ that would mean mass loss instead. Such an artificial mass loss is due to the fact that we assume a hot flow and/or a pair dominated jet base with $\eta_e \gg 1000$. To avoid such a condition, we first calculate the value of $\mu$ from the 5$^{\rm th}$ order polynomial
\begin{equation}\label{app:eq: polynomial for mu}
    \begin{split}
        \log_{10}(\mu) =
0.2231\, x^5 - 0.7242\, x^4 + 0.4546\, x^3 + 0.104\, x^2\\ - 0.09267\, x + 1.031,
    \end{split}
\end{equation}
based on the results of \cltm, and then we calculate $h$ from the equation $h=\mu/\gamma - (\sigma+1)$, where the values of $\gamma$ and $\sigma$ are from equations~\ref{eq: polynomial for gamma} and~\ref{eq: polynomial for sigma}, respectively. In the cases that the specific enthalpy $h$ would drop to zero (see for instance Fig.~\ref{app:fig: constrain false increase for eta>>1000} and ~\ref{app:fig: constrain false increase for gacc 10}, we choose to set $h$ to a very low value, namely one 100th of the magnetisation.  In the above equation, $x$ is the same as in Section~\ref{sec: mass loaded jets}.

Finally, in Fig.~\ref{app:fig: constrain false increase for gacc 10} we plot the scenario where the artificial mass loss is due to a combination of a large Lorentz factor ($\gamma_{\rm acc}=10$) and the profile of $h$ from equation~\ref{eq: h specific enthalpy polynomial}. The above assumption allows forcing a mass-loading scenario.

\section{SED components}\label{app: sed components}
\begin{figure*}
    \centering
    \includegraphics[width=1.1\columnwidth]{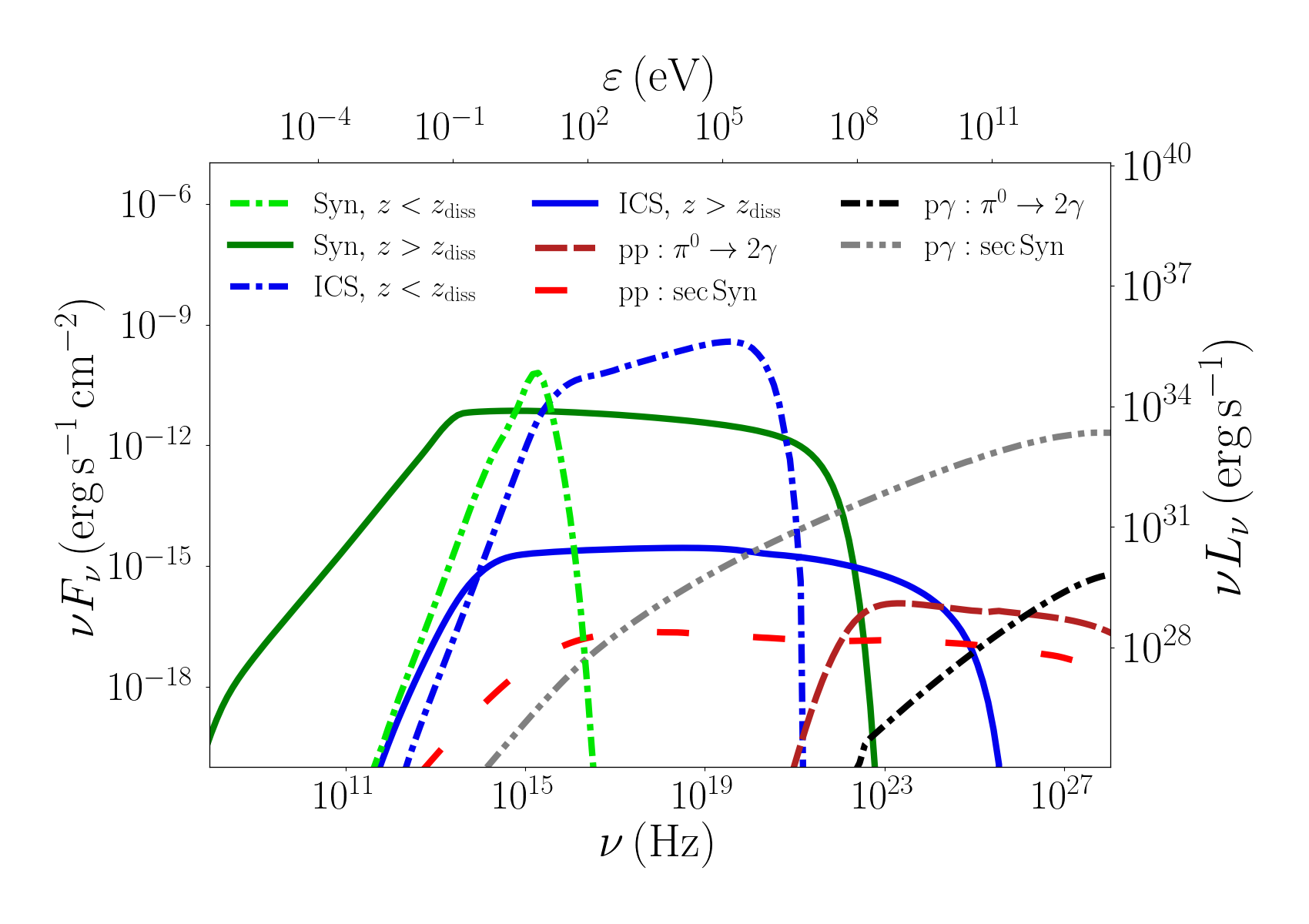}
    \caption{Identical to the right subplot of Fig.~\ref{fig: SEDs for mass loading jets} but in this plot we show the individual radiative components as indicated in the legend. 
}
    \label{app:fig: sed with components}
\end{figure*}

In Fig.~\ref{app:fig: sed with components} we show the spectrum as presented on the right subplot of Fig.~\ref{fig: SEDs for mass loading jets} but with the individual components instead. 

\section{Proton power}\label{app: proton power}
\begin{figure*}
    \centering
	\begin{minipage}{\columnwidth}
    \includegraphics[width=1.05\columnwidth]{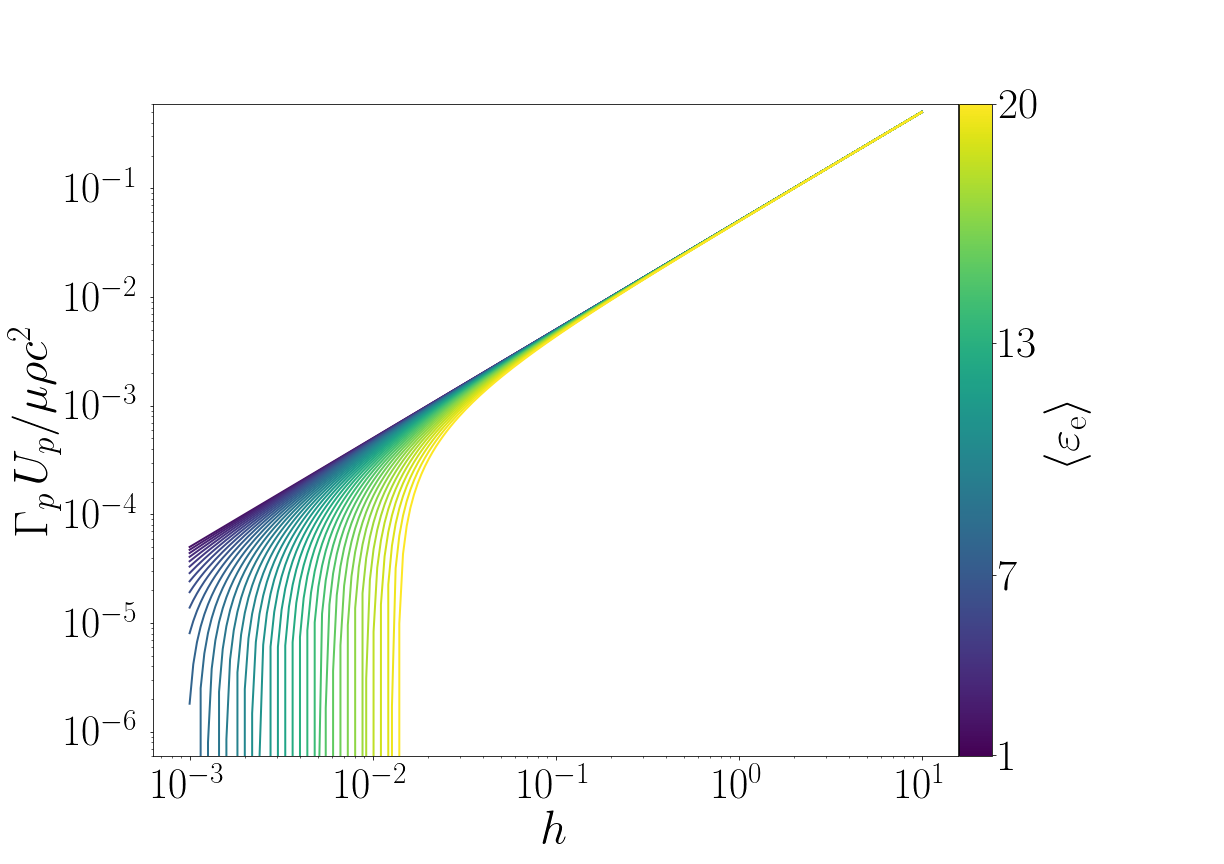}
    \end{minipage}
    \begin{minipage}{\columnwidth}
    \includegraphics[width=1.05\columnwidth]{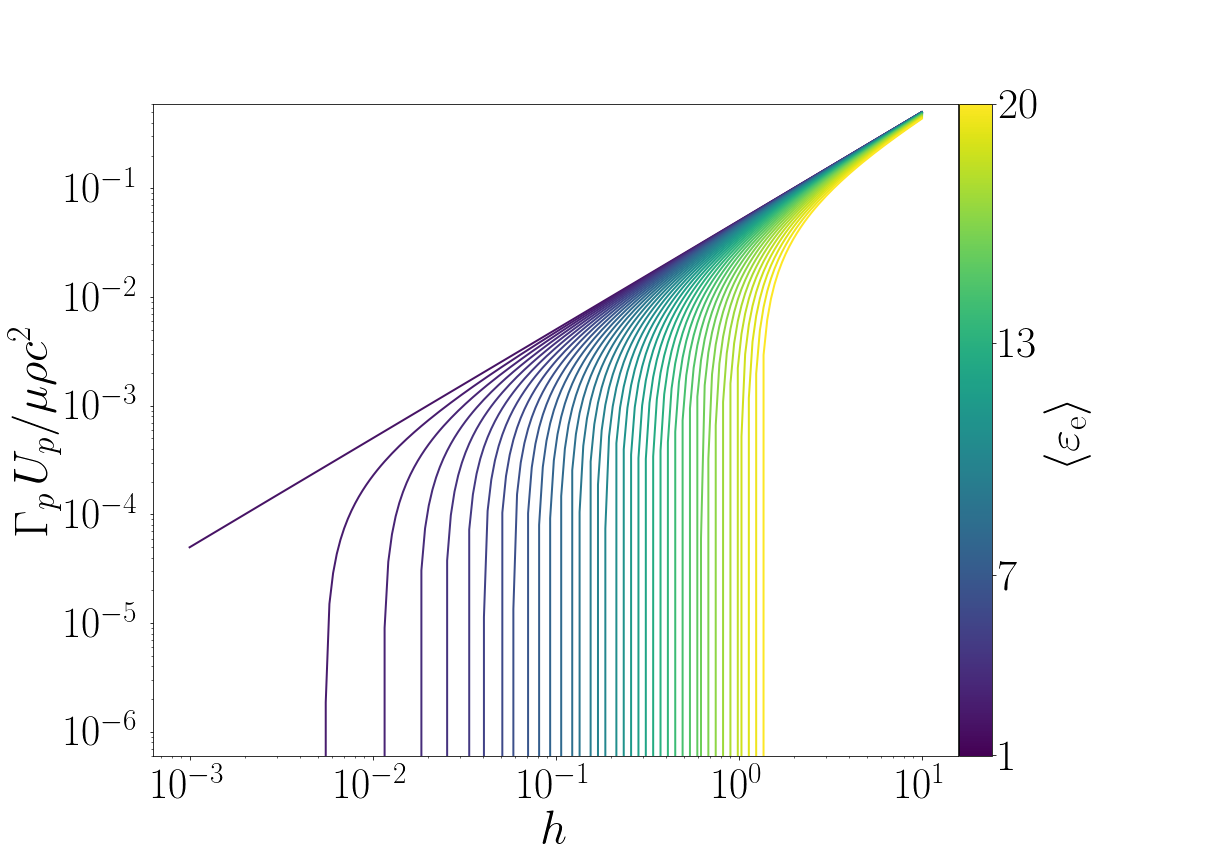}
    \end{minipage}
    \caption{
    The specific enthalpy of the protons $\Gamma_p U_p/\rho c^2$ divided by $\mu$ shows the total energy that is allocated to protons with respect to the total available jet energy, as a function of the jet specific enthalpy $h$. We plot the proton energy density for a number of different electron energy densities that correspond to different values of $\langle \varepsilon_e \rangle$ as shown in the colormap, and we use $\eta_e=1$ in the \textit{left}, and $\eta_e = 100$ in the \textit{right}. 
} 
    \label{app:fig: proton power for different etas}
\end{figure*}

In Fig.~\ref{app:fig: proton power for different etas}, we plot the fraction of the energy that is allocated to proton acceleration with respect to the total available energy flux of the jet $\mu$. We plot this quantity versus the total specific enthalpy of the jet $h$ for different average Lorentz factors $\langle \varepsilon_e \rangle$ of the electrons. In the main text, we included the case where $\eta_e=10$ and here we plot the cases where $\eta_e=1$ (\textit{left}) and $\eta_e =100$ (\textit{right}), for completeness. See Section~\ref{sec: discussion on proton power} for further information.

\bsp	
\label{lastpage}
\end{document}

%% file: Sections/Introduction.tex
\section{Introduction}

Accreting black holes can efficiently launch relativistic outflows, known as astrophysical jets, by converting gravitational energy to kinetic energy. Large-scale jets launched by supermassive black holes (SMBH) share some common physical laws to the small-scale jets launched by stellar-mass black holes in X-ray binaries (\bhs; \citealt[][]{Heinz2003dependence,Merloni2003fundamental,Falcke2004FP}), and hence black hole jets appear to be scale invariant in some of their properties. For example, SMBHs with masses of the order of $\sim 10^6-10^9\,M_{\rm \odot}$ power jets that remain collimated up to Mpc scales \citep[][]{waggett1977ngc}, whereas \bhs with mass of the order of a few solar masses display jets that remain collimated up to sub-pc scales \citep{mirabel1994superluminal}. Galactic \bhs are of particular importance because they transition between different jetted and non-jetted states over human-like timescales, giving us the chance to understand plasma evolution in extreme conditions and better probe jet physics \citep[see, e.g.,][]{markoff2001jet,markoff2003exploring,Markoff2005,Reig2003energy,Giannios2004spectral,maitra2009constraining,Vila2010gx339,zdziarski2014jet,Connors2019combining,Lucchini2021correlation}. 

The exact physical mechanism responsible for jet launching is not clear yet. On one hand, the Blandford-Znajek mechanism \citep{blandford1977extraction} describes a way to extract the rotational energy of a spinning black hole and power relativistic jets that can be pair-plasma dominated \citep[see, e.g.,][]{Broderick_2015,Parfrey2019}
On the other hand, magnetic fields anchored in the accretion disc can launch baryon/proton/ion-dominated jets via the Blandford-Payne mechanism \citep[][]{Blandford1982hydromagnetic}. The difference in jet composition from the two launching mechanisms would have an important impact on the interpretation of the spectral energy distribution (SED) observed from such black hole systems as well as the consideration of relativistic jets as candidate sources of cosmic rays (CRs). 

CRs are charged particles that exhibit a large range of energies going up to ultra-high energies of the order of $10^{20}\,\rm eV$ \citep[][]{Auger201818,Abbasi2020TA}. The detected CR spectrum shows two very prominent features, known as the ``knee''and the ``ankle'' where the spectrum steepens and hardens, respectively. The ``knee'' is observed at $10^{15}\,\rm eV$ (PeV) and is likely to be the maximum energy that CR protons accelerated in Galactic sources can reach, but the identification of these particular sources remains a mystery despite the decades of studies. The ``ankle'', located at $\sim 10^{18}\,\rm eV$ (EeV), is where extragalactic sources are thought to start dominating the spectrum. The exact CR composition is not clear and strongly depends on the particle energy. 
GeV CRs primarily comprise of protons ($\sim99$~per~cent; \citealt[][]{SHIKAZE2007154}), with electrons and positrons mainly contributing to the rest of the spectrum. It is likely that heavier elements/ions accelerated in Galactic sources start dominating the CR spectrum between the ``knee'' and the ``ankle'' \citep[][]{ALOISIO2012129}, beyond which the composition is unclear \citep[][]{TA2019mass,yushkov2019mass,Corstanje2021depth}.

Similar to large-scale jets of active galactic nuclei (AGN), which are among the dominant candidate sources of the extragalactic CRs \citep[][]{Protheroe1983}, recent studies suggest the small-scale jets of \bhs as potential CR acceleration sites \citep[][]{romero2003hadronic,fender2005CRXRBs,cooper2020xrbcrs}. There are currently only a few tens of Galactic \bhs detected in the Milky Way \citep[][]{Tetarenko2016}, but population-synthesis simulations \citep[see, e.g.,][]{Olejak2019synthesis} suggest that a few thousand black holes likely reside in the Galactic disc, in agreement with the recent X-ray observations of the Galactic centre by \cite{Hailey2018cusp} and \cite{Mori_2021}. Based on such observations, \cite{cooper2020xrbcrs} proposed that a few thousand \bhs are capable of contributing to the observed CR spectrum above the ``knee''.

Whether or not \bhs jets can indeed accelerate CRs up to the ``knee'', and AGN jets beyond the ``ankle'', strongly depends on two further issues: (1) can astrophysical jets, in general, accelerate particles to high energies, and (2) are astrophysical jets actually comprised of protons and/or heavier elements? On the former, observations of non-thermal emission from radio bands \citep[see, e.g.,][]{Lister2016mojave} up to GeV/TeV \grs from both SMBHs \citep[see, e.g.,][]{Lister_2009} and \bhs \citep[see, e.g.,][]{zanin2016detection}, suggest that both classes of jets can efficiently accelerate particles. Numerous numerical studies, moreover, suggest that jets can indeed be viable sites of particle acceleration either via shocks \citep[][]{hillas1984origin}, or via magnetic reconnection \citep[][]{Drenkhahn2002,Guo2014relativisticreconnection,Sironi_2014,matthews2020particle}. 

The jet composition however remains an open question. The two different proposed launching mechanisms mentioned above yield an entirely different jet content at the base that significantly alters not only the jet dynamics, but the emitted spectrum as well \citep[][]{Petropoulou_2019}. A pair-dominated jet would allow only for leptonic processes, such as synchrotron and inverse Compton scattering (ICS; \citealt[][]{blumenthal1970bremsstrahlung}). A leptonic plus hadronic jet, on the other hand, allows for further non-thermal processes, when inelastic collisions occur between the accelerated protons and the cold flow or radiation \citep[e.g.,][]{mannheim1993proton,rachen1993extragalactic,mannheim1994interactions,rachen1998photohadronic}. Such hadronic processes can lead to the production of astrophysical neutrinos, but usually require a much larger jet energy budget than the leptonic ones, sometimes requiring super-Eddington jet powers \citep[][]{boettcher2013leptohadronic,Liodakis2020}. Such super-Eddington powers challenge the accretion paradigm \citep[][]{zdziarski2015hadronic}, but they still seem feasible for relativistic AGN jets \citep[][]{ghisellini2014power}.

 Several \bh jets, such as the peculiar case of SS433 or the prototypical Cygnus~X--1, show evidence of baryonic jet content 
(\citealt[][]{2004ASPRv..12....1F} and \citealt[][]{gallo2005dark,Heinz_2006}, respectively). 
Both the compact objects of SS433 and Cygnus~X--1 are accompanied by a high mass donor star that may be the source of the heavy composition through its stellar wind. 
There is evidence of baryon-loaded jets though, even in the case of a low-mass companion, such as the black hole candidate 4U~1630--47, based on iron emission lines \citep[][]{trigo2013baryons}. The cases of MAXI~J1820+070 \citep[][]{tetarenko2021measuring,Zdziarski_2022_J1820}, MAXI~J1836-194 \citep[][]{Lucchini2021correlation}, 
XTE~J1752--223, MAXI~J1659--152, and XTE~J1650--500 \citep[][]{cao2021evidence}
on the other hand, favour a jet composition of the order of a few to a few tens of pairs per proton based on energetic arguments.

The composition is also difficult to constrain in extragalactic jets. Circular polarisation measurements indicate that the jets of the blazar 3C~279 are pair-dominated \citep[][]{Liodakis2022constraints}, and energetic arguments of the radio galaxy 3C~120 are consistent with a pair-dominated jet \citep[][]{zdziarski2022composition}.
\citet[][]{Celotti1993heavy}, on the other hand, based on very-large baseline interferometry and spectral arguments for numerous sources, support an electron-proton plasma. The blazar TXS~0506+056, finally, due to the correlation with the high-energy neutrino IceCube-170922A, supports a baryon content in its jets as well \citep[][]{Aartsen2018Multimessenger}.

Currently, the state-of-the-art to model jet launching and dynamics in a more \textit{a priori} way are high-resolution simulations that solve the magneto hydrodynamic equations in the general relativistic regime (GRMHD). Such simulations have furthered our understanding of the accretion-launching paradigm and have shown that a Poynting flux dominated outflow can convert a significant amount of its initial magnetic energy into kinetic energy to accelerate the bulk flow \citep[][]{McKinney2006, Komissarov2007magnetic,Tchekhovskoy2008simulations,Tchekhovskoy_2009,Komissarov2009ultrarelativistic}. The same simulations, have established that the accretion disc can significantly impact the spatial evolution of the jets not only at $r_g$-scale distances ($r_g = GM_{\rm bh}/\rm c^2$, where $M_{\rm bh}$ is the mass of the black hole), but also further out. 
In particular, \citet[][hereafter \cltm]{chatterjee2019accelerating} performed a series of high-resolution GRMHD simulations of strongly magnetised systems to better understand the loading of jets with matter from the wind of the accretion disc. When the jets propagate in a medium, pinch instabilities can occur in the interface between the jet and the ambient medium to give rise to eddies that eventually allow for matter to entrain the jet (\citealt[][]{Eichler1993magnetic,Spruit1997,Begelman_1998,giannios2006role}; \cltm; \citealt{Sironi2020}). Such mass entrainment can significantly affect the jet kinematics and hence the non-thermal emission.

Such GRMHD simulations, though, usually make the ideal gas assumption and therefore cannot capture dissipative processes like particle acceleration self-consistently. Kinetic simulations of particles-in-cell (PIC), on the other hand, calculate the trajectories of individual particles based on first principles, allowing for a more detailed and comprehensive understanding of the relativistic outflows. 
Both GRMHD and PIC simulations, however, are very computational expensive, and they cannot easily be compared to observations through statistical methods that explore the full parameter phase space.

In this work, we develop a new treatment for incorporating mass-loading and thus evolving compositions in jets, and apply it to a multi-zone jet model. This treatment is inspired by recent GRMHD simulations such as \cltm, to explore jet composition and its impact on the total jet power as well as its electromagnetic emission. In particular, we build on the multi-zone jet model developed by \cite{Markoff2005} that relies on the pioneering ideas of \cite{blandford1979relativistic}, \cite{hjellming1988radio}, and \cite{falcke1995jet}. 
After many developments, the latest version of the model is \texttt{BHJet} \citep[][]{lucchini2022BHJet}, a multi-zone jet model that better connects the jet acceleration and jet physical quantities to the radiative output. 
For the first time, we connect the physically motivated model \bhjet with hadronic acceleration, accounting for self-consistent energy conservation.
We further present \hadjet, a multi-zone, lepto-hadronic, mass-loaded jet model. In this work, we discuss the main physical properties of both models and how \hadjet can be used to address the jet-power crisis of lepto-hadronic models.

The paper is structured as follows. In Section~\ref{sec: magnetically accelerated jets} we describe the semi-analytical calculations for the magnetically accelerated jet accounting for both leptonic and hadronic acceleration and radiative processes. We present the results of the above jet model in Section~\ref{sec: results on the steady-state jet}. In Section~\ref{sec: mass loaded jets}, we describe the details of the mass-loaded jet model (\hadjet) and present the results in Section~\ref{sec: results on mass-loaded jets}. Finally, in Section~\ref{sec: discussion} we discuss the implication of our new models on the proton power issue and conclude in Section~\ref{sec: summary}.

%% file: Sections/Steady_jet.tex
\section{Magnetically accelerated steady-state jets}\label{sec: magnetically accelerated jets}
We assume two initially cold, Poynting flux dominated jets of either leptonic or lepto-hadronic content, that accelerate up to some maximum velocity because of magnetic energy dissipation \citep[][]{Vlahakis_2003,McKinney2006,Komissarov2007magnetic}. At the region where the bulk velocity reaches the maximum value (acceleration region henceforth, denoted by $z_{\rm acc}$), we further assume that energy is also dissipated to accelerate particles to non-thermal energies \citep[][]{Blandford1974twin_exhaust,Begelman1984theory}. With our formalism, we cannot capture whether the magnetic energy dissipates immediately to particle acceleration (as in the case of magnetic reconnection) or if magnetic energy dissipates to kinetic energy first and this extra kinetic energy dissipates to particle acceleration through shocks \citep[][]{Bogovalov2005shock}.
We assume instead that the total energy of the jet is conserved at the particle acceleration region. From this point outwards along the jets, we assume a constant particle acceleration rate and discuss below how this assumption affects the evolution of both the jet velocity and magnetic field. In Table~\ref{table: definitions and fiducial values}, we define all the parameters and their fiducial values (if applicable) that we use in this section.

\subsection{Jet dynamical properties}
Based on both semi-analytical and numerical calculations, the bulk jet Lorentz factor \g is expected to scale approximately as $z^{1/2}$, where $z$ is the distance along the jet \citep{Beskin2006effective,McKinney2006}. We parametrise the jet Lorentz factor as \citet[][]{lucchini2019breaking} (and see also \citealt[][]{Potter2012accelerating})
\begin{equation}\label{eq: g jet before gmax jet}
    \gamma(z\le z_{\rm acc}) = \gamma_0 + (\gamma_{\rm acc} -\gamma_0) \dfrac{z^{1/2}-z_0^{1/2}}{z_{\rm acc}^{1/2} -z_0^{1/2}}, 
\end{equation}
where $\gamma_0$ is the initial Lorentz factor at the jet base and $z_0$ is the distance of the jet base from the black hole and \g$_{\rm acc}$ is the maximum bulk Lorentz factor at $z_{\rm diss}$. We assume that the jets launch initially with the speed of sound, which for a relativistic flow with adiabatic index 4/3 is equal to 0.43\,c, or $\gamma_0 = 1.11$ \citep[][]{crumley2017symbiosis}.

The jets are thus set to be initially parabolic while they accelerate and become conical when they achieve $\gamma_{\rm acc}$ \citep[][]{Komissarov2009ultrarelativistic}. We express the cross-sectional radius of the jet along the jet axis as
\begin{equation}\label{eq: radius}
    r =r_0 +(z-z_0)\tan (\theta),
\end{equation}
where $r_0$ is the radius of the jet base and $\theta$ is the opening angle of the jets. Based on very long baseline interferometry observations and the Monitoring of jets in AGN with VLBA Experiments (MOJAVE; \citealt[see, e.g.,][]{Pushkarev2009opening,Pushkarev2017shapes}, we set the jet opening angle to be
\begin{equation}\label{eq: opening angle}
    \theta = \dfrac{0.15}{\gamma}.
\end{equation}

While the number of particles along the jet is conserved, we express the number density of leptons as
\begin{equation}\label{eq: number density before gmax}
    n = n_0 \left( \dfrac{\gamma \beta} {\gamma_0 \beta_0} \right)^{-1} \left(\dfrac{r}{r_0} \right)^{-2},
\end{equation}
where $\beta$ is the jet velocity normalized to the speed of light and $n_0$ is the initial number density.  We calculate $n_0$ by the power $L_{\rm jet}$ injected at the jet base in the comoving frame
\begin{equation}\label{eq: injected power at the jet base}
    L_{\rm jet} = 2\beta_0 \gamma_0 {\rm c\pi}r_0^2 \omega_0 
\end{equation}
where we account for two identical jets (hence the factor of 2), and
$n_0$ depends on $L_{\rm jet}$ and the initial conditions of the jet base as written out below. 
We write the jet enthalpy $\omega$ as \citep[][]{falcke1995jet,crumley2017symbiosis}
\begin{equation}\label{eq: enthalpy}
    \omega = \rho {\rm c^2} +U_j +P_j = \rho c^2 + U_{\rm p} +P_{\rm p} + U_{\rm e}+P_{\rm e}+ U_B+P_B,
\end{equation}
where $U_j=U_{\rm p}+U_{\rm e} +U_B$ is the total internal jet energy density and $P_j=P_{\rm p}+P_{\rm e} +P_B$ is the total jet pressure. In the above equation, $\rho$ is the jet mass density
\begin{equation}\label{eq: mass density}
    \rho = n_{\rm p}\, {\rm m_p} + n_{\rm e}\,\rm m_e.
\end{equation}

We express the number of protons in terms of the number of leptons as $n_{\rm p} = n_{\rm e} /\eta_e$, where $n_{\rm e/p}$ is the number density of leptons/protons, respectively, and $\eta_e\ge 1$ is a free parameter that remains constant unless the jets are mass-loaded (see below).

For an ideal gas, we can write the pressure terms as
\begin{equation}\label{eq: pressure}
    P_{\rm e,p} = (\Gamma_{\rm e,p}-1)\,U_{\rm e,p},
\end{equation}
where $\Gamma_{\rm e,p}$ is the adiabatic index. For the rest of the paper, we assume a relativistic pair content ($\Gamma_{\rm e}=4/3$) at the jet base and a cold proton population ($\Gamma_{\rm p}=5/3$) until the particle acceleration region (see below). For the pair temperatures we are interested in this work, the flow remains cold even if is dominated by pairs at the base.
For $U_B = P_B = B^2/8\pi$, we 
write the jet enthalpy as
\begin{equation}\label{eq: jet enthalpy for ideal gas}
    \omega = \rho{\rm c^2}+ \Gamma_{\rm p} U_{\rm p}+\Gamma_{\rm e} U_{\rm e} +\dfrac{B^2}{4\pi}.
\end{equation}

We define the specific enthalpy of the gas as
\begin{equation}\label{eq: specific enthalpy}
    h = \dfrac{U_g +P_g}{\rho{\rm{c^2}}} = \dfrac{\Gamma_{\rm p} U_{\rm p}+ \Gamma_{\rm e} U_{\rm e}}{\rho{\rm{c^2}}}
\end{equation}
where we used equation~(\ref{eq: pressure}). We calculate $U_{\rm e,p}$ by computing the integral
\begin{equation}\label{eq: U internal energy density integral}
    U_{\rm e,p} = \int \dfrac{dn_{\rm e,p}}{d\varepsilon_{\rm e,p}} \varepsilon_{\rm e,p} \rm  m_{e,p}c^2 d\varepsilon_{\rm e,p}.
\end{equation}
where $\varepsilon_{\rm e,p}$ is the Lorentz factor of the particles, but we can also express the internal energy density in terms of the average total energy of the particles
\begin{equation}\label{eq: U internal energy density}
    U_{\rm e,p} \simeq (\langle\varepsilon_{\rm e,p}\rangle-1)\,n_{\rm e,p} \rm m_{e,p}c^2,
\end{equation}
where $\langle\varepsilon_{\rm e,p}\rangle$ is the average Lorentz factor of the pairs/protons of the jet segment (see below for calculation). This equation is more convenient than equation~(\ref{eq: U internal energy density integral}) for the following discussion, however we note that it might not be accurate enough if a significant fraction of the leptons accelerate to non-thermal energies, in particular in a hard power law with slope $<2$. 

A useful parameter to characterise the jets is the magnetisation. We define the magnetisation of a flow as the Poynting flux over the total energy flux \citep[][]{Nokhrina2015Intrinsic}
\begin{equation}\label{eq: sigma magnetisation with h}
\begin{split}
    &\sigma  = \dfrac{B^2}{4\pi \left( \rho\rm{c^2} + U_g + P_g\right)}    \Rightarrow \\
    & \sigma = \dfrac{B^2}{4\pi \rho {\rm c^2} \left( 1+h \right)}.
\end{split}
\end{equation}

When the flow is cold ($h\ll 1$), the above definition reduces to the well-known expression of
\begin{equation}\label{eq: sigma magnetisation cold flow}
    \sigma_c \simeq \dfrac{B^2}{4\pi  \rho\rm{c^2} }.
\end{equation}

We write the enthalpy of equation~(\ref{eq: jet enthalpy for ideal gas}) of a flow from equations~\ref{eq: specific enthalpy} and \ref{eq: sigma magnetisation with h} as
\begin{equation}\label{eq: omega simplified for general flow}
    \omega = \rho{\rm{c^2}}(1+\sigma)(1+h).
\end{equation}

We can plug this equation into equation~(\ref{eq: injected power at the jet base}) to calculate the particle number density at the jet base
\begin{equation}\label{eq: n0 at the jet base final}
 n_0 = \frac{L_{\rm{\rm jet}}}{2\beta_{0}\gamma_{0}c \pi r_0^2\,({\rm m_p}/\eta_e + {\rm m_e}) c^2 (1+\sigma_{\rm c}) }. 
\end{equation}

We further use the relativistic Bernoulli's equation to express the conservation of energy flux along the jet axis \citep{Koenigl1980}
\begin{equation}\label{eq: Bernoulli's equation}
    \gamma \dfrac{\omega}{\rho}=\rm constant,
\end{equation}
and from equation~(\ref{eq: omega simplified for general flow}) we rewrite the above equation such as to define:
\begin{equation}\label{eq: mu}
    \mu \equiv \gamma\,(1+\sigma )(1+h), 
\end{equation}
where $\mu$ is the normalised total energy flux and is conserved along the jets (unless the jets entrain mass; see below). In a cold jet where the specific enthalpy $h$ is negligible, equation~(\ref{eq: mu}) simplifies to $\mu \simeq \gamma\, (1+\sigma_c)$. This is a very well-known equation to express the maximum jet Lorentz factor when the majority of the Poynting flux has been converted to kinetic energy ($\gamma_{\rm max} \simeq \mu$). In this work, we keep this term in our calculations because $h$ is an estimate of the energy that the accelerated particles carry in each jet segment, and in numerous instances can dominate both the magnetisation and the jet Lorentz factor.

While the jets accelerate between the launching point and the acceleration region $z_{\rm acc}$, $\mu$ remains constant. We write equation~(\ref{eq: mu}) at the jet base and equate it to the acceleration region and solve for the initial magnetisation
\begin{equation}\label{eq: initial magnetisation}
\begin{split}
    & \gamma_0(1+\sigma_0)(1+h_0) = \gamma_{\rm acc}(1+\sigma_{\rm acc})(1+h_{\rm acc}) \Rightarrow \\
    & \sigma_{0} = \frac{\gamma_{\rm acc}}{\gamma_0}\left( 1+\sigma_{\rm acc}\right) \left( \dfrac{1 + h_{\rm acc}}{1+ h_0} \right) - 1,
\end{split}
\end{equation}
and in general for every z below the acceleration region
\begin{equation}\label{eq: initial magnetisation below zacc general}
    \sigma (z \leq z_{\rm acc}) = \frac{\gamma_0}{\gamma}\left( 1+\sigma_{0}\right) \left( \dfrac{1 + h_0}{1+ h} \right) - 1,
\end{equation}
or
\begin{equation}\label{eq: initial magnetisation below zacc general second}
    \sigma (z \leq z_{\rm acc}) = \frac{\gamma_{\rm acc}}{\gamma}\left( 1+\sigma_{\rm acc}\right) \left( \dfrac{1 + h_{\rm acc}}{1+ h} \right) - 1.
\end{equation}

With the magnetisation and the specific enthalpy at the acceleration region as free parameters ($\sigma_{\rm acc}$ and $h_{\rm acc}$, respectively), we set the initial magnetisation $\sigma_0$ required for the flow to be Poynting flux dominated and to carry enough energy to efficiently accelerate particles to non-thermal energies. In particular, we use $\sigma_{\rm acc}$ as a free parameter because this is the simplest way to force our semi-analytical model to have dissipated the majority of the magnetisation at the acceleration region, and we set $h_{\rm acc}$ from equation~(\ref{eq: specific enthalpy}) (see also the discussion on particle acceleration below). The initial specific enthalpy $h_0$ is set by the free parameters at the jet base, and as we discuss below, it is negligible for the standard case of an initially cold jet that we study here (see subsection~\ref{sec: specific enthalpy}).

Above the acceleration region, we assume the toroidal component dominates the poloidal component of the magnetic fields similar to \cite{blandford1979relativistic}, so
\begin{equation}\label{eq: magnetic field above the acceleration region}
    B(z>z_{\rm acc}) = B_{\rm acc}\left(\frac{z}{z_{\rm acc}}\right)^{-1},
\end{equation}
where $B_{\rm acc}$ is the magnetic field strength at the acceleration region.

Based on equation~(\ref{eq: sigma magnetisation with h}), we generalize the expression of $\sigma$ for every $z$ above the acceleration region
\begin{equation}\label{eq: sigma magnetisation above acceleration region}
    \sigma(z\geq z_{\rm acc}) = \sigma_{\rm acc} \frac{\rho_{\rm acc} (1+h_{\rm acc})}{\rho (1+h)}\left( \dfrac{z}{z_{\rm acc}} \right)^{-2}.
\end{equation}

\subsection{The acceleration region and particle acceleration}
We assume that the pairs at the jet base follow a Maxwell-J\"uttner distribution (MJ; the relativistic regime of the Maxwell-Boltzmann distribution) with a peak energy $k_{\rm B}T_{\rm e}$ that is a free parameter. The population of protons on the other hand is cold, making the flow cold at the launching point. 

By the time the flow reaches the acceleration region the Poynting flux dominated flow has dissipated the magnetic energy, hence the magnetisation has dropped to a value $\sigma_{\rm acc}$. At the same region, we assume a constant fraction $f_{\rm pl}\sim 0.1$ of particles accelerates to a non-thermal power law between a minimum and a maximum energy. For the leptonic scenario, we assume that only pairs accelerate in a power law from an energy $\varepsilon_{\rm min}{\rm m_ec^2} = k_{\rm B}T_{\rm e}$ to some $\varepsilon_{\rm max}$ that we calculate self-consistently by equating the acceleration timescale $4\varepsilon {\rm m_ec^2}/(3f_{\rm sc}ecB)$ to the escape timescale \citep{jokipii1987rate,aharonian2004very}. The acceleration efficiency $f_{\rm sc}$ depends on the particle acceleration mechanism, but we fix it at a value between 0.01 and 0.1 leading to a maximum electron energy of the order of GeV for the case of a \bh. 
For the lepto-hadronic scenario, we assume that protons accelerate as well in a power law from an $\varepsilon_{\rm min}$= 1 to some $\varepsilon_{\rm max}$ that we calculate by equating the acceleration timescale to the (lateral) escape timescale $r$/c of the jet segment and for the case of \bhs it may attain values of the order of 100\,TeV and above \citep[][]{pepe2015lepto,kantzas2020cyg,kantzas2022gx}. We constrain the non-thermal particle distributions by assuming that they extend up to the maximum energy, and then they drop exponentially
\begin{equation}\label{eq: particle distribution}
  \dfrac{{\rm{d}}n\left(\varepsilon\right)}{{\rm{d}}\varepsilon} = K \varepsilon^{-p}\, \exp{\left(-\varepsilon/\varepsilon_{\rm{max}}\right)},
\end{equation}
where $n$ is the particle number density for any species, $K$ is the normalisation, and the slope $p$ of the power law depends on the particle acceleration mechanism, but we use it as a free parameter between 1.7 and 2.4, assuming it remains the same between electrons and protons.

Finally, we derive the average Lorentz factor for every species from the equation
\begin{equation}\label{eq: average Lorentz factor of particles}
    \langle \varepsilon \rangle = \dfrac{\bigintss \varepsilon \dfrac{{\rm d}n}{{\rm d}\varepsilon}{\rm d}\varepsilon}{\bigintss \dfrac{{\rm d}n}{{\rm d}\varepsilon}{\rm d}\varepsilon}.
\end{equation}

\subsection{Jet evolution and particle acceleration}
Beyond the acceleration region where particles accelerate to non-thermal energies as well, the specific enthalpy can become important because the average Lorentz factors of pairs and/or protons may have significantly increased (see equation~\ref{eq: specific enthalpy}). 
We write the bulk Lorentz factor for every jet segment above the acceleration region for an outflow from equation~(\ref{eq: mu}):
\begin{equation}\label{eq: bulk Lorentz factor above acceleation}
    \gamma (z) = \gamma_{\rm acc} \left(\dfrac{1+h_{\rm acc}}{1+h}\right) \left(\dfrac{1+\sigma_{\rm acc}}{ 1+ \sigma}\right).
\end{equation}

\begin{table*}
\begin{center}
	\setlength{\tabcolsep}{6pt} 
	\renewcommand{\arraystretch}{1.2} 
	\begin{tabular}[b]{lcclc}\hline\hline
		Parameter & Units & Fiducial value(s) & Definition & Equation \\ \hline
        $z$   &   $r_g$   & $-$          & distance from the black hole along the jet axis & $-$\\
        $z_0$   &   $r_g$   & 6          & distance of the jet base from the black hole & $-$\\

        $\gamma$ &$-$     & $1-3$                  & bulk Lorentz factor of the flow & \ref{eq: g jet before gmax jet}\\
        $\gamma_0$ &$-$     & 1.1                  & bulk Lorentz factor at the jet base & $-$ \\        
        $r$   &   $r_g$   & $-$          & cross-sectional radius of the flow & \ref{eq: radius}\\
        $\theta$   &   rad   & $-$          & jet opening angle & \ref{eq: opening angle}\\
        $n$   &   $\rm cm^{-3}$   & $-$  & jet (total) particle number density & \ref{eq: number density before gmax}\\
        $n_0$   &   $\rm cm^{-3}$   & $-$  & jet number density at the jet base & \ref{eq: n0 at the jet base final}\\        
        $n_{\rm e}$   &   $\rm cm^{-3}$   & $-$  & jet pair number density & $-$\\
        $n_{\rm p}$   &   $\rm cm^{-3}$   & $-$  & jet proton number density & $-$\\
        $\rho$   &   $\rm g\,cm^{-3}$   &$-$  & jet mass density & \ref{eq: mass density}\\
        $\omega$ & $\rm erg\,cm^{-3}$  & $-$ & total jet enthalpy& \ref{eq: jet enthalpy for ideal gas}\\   
        $h$ & $-$ & $-$         & jet specific enthalpy & \ref{eq: specific enthalpy}\\
        $\sigma$ & $-$ & $-$            & magnetisation of the flow & \ref{eq: sigma magnetisation with h}\\
        $\sigma_0$ & $-$ & $1-100$            & magnetisation of the flow at the jet base & \ref{eq: initial magnetisation}\\
        $\mu$ &$-$ & $1-100$                  & normalised total jet energy flux & \ref{eq: mu}\\
        $\langle \varepsilon_{\rm e,p}\rangle$&$-$&$1-100$& particle average Lorentz factor & \ref{eq: average Lorentz factor of particles}\\
\hline
        $z_{\rm acc}$ &$r_g$& $10^3$      & location where jet acceleration reaches the max value & free parameter\\
        $\gamma_{\rm acc}$&$-$& 3             & maximum Lorentz factor of the flow at $z_{\rm acc}$ & free parameter\\
        $r_0$   &   $r_g$   & $10-10^2$          & jet base radius & free parameter\\
        $L_{\rm jet}$& $L_{\rm Edd}$& 0.002-0.02& injected jet power at the jet base&free parameter\\
        $\eta_e$    &   $-$               & $1-10^6$ &       jet pair-to-proton content & free parameter\\
        $\sigma_{\rm acc}$ & $-$ & 0.1        & magnetisation of the flow at the acceleration region & free parameter\\
        $k_BT_e$ &keV & $-$         & electron peak energy at the jet base & free parameter\\
        \hline

		\hline
	\end{tabular} 
	\caption{The definition of the jet quantities we use in this work with their units, some fiducial values (if applicable), the equation number where we define the parameter or whether it is a free parameter. See Sections~\ref{sec: magnetically accelerated jets} and \ref{sec: mass loaded jets} for further information.
	}\label{table: definitions and fiducial values}
\end{center}
\end{table*}

\subsection{Radiative Processes}\label{section: radiative processes}
We suggest the interested readers to seek for further details on the radiative processes in \cite{lucchini2022BHJet} for the leptonic processes, and in \cite{kantzas2020cyg} for the hadronic processes. We nevertheless briefly discuss the main processes here for completeness.

\subsubsection{Leptonic processes}
The main three radiative processes of leptonic nature that we require in our analysis here are: synchrotron radiation, inverse Compton scattering (ICS) and pair production. In particular, the thermal pairs of the MJ distribution and the non-thermal 
power-law tail above the dissipation region, lose energy due to cyclo-synchrotron radiation \citep[][]{blumenthal1970bremsstrahlung,rybicki2008radiative}. We only account for the average magnetic field strength of the particular jet segment and assume an isotropic distribution of pitch angles that we average over.

We further account for the ICS between the pairs and the radiation fields of the outflow \citep[][]{blumenthal1970bremsstrahlung,rybicki2008radiative}. In particular, in this work we neglect any external photon field and only allow for ICS between the emitting pairs and the synchrotron photons (synchrotron self Compton; SSC). Plausible external photon fields may be important in the case of AGN jets but for the study-cases as \bhs we discuss in this work, we have shown in previous works that the external photon fields are not critical (see e.g. \citealt{Lucchini2021correlation,kantzas2020cyg}, however, see also \citealt{zdziarski2014jet} and \citealt{Zacharias2022exhale} for cases where the external photon fields may be important to explain the \gr spectrum). For simplicity, we also neglect any accretion disc in the following discussion, but we do account for it when examining particular sources, following \cite{lucchini2022BHJet}. For the ICS processes, we account for the Klein-Nishina regime when necessary, and allow for multiple scatterings to better capture the evolution of the exponential cutoff. This particular process is the most computationally expensive amongst the leptonic ones, we hence choose to neglect it when the radiative output becomes $10^4$ times smaller than the synchrotron counterpart for the particular segment.

The final process of leptonic nature we account for is the photon annihilation to pair production and vice versa \citep[][]{coppi1990reaction}. These two processes are usually negligible, so we do not mention them unless we discuss their impact on the particle population or the spectrum \citep[see, e.g.,][]{Connors2019combining}.

\subsubsection{Hadronic Processes}\label{hadronic processes}

We account for both proton-proton (pp) and proton-photon (\pg) processes when accelerated protons interact with the cold protons of the flow and the jet radiation, respectively. In particular, we use the semi-analytical parametrisation of \cite{kelner2006energy} for the pp interactions, and \cite{kelner2008energy} for the \pg. The above analysis provides the resulted distributions of secondary particles (pions that decay into muons, and the muons decay into neutrinos, pairs and \grs) and hence cannot account for any synchrotron radiation of muons and/or pions, but for the current systems we examine, we see that it is not required. 
We do however consider the cyclo-synchrotron radiation of secondary pairs due to the presence of the magnetic field.

In our particular analysis, we find that the synchrotron photons produced by the primary pairs act as the target for the \pg interactions. 
Based on this analysis, we can also produce the neutrino counterpart in a self-consistent manner (Kantzas et al. in prep).

%% file: Sections/Steady_jet_results.tex
\section{Results for the steady-state jets}\label{sec: results on the steady-state jet}
We first present the results of the analysis of the model where we do not account yet for any mass entrainment. In this flavour of the model, we try to better understand and constrain the number of leptons in the jets with respect to the number of protons $\eta_e$. We further present the jet dynamical properties and their corresponding multiwavelength spectra before we compare them to ones when we account for mass-loading.

\subsection{Specific enthalpy and particle acceleration}\label{sec: specific enthalpy}

\begin{figure*}
    \centering
	\subfigure[Purely leptonic acceleration with $\varepsilon_{\rm e,min}=1.5$.]{
        \includegraphics[width=1.0\columnwidth]{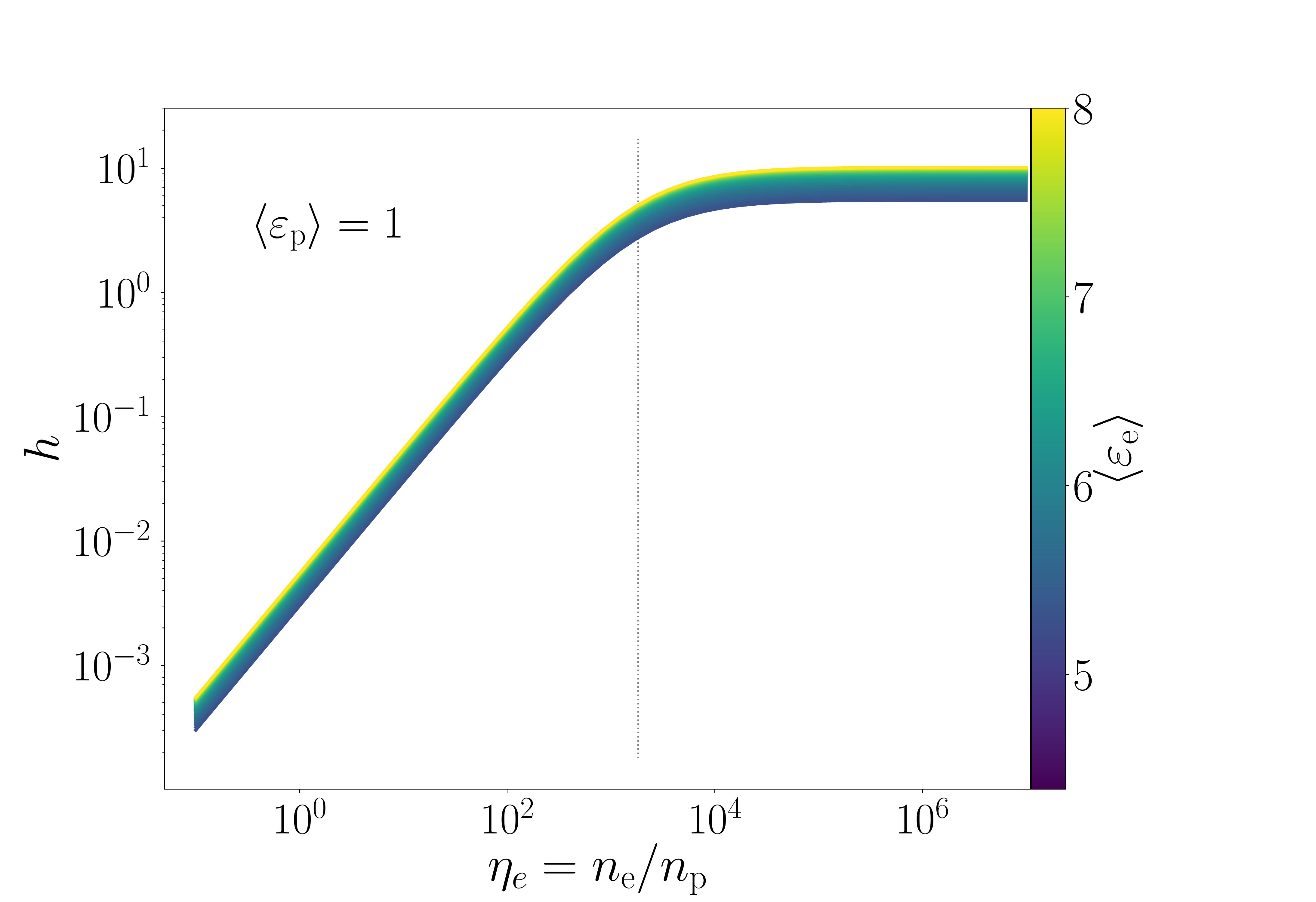}
    }
    \subfigure[Purely leptonic acceleration with $\varepsilon_{\rm e,min}=10$.]{
        \includegraphics[width=1.0\columnwidth]{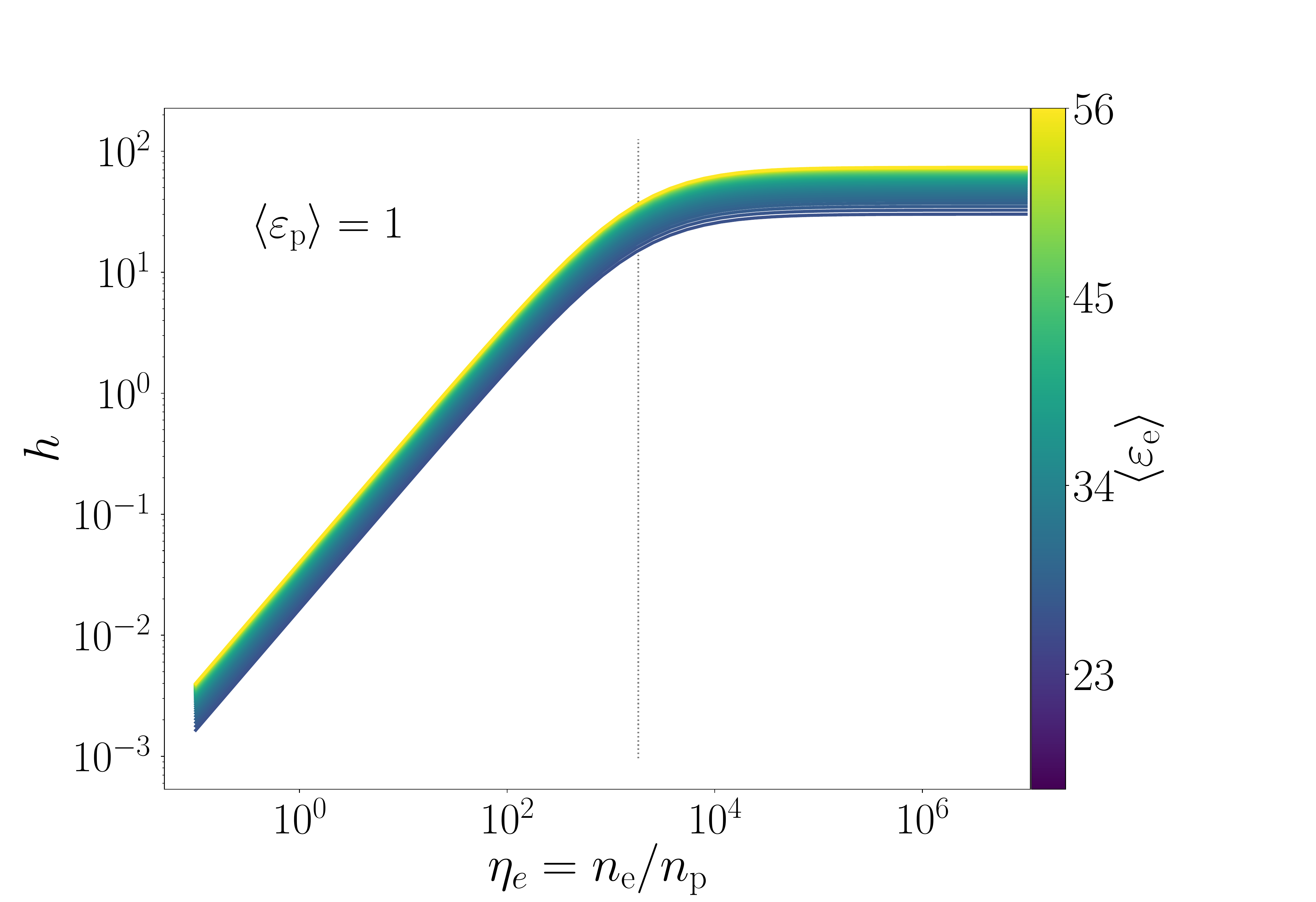}
    } 
	\subfigure[Leptohadronic acceleration with $\varepsilon_{\rm e,min}=1.5$.]{
        \includegraphics[width=1.0\columnwidth]{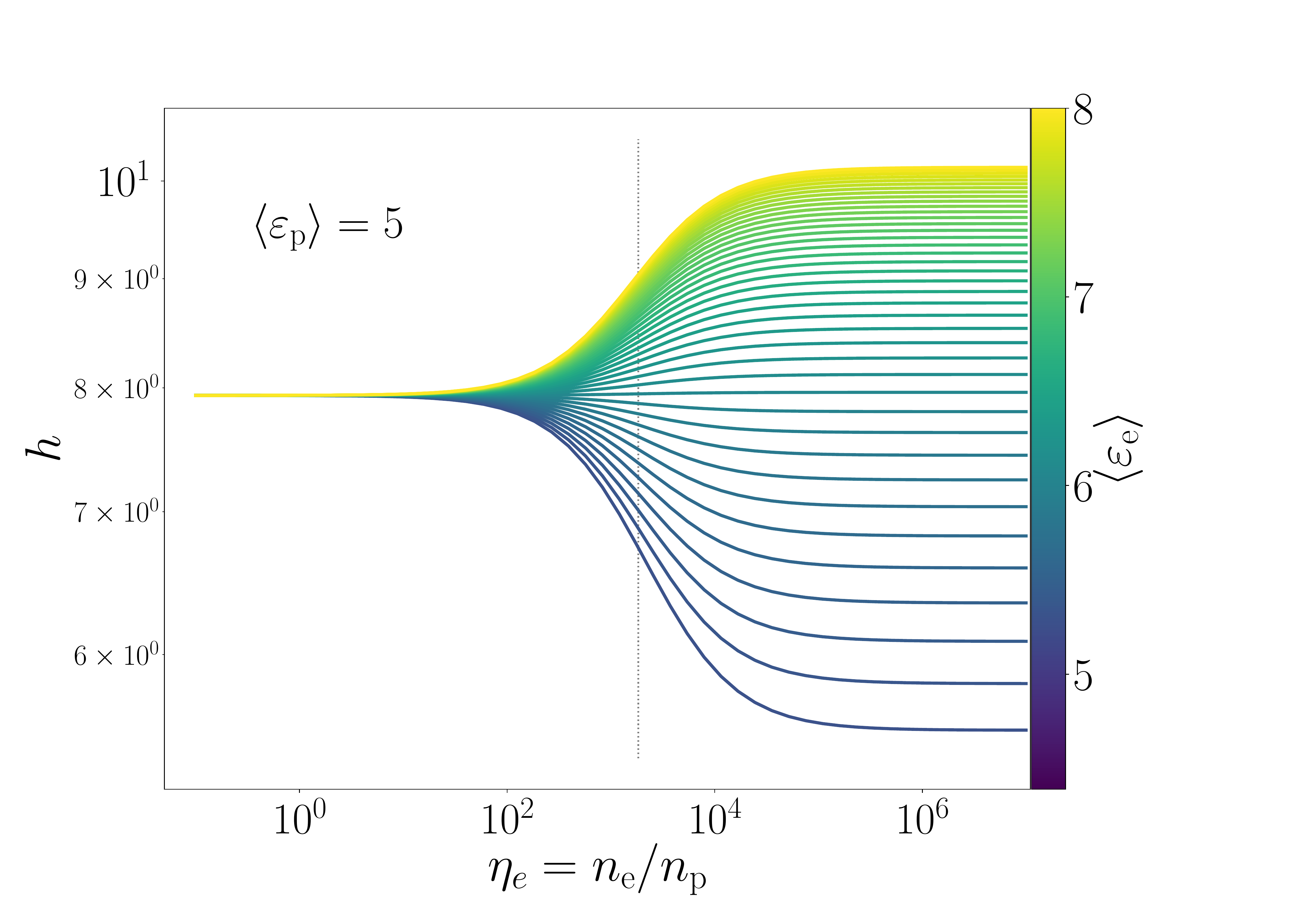}
    }
	\subfigure[Leptohadronic acceleration with $\varepsilon_{\rm e,min}=10$.]{
        \includegraphics[width=1.0\columnwidth]{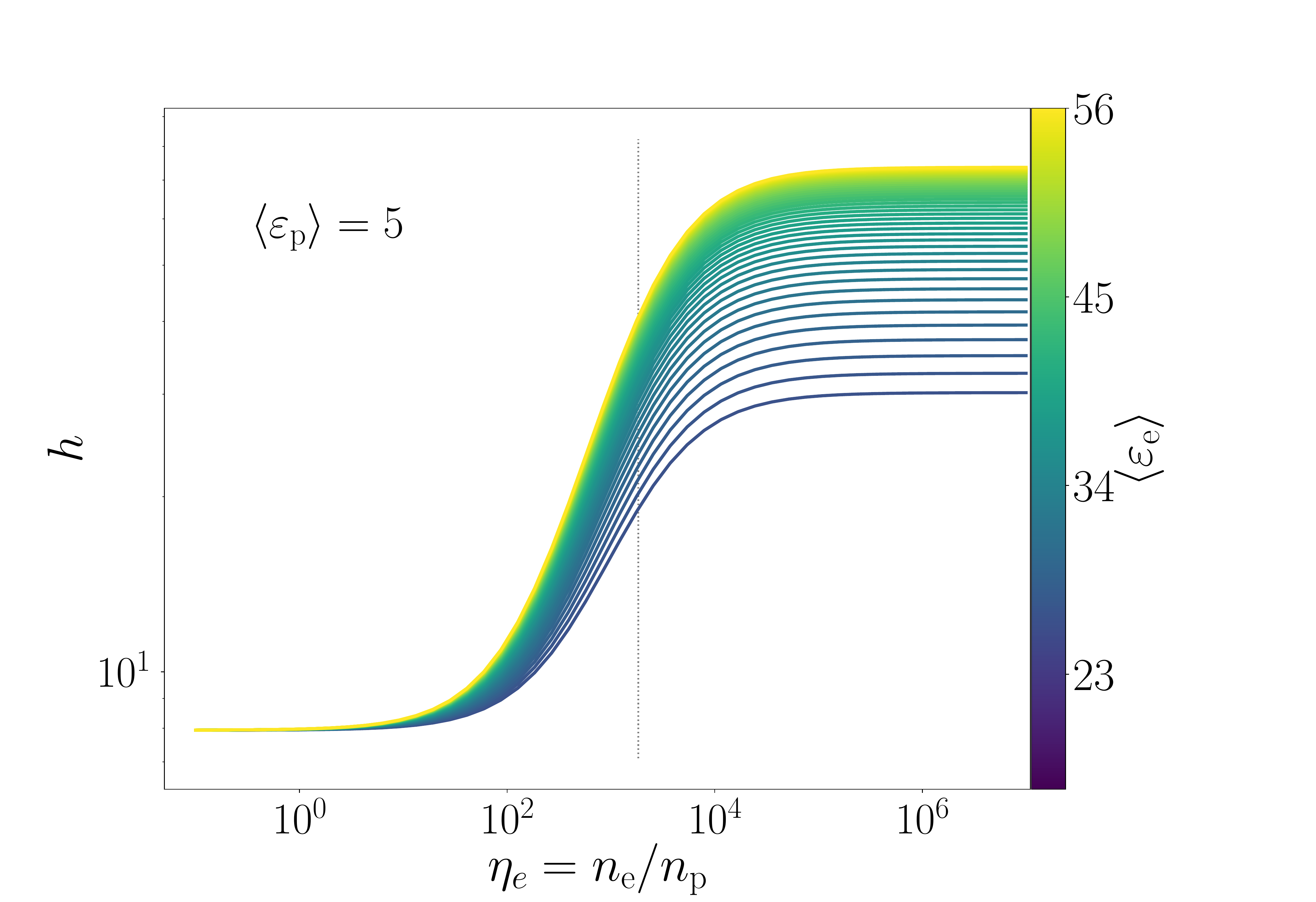}
    }
	\subfigure[More efficient hadronic, and leptonic acceleration with $\varepsilon_{\rm e,min}=1.5$.]{
        \includegraphics[width=1.0\columnwidth]{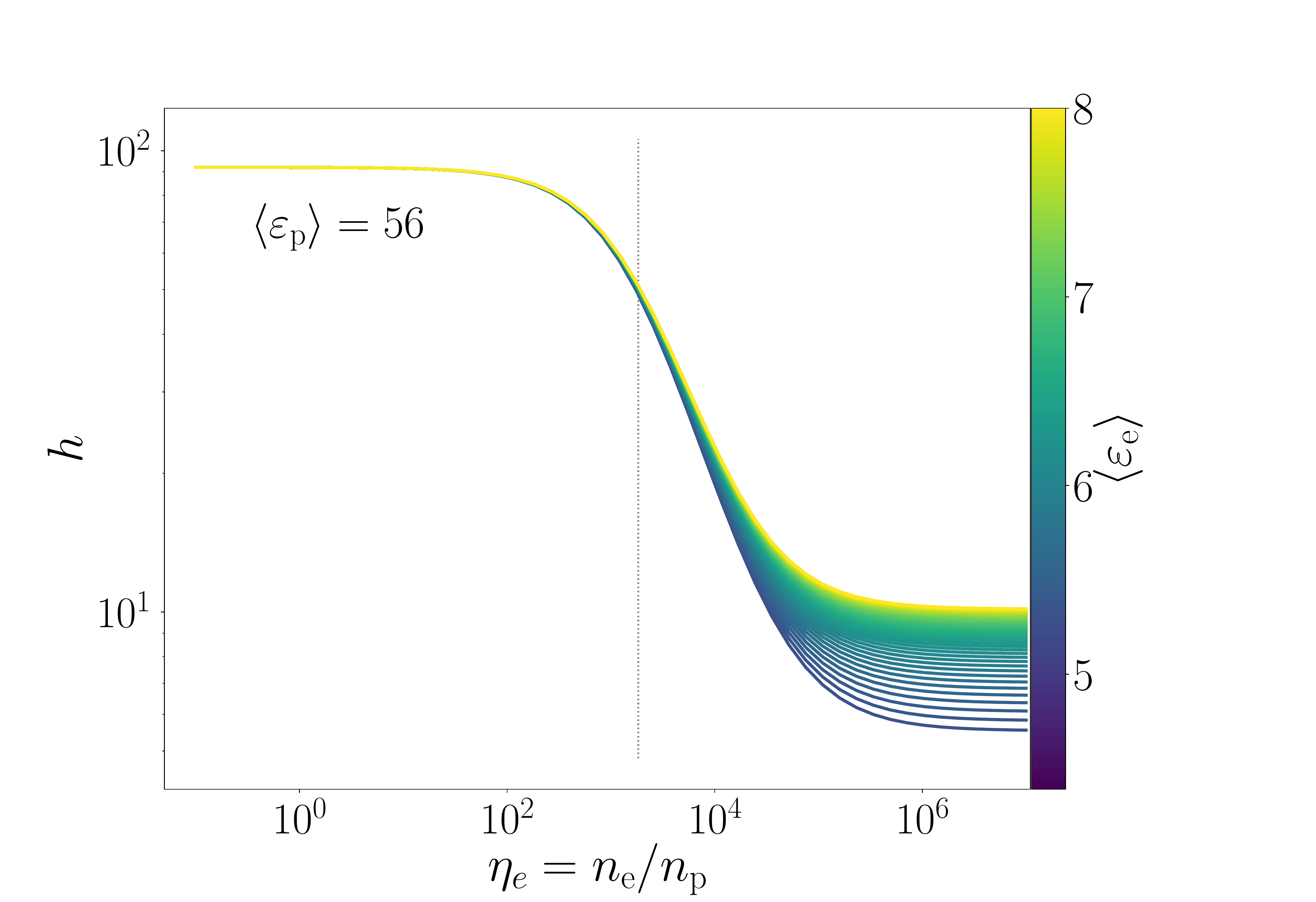}
    }
	\subfigure[More efficient hadronic, and leptonic acceleration with $\varepsilon_{\rm e,min}=10$.]{
        \includegraphics[width=1.0\columnwidth]{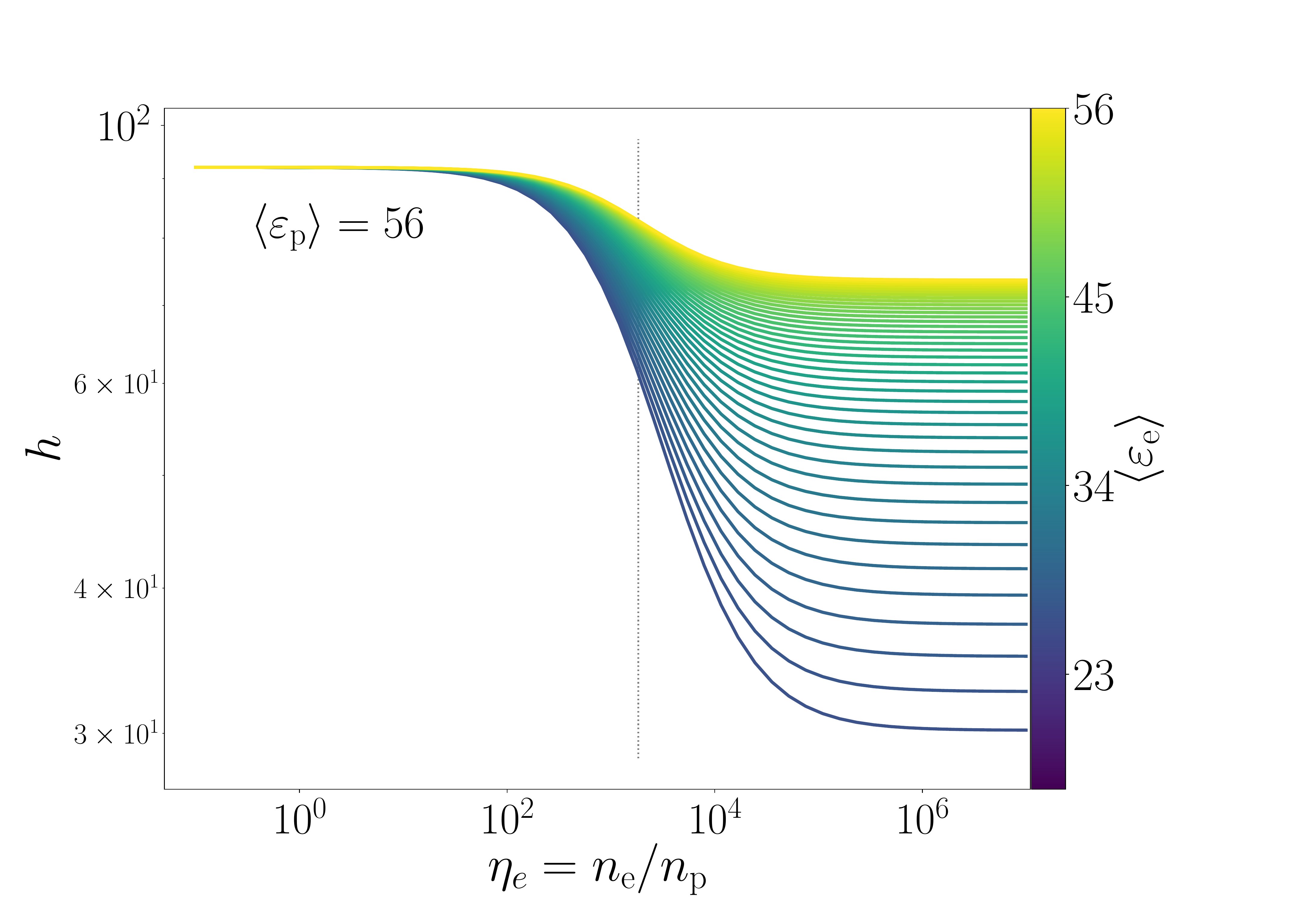}
    }
    \caption{The jet specific enthalpy $h$ as a function of the jet content $\eta_e=n_{\rm e}/n_{\rm p}$. In all plots, we assume a soft non-thermal power law with $p=2.2$ to derive the average particle Lorentz factors from equation~(\ref{eq: average Lorentz factor of particles}). The color-map corresponds to the average Lorentz factor of electrons, with lighter colors to indicate more efficient acceleration. In the \textit{left} column and for a less efficient electron acceleration, the minimum Lorentz factor of the pairs is $\varepsilon_{\rm e,min}=1.5$, whereas in the \textit{right} column with a more efficient electron acceleration, $\varepsilon_{\rm e, min}=10$. In the \textit{top} panels, we assume only leptonic acceleration, in the \textit{middle}, we assume non-efficient hadronic acceleration with $\varepsilon_{\rm p, min} = 1$ and $\varepsilon_{\rm p, max} = 100$, and in the \textit{bottom} panels, we assume efficient hadronic acceleration with $\varepsilon_{\rm p, min} = 10$ and $\varepsilon_{\rm p, max} = 10^7$. The vertical lines correspond to $\eta_e=\rm m_p/m_e$. Overall, the specific enthalpy $h$ may attain values greater than unity and may hence significantly alter the jet kinematics. See the text for more details.
    }
    \label{fig: specific enthalpy}
\end{figure*}

We can express equation~(\ref{eq: specific enthalpy}) as
\begin{equation}\label{eq: h specific enthalpy versus eta}
    h = \dfrac{\Gamma_{\rm e} (\langle \varepsilon_{\rm e}\rangle -1) + \Gamma_{\rm p} (\langle \varepsilon_{\rm p}\rangle -1) \dfrac{{\rm m_p/m_e}}{\eta_e} }{1+ \dfrac{{\rm m_p/m_e}}{\eta_e}},
\end{equation}
where we used equations~(\ref{eq: mass density}), (\ref{eq: U internal energy density}), and $n_{\rm p} = n_{\rm e}/\eta_e$. 

From the above equation, we see that the specific enthalpy 
depends merely on the ratio between pairs and protons. Moreover, we see that $h$ strongly depends on any mechanism (acceleration or cooling) that would significantly change the average Lorentz factor of the particles.

In Fig.~\ref{fig: specific enthalpy}, we plot the specific enthalpy $h$ as a function of the pair-to-proton ratio $\eta_e$ for various values of $\langle \varepsilon_{\rm e}\rangle$ and $\langle \varepsilon_{\rm p}\rangle$. Both $\langle \varepsilon_{\rm e}\rangle$ and $\langle \varepsilon_{\rm p}\rangle $ depend on the power law slope of the accelerated particles, as well as the minimum and the maximum particle energy. We let $\eta_e$ to scale between a few and $10^6$ although the latter values are extreme and perhaps not physically expected. A jet with more protons than leptons ($\eta_e<1$) would be positively charged and hence is unphysical. On the other hand, a very large number of pairs per protons would be difficult to explain the observed Lorentz factors on parsec scales \citep[][]{Ghisellini2010rocket}.

In the top left plot of Fig.~\ref{fig: specific enthalpy} where no protons accelerate at all, and in particular in the case of approximately equal amount of pairs and protons ($\eta_e \sim 1$), we see that the specific enthalpy is significantly smaller than unity ($ h \ll 1$). This is in agreement with the initial setups of GRMHD simulations where the specific enthalpy is usually neglected \citep[][]{McKinney2006,Komissarov2007magnetic}. In the other regime, where the flow is dominated by pairs ($\eta_e\gtrsim 10^3$), we see that $h \sim \Gamma_e \langle \varepsilon_{\rm e}\rangle$ (equation~\ref{eq: h specific enthalpy versus eta}). In the top right plot of Fig.~\ref{fig: specific enthalpy} where we assume $\varepsilon_{\rm e, min}=10$, we see a similar evolution of $\eta_e$. The main difference is that $\langle \varepsilon_{\rm e}\rangle$ goes to larger values, hence $h$ goes to larger values as well. From both plots, we see that for a purely leptonic flow, the specific enthalpy is not negligible and in fact, it can be as important as the magnetisation and the kinetic energy in the evolution of the jets (as discussed below).

In the middle plots of Fig.~\ref{fig: specific enthalpy}, where protons accelerate in a similar power law as the accelerated pairs, we see a significantly different evolution of $h$ for different jet content. In particular, in the case where $\varepsilon_{\rm e,min}=1$ and $\varepsilon_{\rm p,min}=1$ (middle left plot), we see that for an equal pair-to-proton jet content ($\eta_e =1$), $h$ is driven by the accelerated protons and in fact, $h\sim \Gamma_p \langle \varepsilon_{\rm p}\rangle $ (see equation~\ref{eq: h specific enthalpy versus eta}). In the regime of a purely leptonic flow ($\eta_e \gg 1$), we see that $h \sim \Gamma_e \langle \varepsilon_{\rm e}\rangle $ and depending on whether $\langle \varepsilon_{\rm e}\rangle > \langle \varepsilon_{\rm p}\rangle $ or  $\langle \varepsilon_{\rm e}\rangle < \langle \varepsilon_{\rm p}\rangle $, $h$ will increase or decrease, respectively. In the right-hand-side of the middle panels of Fig.~\ref{fig: specific enthalpy}, we get larger values of $\langle \varepsilon_{\rm e}\rangle $ because of the larger value of $\varepsilon_{\rm e, min}$ (for the particular $p=2.2$), and hence the specific enthalpy may attain significantly larger values reaching values of the order of $\Gamma_{e}\langle \varepsilon_{\rm e}\rangle $. 

In the bottom plots of Fig.~\ref{fig: specific enthalpy} where protons accelerate in a power law from a $\varepsilon_{\rm p,min}=10$, we see that a flow of $\eta_e\sim 1$ has a significant fraction of energy in the specific enthalpy because $h \sim \Gamma_p \langle \varepsilon_{\rm p}\rangle \sim 90$. In the purely leptonic regime ($\eta_e \gg 1$), we see that $h$ can drop to values smaller than 10 depending on the average Lorentz factor of the pairs. In the case where pairs accelerate in a power law from a high energy as $10\,\rm m_ec^2$ (right-hand-side plot of lowermost panels of Fig.~\ref{fig: specific enthalpy}), the energy content in the specific enthalpy remains significant for both $\eta_e \sim 1$ and $\eta_e \sim 10^6$.

From Fig.~\ref{fig: specific enthalpy}, we overall see that the specific enthalpy of a flow that accelerates particles can be important in the evolution of the flow (see also discussion below). In the case where only pairs accelerate in the jets and for an equal amount of electrons-to-protons as is commonly assumed in GRMHD (left-hand-side of the uppermost panels, and in particular in the case of one), we see that the specific enthalpy is indeed negligible ($h\ll 1$). In any other case where both pairs and protons accelerate in the jets, and regardless of the jet content (either pair-dominated or equal pair-to-proton content), the specific enthalpy of the flow might be of the order of a few-to-tens, and hence it is important for the evolution of the flow (see also discussion of \cltm). 

In the Appendix~\ref{app: h specific enthalpy for p=1.7} we discuss the evolution of $h$ in the case of a hard power law of accelerated particles with $p=1.7$ power law index. Such hard values, resulting from efficient particle acceleration e.g., in magnetic reconnecting regions \citep[][]{sironi2015relativistic,Ball2018} or relativistic shocks \citep[see e.g.][]{Boettcher_2019}, lead to even larger values of $h$ of the order of thousands. Such large values of $h$ along with large bulk Lorentz factors as observed in relativistic outflows in AGN and GRBs, would lead to significantly larger values of total energy flux $\mu$ compared to those in the literature \citep[][]{Komissarov2007magnetic,Komissarov2009ultrarelativistic,petropoulou2022baryon}. Furthermore, equation \ref{eq: mu}, which has broadly been used to provide an estimate for the maximum bulk Lorentz factor when the magnetic energy has been converted into kinetic energy, would not hold anymore and a more careful treatment where the specific enthalpy is calculated from first principles is needed.

\subsection{Total energy flux evolution for steady state jets}\label{sec: mu plots subsection}
\begin{figure*}
    \centering
\subfigure[Pair-proton outflow with only leptonic acceleration]{
        \includegraphics[width=1.0\columnwidth]{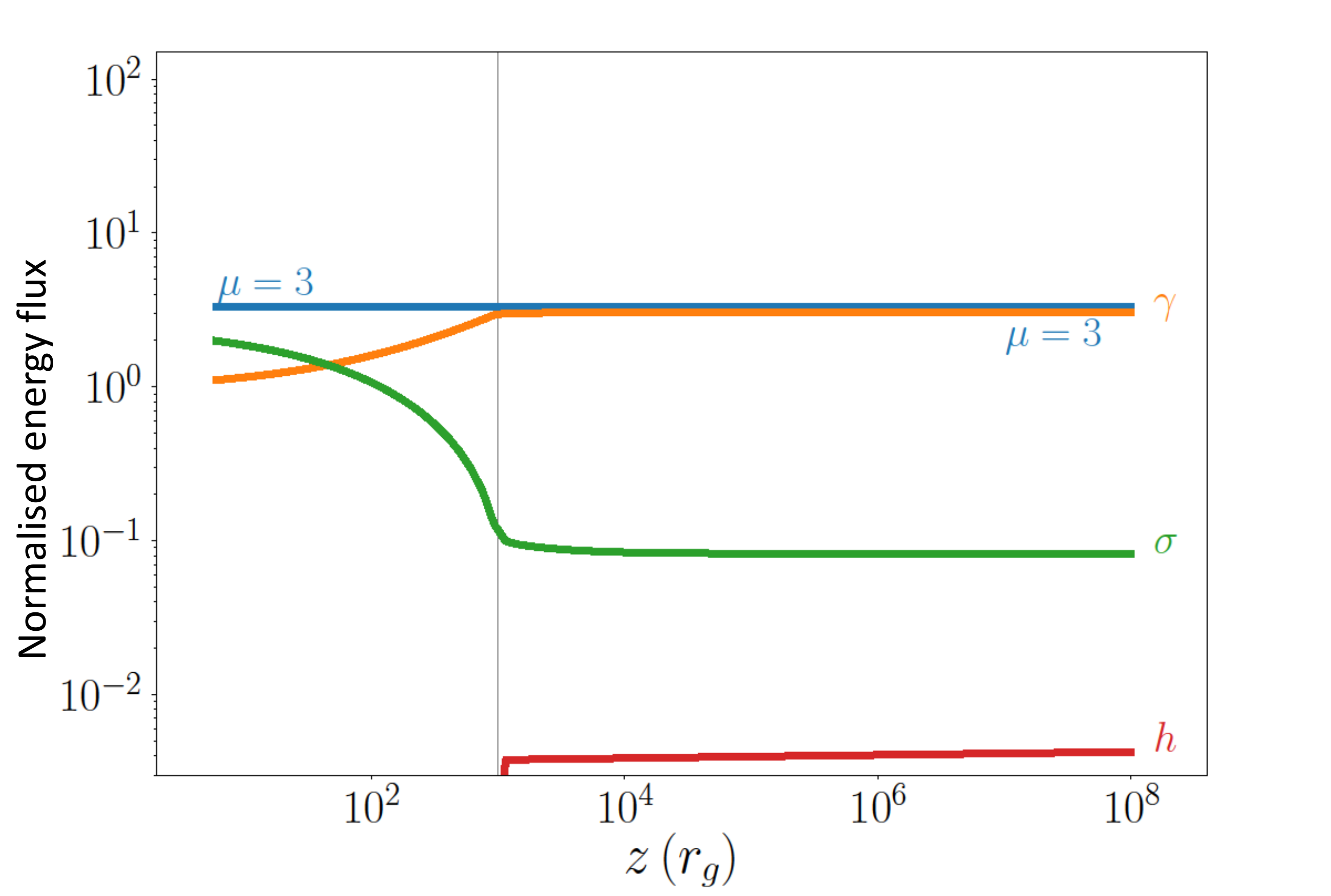}
        \label{fig: mu and components along the jet A}
}
\subfigure[Pair-dominated outflow with only leptonic acceleration]{
        \includegraphics[width=1.0\columnwidth]{figs/mu/eta_10000__geav_6__gpav_1__pspec_2.2__gacc_3__dge_1.001.pdf}
        \label{fig: mu and components along the jet B}
}
\subfigure[Pair-proton outflow with lepto-hadronic acceleration]{
        \includegraphics[width=1.0\columnwidth]{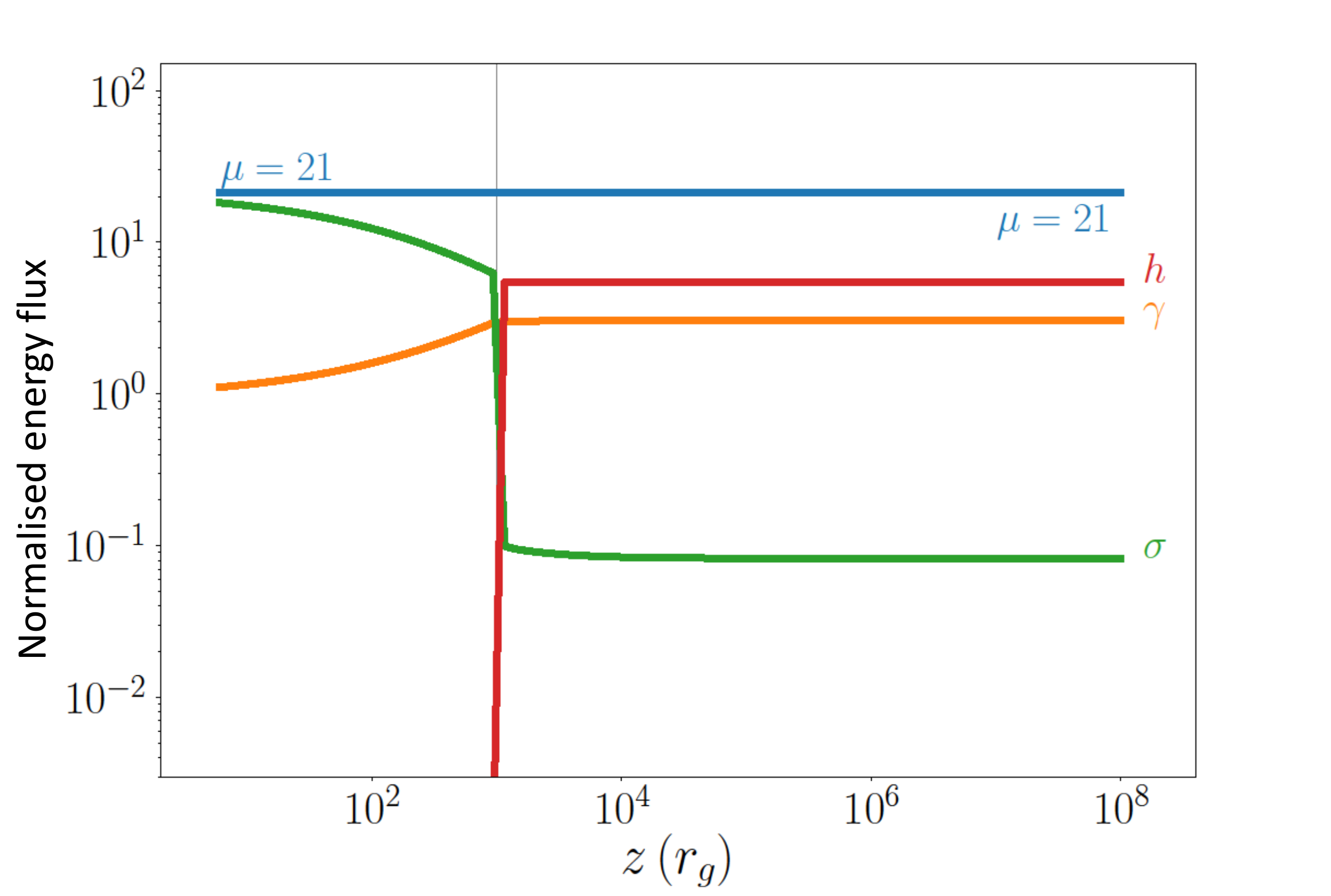}
        \label{fig: mu and components along the jet C}
}
\subfigure[Pair-dominated outflow with  lepto-hadronic acceleration]{
        \includegraphics[width=1.0\columnwidth]{figs/mu/eta_10000__geav_6__gpav_4__pspec_2.2__gacc_3__dge_1.001.pdf}
        \label{fig: mu and components along the jet D}
    }
    \caption{The energy jet components \g (the bulk Lorentz factor), $\sigma$ (the magnetisation), and $h$ (the specific enthalpy) that follow the relation $\mu = \gamma (1+\sigma)(1+h)$ (equation~\ref{eq: mu}). In all plots we use $z_0=6\,r_g$, $z_{\rm acc}=10^3\,r_g$, $\gamma_{\rm acc}=3$, $\sigma_{\rm acc}=0.1$ and $\langle \varepsilon_{\rm e}\rangle =6$ (see Table~\ref{table: definitions and fiducial values} for definitions). We show a pair/proton flow with $\eta_e=1$ in the \textit{left} column and a pair-dominated flow with $\eta_e=10000$ in the \textit{right} column. In the \textit{top} panels, we only account for leptonic acceleration and in the \textit{bottom} panels, we consider hadronic acceleration as well with $\langle \varepsilon_{\rm p}\rangle =4$. The specific enthalpy $h$ leads to different jet dynamical quantities based on whether hadronic acceleration takes place and jet content. 
    }
    \label{fig: mu and components along the jet}
\end{figure*}

In Fig.~\ref{fig: mu and components along the jet}, we plot the evolution of $\mu$ along the jets with the different components: magnetisation ($\sigma$), bulk Lorentz factor ($\gamma$) and specific enthalpy ($h$). In the left plots of Fig.~\ref{fig: mu and components along the jet}, we assume a jet content of equal number of leptons and protons ($\eta_e=1$) and in the right plots we assume a pair-dominated outflow ($\eta_e=10000$). In the top panels, we assume that only leptons accelerate to non-thermal energies, whereas in the bottom panels, we assume that hadrons accelerate as well in a power law with the same index. 

In the top left panel, where we 
account only for leptonic acceleration with $\langle \varepsilon_{\rm e} \rangle =6$, we see that the initial magnetisation of the outflow converts to bulk kinetic energy whereas the magnetisation drops to $\sigma_{\rm acc} =0.1$ (a free parameter). The specific enthalpy starts as negligible at the cold jet base ($h_{0}\ll 10^{-2}$) and remains insignificant for the jet evolution above the particle acceleration region $z_{\rm acc}$. This particular regime where the specific enthalpy is insignificant and the jet composition is one lepton per proton, is the regime considered by most GRMHD simulations (see also Section~\ref{sec: mass loaded jets}), and in fact, is the only regime that \bhjet can probe self-consistently so far \citep[][]{lucchini2022BHJet}. With the current improvement of this work, we can now further explore the jet kinematics to other regimes where the distribution of the internal energy density is important in the evolution of the jet dynamics and the electromagnetic spectrum.

In the top right panel, where we assume a pair-dominated jet ($\eta_e\gg 1$) that accelerates only leptons, we see that the initial magnetisation converts almost equally to bulk kinetic energy and internal energy ($h$ is now comparable to $\gamma$). The initial specific enthalpy at the jet base is larger compared to the previous case and based on equation~(\ref{eq: mu}), we see that also $\mu$ has significantly increased (see also section~\ref{sec: specific enthalpy}). 

In the bottom left panel of Fig.~\ref{fig: mu and components along the jet} where we account for hadronic acceleration with $\langle \varepsilon_{\rm p}\rangle=4$, we see that the initial magnetisation dissipates almost equally to kinetic and internal energy. The initial specific enthalpy is negligible at the cold jet base but when particles accelerate at the acceleration region, $h$ increases to values comparable to \g. Finally, in the bottom right panel where the jet is pair-dominated, we see that the specific enthalpy at the jet base is of the order of 1 but still much smaller than the initial magnetisation. 

In Fig.~\ref{fig: mu and components along the jet}, according to the approach we follow here, $h$ can overall be significant for the jet evolution depending on the hadronic acceleration and the jet content. The former, in particular, strongly depends on the jet properties, but we cannot capture this non-linear behaviour of the jet evolution, its effect on the particle acceleration and the consequent feedback of particle acceleration back to the jet evolution without significantly increasing the computational cost of the model. However, we can still investigate the jet properties to gain a better insight on jet physics.

In Appendix~\ref{app: mu all plots} we present a more detailed series of jet evolution for various jet quantities and different average particle Lorentz factors. Overall, we find that for many physical scenarios, the specific enthalpy becomes important for the jet evolution, especially in the case where hadrons accelerate in the jets as well, and for pair-dominated outflows (see also section~\ref{sec: specific enthalpy}).

\begin{figure*}
    \centering
\subfigure[Pair-proton outflow with only leptonic acceleration]{

        \includegraphics[width=1.\columnwidth]{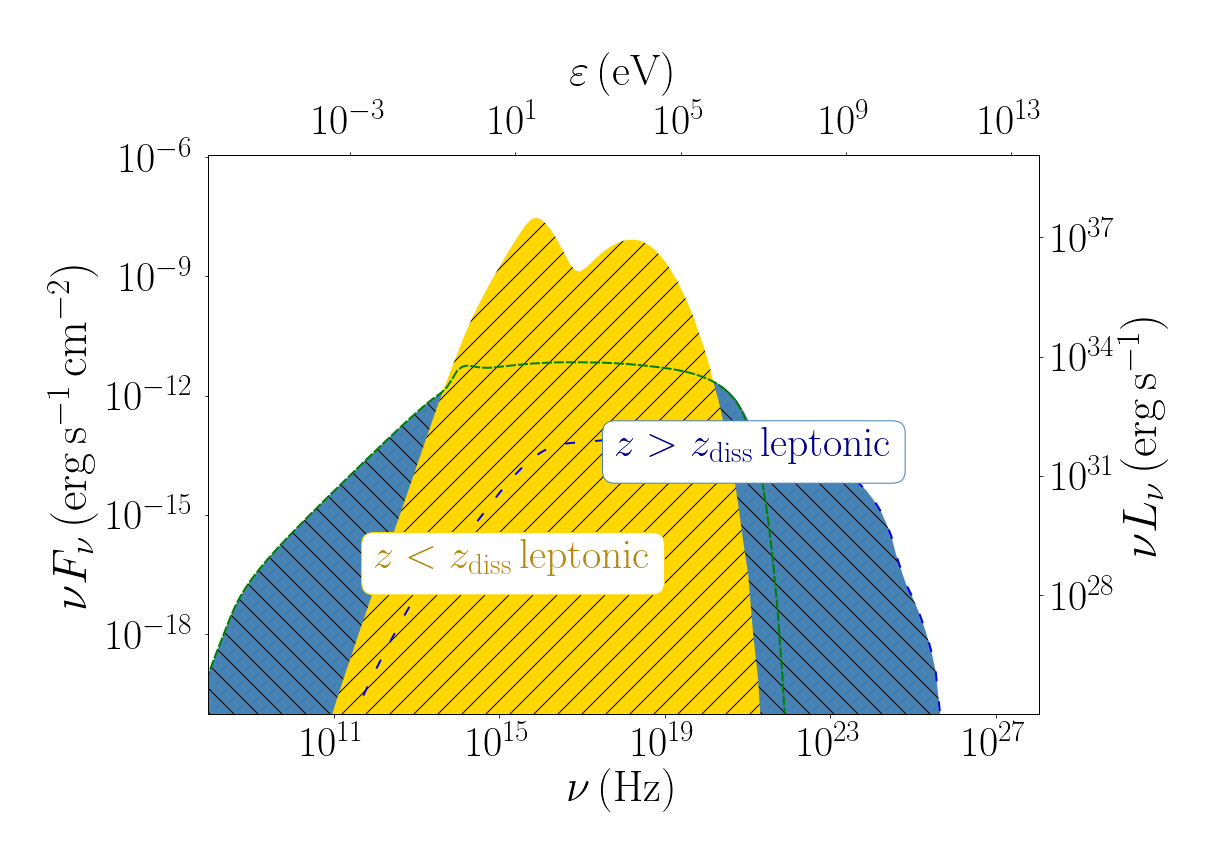}
    }
\subfigure[Pair-dominated outflow with only leptonic acceleration]{
    
        \includegraphics[width=1.\columnwidth]{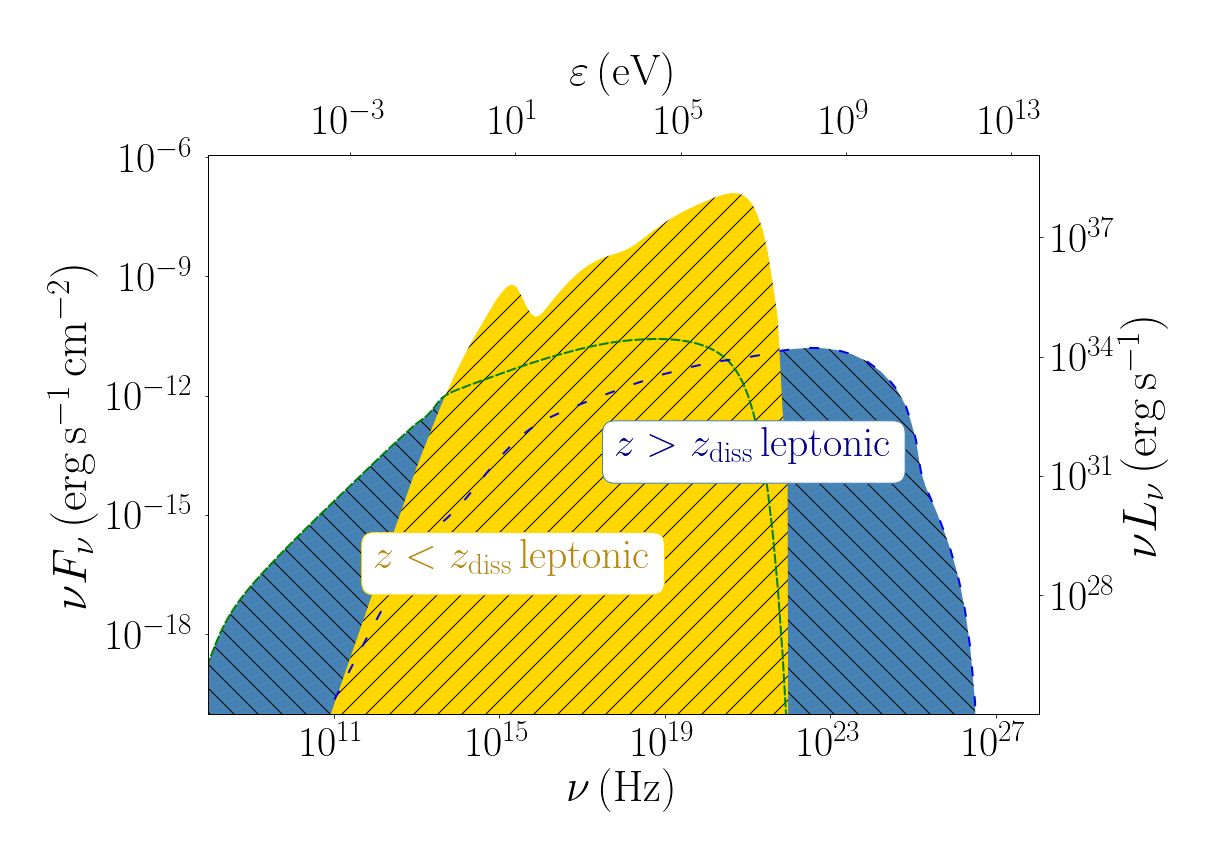}
}
\subfigure[Pair-proton outflow with lepto-hadronic acceleration]{

        \includegraphics[width=1.\columnwidth]{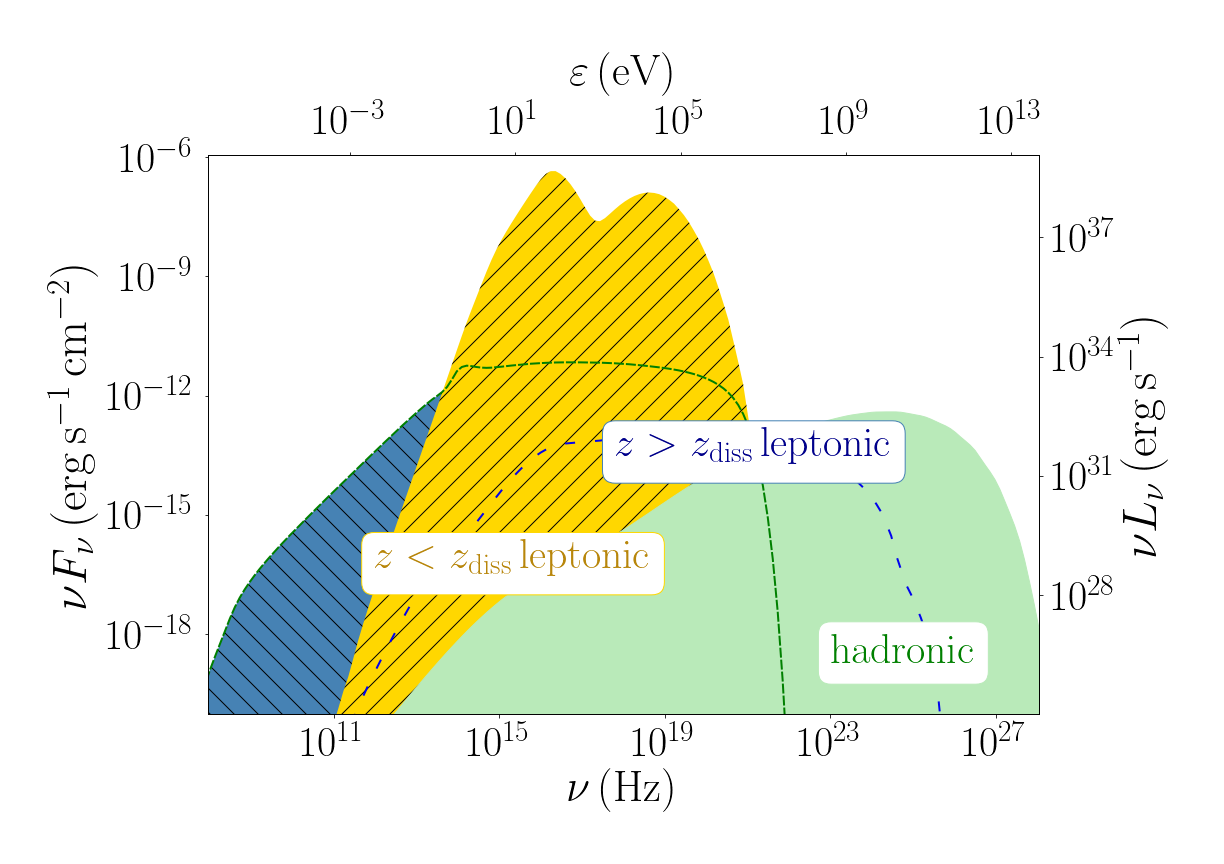}
}
\subfigure[Pair-dominated outflow with  lepto-hadronic acceleration]{

        \includegraphics[width=1\columnwidth]{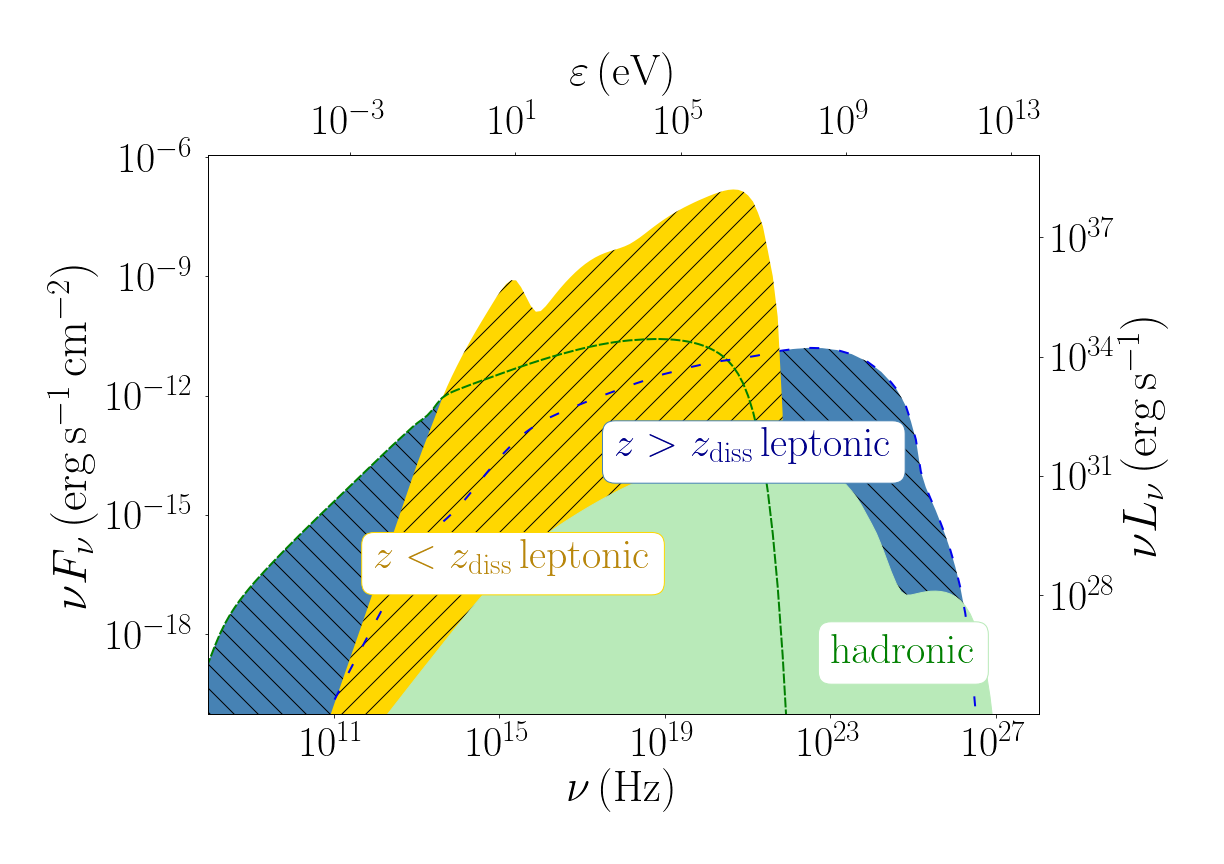}
    }
    \caption{The predicted spectral energy distributions for the four models of Fig.~\ref{fig: mu and components along the jet}. In the \textit{top} panels, we only account for leptonic acceleration, and in the \textit{bottom} ones, we consider both leptonic and hadronic. In the two \textit{left} plots, we assume one proton per electron ($\eta_e=1$) and in the \textit{right} ones we assume $\eta_e=10^4$. In all four panels, we use $k_BT_e = 500\,\rm keV$, and $z_{\rm diss} = 1000\,r_g$ for a 10\,$\rm M_{\odot}$ \bh at 3\,kpc. We also assume $L_{\rm jet} = 2\times 10^{-2}\,L_{\rm Edd}$ for the leptonic scenarios and $L_{\rm jet}=2\times 10^{-3}\,L_{\rm Edd}$ for the hadronic. The aforementioned values lead to $\langle\varepsilon_{\rm e}\rangle = 5$. We highlight the contribution of the jet-segments before the dissipation region (yellow shaded) and that of the jet-segments above the dissipation region (blue shaded). We show the synchrotron emission with densely-dashed green line, and the contribution of the ICS with loosely-dashed blue line. Finally, the green shaded region is the hadronic contribution where we include both neutral pion decay and the synchrotron radiation of the secondary electrons.   }
    \label{fig: SEDs for both leptonic and hadronic}
\end{figure*}

\subsection{Electromagnetic spectrum of steady state jets}\label{sec: resutls of EM spectrum of steady jets}

We plot in Fig.~\ref{fig: SEDs for both leptonic and hadronic} the multiwavelength spectra that correspond to the four different models of Fig.~\ref{fig: mu and components along the jet}. In particular, in the top panels we plot the purely leptonic scenarios, whereas in the bottom we plot the lepto-hadronic models. For the left plots, we assume one proton per electron ($\eta_e=1$), whereas on the right plot we examine the extreme case of $\eta_e=10^4$. 

For all four panels, we assume a quite ``warm'' MJ distribution of leptons with $k_BT_e=1000\,\rm keV$, an initial jet-radius of $10\, r_g$ in which we inject some power equal to $10^{-2}\,L_{\rm Edd}$ for the leptonic models, and $10^{-3}$ for the lepto-hadronic ones. The particle acceleration that happens at $1000\, r_g$ leads to a power-law of particles with an index of 2.2. In all panels, we show the contribution to the spectrum of the jet segments before the dissipation region (yellow-shaded) and above (blue-shaded). For the lepto-hadronic model of the bottom panels, we include the hadronic contribution as green-shaded. Finally, the densely-dashed line shows the synchrotron contribution, whereas the loosely-dashed line corresponds to the ICS. 

In the top left panel of Fig.~\ref{fig: SEDs for both leptonic and hadronic}, we see the emission from the thermal electrons dominates in the UV and X-ray bands, whereas the outer jets dominate in the radio bands via synchrotron radiation, and in the GeV with ICS. In the case where we assume an increased ratio of pairs (top right panel), for the same initial conditions we see once more the emission from the thermal pairs to dominate the UV/X-ray bands but the X-ray luminosity is increased because the initial pair number density has increased (see equation~\ref{eq: n0 at the jet base final}).

In the lepto-hadronic cases of the bottom panels of Fig.~\ref{fig: SEDs for both leptonic and hadronic}, we see that the pair content may significantly affect the SED, and in particular the high-energy part. For the case of one proton per lepton, we see that the GeV-to-TeV spectrum first drops exponentially due to the synchrotron emission of the primary pairs, but later increases due to the hadronic contribution of the \pg interactions. The ICS contribution in this particular case is well below the hadronic contribution (loosely-dashed line). In the pair-dominated jet of the right-hand panel, we see that the increased number of pairs leads to a stronger GeV-to-TeV flux that dominates over the hadronic contribution.

%% file: Sections/Mass_loaded_jets.tex
\section{Mass loaded jets}\label{sec: mass loaded jets}

\begin{table*}
\begin{center}
	\setlength{\tabcolsep}{6pt} 
	\renewcommand{\arraystretch}{1.2} 
	\begin{tabular}[b]{lclc}\hline\hline
		Parameter & Fiducial value(s) & Definition & Status \\ \hline
        $\gamma_{\rm 0}$& 1.11             & bulk Lorentz factor at the jet base $z_{\rm 0}$ & fixed\\
        $\sigma_0$ & $10-50$            & magnetisation of the flow at the jet base & free\\
        $k_BT_e\,/\rm keV$ & 500         & electron peak energy at the jet base & free\\
        $\gamma_{\rm acc}$& $2-10$             & bulk Lorentz factor at $z_{\rm acc}$ & free\\
        $h_{\rm acc}$& $h_0$$^{\dagger}$             & jet specific enthalpy at $z_{\rm acc}$ & fixed\\

        $f_{\rho}$   & 10  & jet mass density increase factor & fixed\\

        $z_{\rm diss}/r_g$&$100$& region where the mass entrainment initiates$^{\ddagger}$ & free\\
        $z_{\rm load,end}/z_{\rm diss}$&$100$& region where the mass entrainment finishes & fixed\\
		\hline
	\end{tabular} 
	\caption{The fixed and the free (fitted) parameters that drive the mass-loading jet dynamics. See Section~\ref{sec: mass loaded jets} for further information.\\
	$^{\dagger}$calculated by the temperature of the electrons at the jet base (see equation~\ref{eq: specific enthalpy}),\\
	$^\ddagger$same as $z_{\rm acc}$.
	}\label{table: mass loading parameters}
\end{center}
\end{table*}

High-resolution GRMHD simulations of accreting black holes that launch highly collimated jets suggest that a significant portion of the wind from the accretion disc might end up in the jet via entrainment. While the jets accelerate in a dense surrounding medium, they are subject to lateral pressure from the wind of the accretion disc that results in jet-wind collisions, causing the jet to wobble. Pinch instabilities form at the jet-wind interface close to the black hole, almost independently of the initial magnetisation of the jet, as long as it starts out Poynting flux dominated. These instabilities dissipate magnetic energy to heat and increases the specific enthalpy of the jet \citep[see, e.g.,][]{Eichler1993magnetic, Bowman1996deceleration, Spruit1997, Begelman_1998, giannios2006role, Bromberg2015Relativistic}. 

Interestingly, two properties of a collimated jet change at distances $\sim10^2-10^3\,r_g$: (1) the toroidal component of the magnetic field starts to dominate over the poloidal component, and (2) the jet speed exceeds the local fast magnetosonic wave speed, i.e., becomes superfast, the magnetic analogue of the fluid becoming supersonic \cltm. Beyond this region, the jet becomes more susceptible to instabilities forming at the interface between the flow and the ambient medium. In particular, magnetic pinch instabilities lead to the formation of eddies that trap matter from the wind and drive it inwards through the jet-wind interface, allowing for mass entrainment \citep[][]{Mignone2013kinked,gourgouliatos2018reconfinement,Bodo2021kink}. Without such eddies, significant mass entrainment into the jet from the external medium may not be possible due to the jet's strong magnetic field. Hence, we link the region where the mass-loading becomes important because of instabilities explicitly to the region where non-thermal particle acceleration occurs. 
Following the results of \cltm, we connect this region to the first particle acceleration region of jets as originally proposed by \cite[][]{Markoff2005,polko2014linking}.

In this work, we parametrise the fiducial model B10 of \cltm to derive a semi-analytical formalism that connects the mass loading region to the particle acceleration region, and study its impact on the emitted electromagnetic spectrum by studying both the leptonic and the hadronic processes we discussed above. We consider B10 for our problem because the jet undergoes strong collimation out to very large scales.  Other models explored in \cltm either have too small an accretion disk, such that there is hardly any lateral pressure from the disk wind. The jets therefore become uncollimated and thus conical within $1000\,r_g$, and hence do not properly represent the highly collimated, parabolic, large-scale jets we are targeting. Further, the fast lateral expansion of the uncollimated jet suppresses pinch instabilities \citep[][]{Moll2008kink,Granot2011impulsive,Porth2015causality} and thus exhibits little to no mass-loading (\cltm).

\cltm confirm that the magnetic energy converts to kinetic energy, accelerating the jets similar to what was found in previous works \citep[][]{McKinney2006,Komissarov2007magnetic,Komissarov2009ultrarelativistic}. When matter is entrained by the jets, further magnetic energy is dissipated to heat up the jet, and the inertia of the entrained gas slows down the jet. The mass entrainment leads to a decrease of the total (specific) energy flux $\mu$ along the jets up to the distance where the mass loading stops. Beyond distances of a few $10^4\,r_g$, the \cltm jet properties have not achieved steady-state as the slow, mass-loaded jet is still punching through the ambient medium at this point of time in the GRMHD simulation. Indeed, the simulations suggest that as the jet slowdown due to massloading suppresses pinch instabilities further along the jet, and therefore, massloading becomes considerably weaker beyond $10^4\, r_g$. As a result, when the simulated jets attain steady-state out to $\gtrsim10^5\, r_g$, we expect that $\mu$ should be conserved for the rest of the jets and there would be jet re-acceleration while both the magnetisation and the specific enthalpy decrease.

\begin{figure}
    \centering
    \includegraphics[width=1.1\columnwidth]{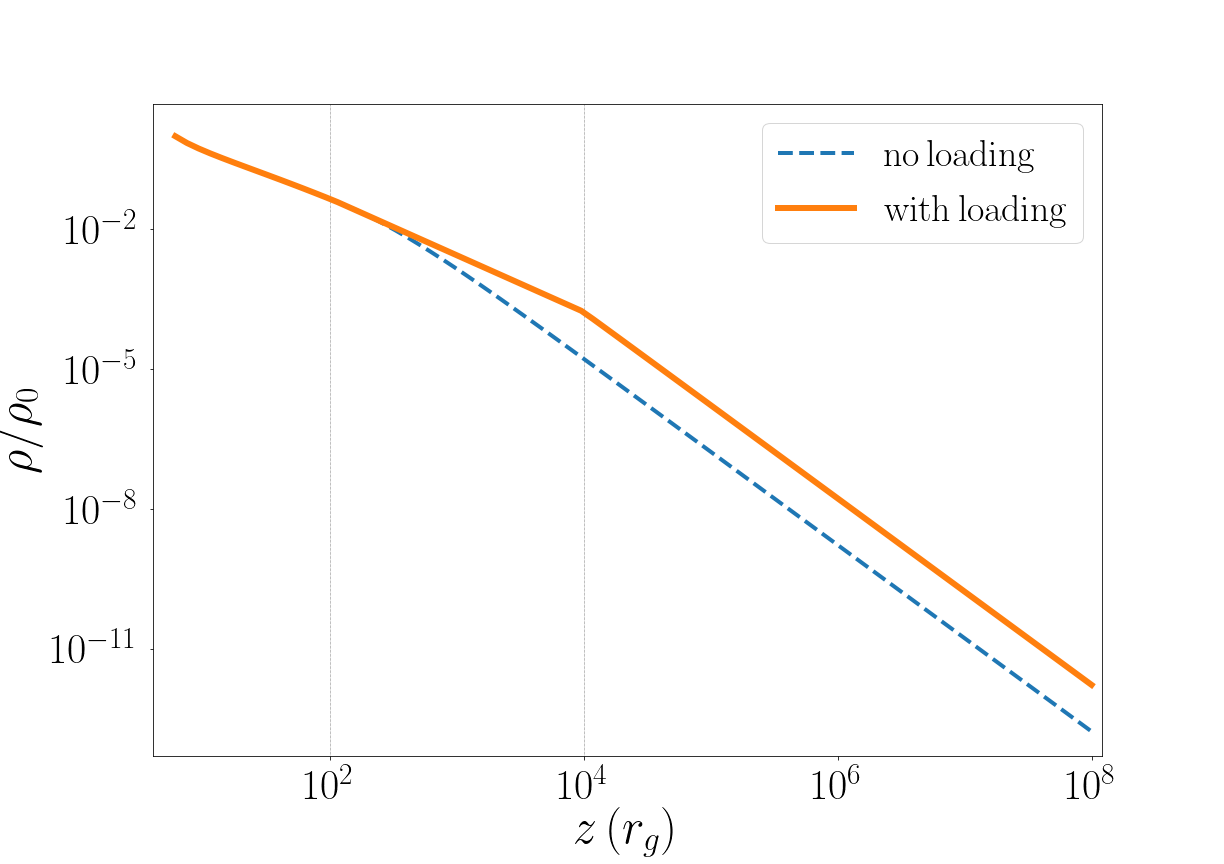}
    \caption{The mass density profile of a mass loaded jet (solid line) compared to a steady state jet without mass loading (dashed line). Both profiles are normalized to the initial mass density at the jet base. The mass loading initiates at a distance $z_{\rm diss}$ and at $100\,z_{\rm diss}$ the mass density has increased by a factor of 10 compared to a non-loading, steady-state jet. 
    }
    \label{fig: mass density with loading}
\end{figure}

\begin{figure}

    \centering
    \includegraphics[width=1.1\columnwidth]{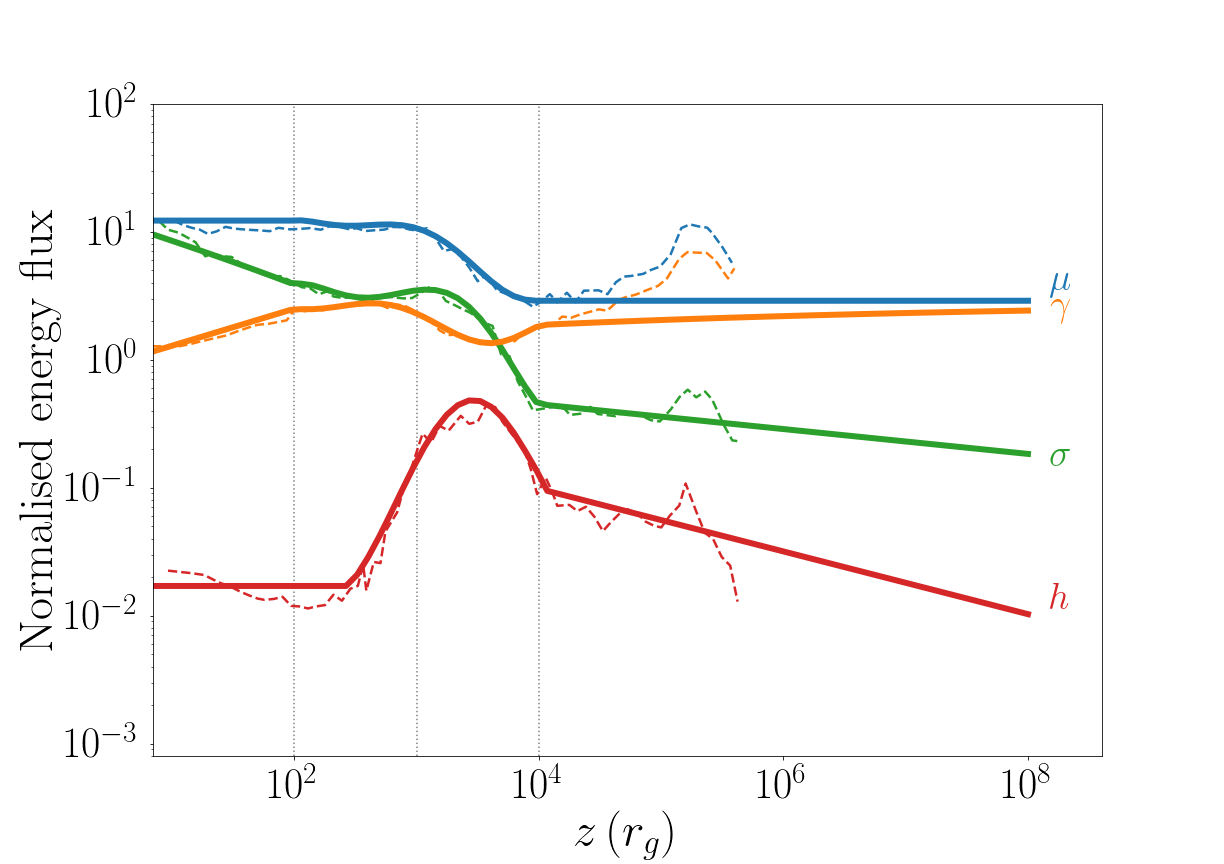}
    \caption{The energy flux components of a mass loaded jet, where $\mu$ is the ratio between the total energy flux and the rest-mass flux, \g is the bulk Lorentz factor, $\sigma$ is the magnetisation, and $h$ is the specific enthalpy. The mass entrainment occurs between $10^2$ and $10^4\,r_g$ (vertical lines), but the entrained matter becomes comparable to the mass of the jet at a distance of $10^3\,\rm r_g$ (middle vertical line). Finally, we over-plot with dashed lines the fiducial model B10 of \cltm on which we base our analysis (see Section~\ref{sec: mass loaded jets}). 
    }
    \label{fig: jet evolution with mass loading and comparison to CLTM19}    
\end{figure}

Inspired by the simulation results, our semi-analytical ``mass-loaded'' jet model assumes that the mass loading initiates at a distance $z_{\rm diss} \sim 100\,r_g$ and ends at $100\,z_{\rm diss}$, with a net increase of $f_{\rho}=10$ in the jet mass density. Beyond $100\,z_{\rm diss}$, we assume a constant $\mu$ and steady jet acceleration. In Fig.~\ref{fig: mass density with loading}, we plot the mass density of a mass loaded jet (solid line) and compare it to a non-loaded steady state jet, assuming one proton per lepton. We show the resulting energy components (\g, $\sigma$, and  $h$) of the B10 model of \cltm in Fig.~\ref{fig: jet evolution with mass loading and comparison to CLTM19} with dashed lines, and below, we discuss the way we parametrise these quantities.


\subsection{Mass loading region}
In this section, we present the parametrisation of the values of $\sigma$, \g\ and $h$ of the mass loading region based on the B10 model described above. In particular, we fit a polynomial to the \cltm profiles along the jet between $z_{\rm diss}=100\,r_g$ and $z_{\rm load,\, end}=10^4\,r_g$. 
The profiles of the three quantities $\sigma$, $\gamma$ and $h$ are hard to predict in such a complex and non-linear system, we hence decide to fit only for these three quantities and derive $\mu$ from equation~(\ref{eq: mu}).
\begin{equation}\label{eq: polynomial for sigma}
    \begin{split}
        \log_{10}(\sigma) =
0.621\, x^5 - 3.005\, x^4 + 4.599\, x^3\\ - 2.502\, x^2 + 0.242\, x + 0.563,
    \end{split}
\end{equation}
\begin{equation}\label{eq: polynomial for gamma}
    \begin{split}
        \log_{10}(\gamma) = 
-0.276\, x^6 + 1.412\, x^5 - 2.207\, x^4 + 0.853\, x^3\\ + 0.257\, x^2 - 0.075\, x + 0.394,
    \end{split}
\end{equation}
\begin{equation}\label{eq: h specific enthalpy polynomial}
    \begin{split}
        \log_{10}(h) = 
        0.467\, x^5 - 1.903\, x^4 + 1.100\, x^3\\ + 2.482\, x^2 - 1.171 \,x - 1.826,
    \end{split}
\end{equation}
where $x=\log_{10} \left(z/z_{\rm diss}\right)$ and $1\le x \le \log_{10} \left(z_{\rm diss}/z_{\rm load,\, end}\right)$.

We connect the jet base to the mass-loading region assuming that the specific enthalpy is constant to its initial value at the jet base as we calculate it with equation~(\ref{eq: specific enthalpy}). We assume that the flow is launched at some speed equal to the speed of sound (see equation~\ref{eq: g jet before gmax jet}) and reaches a value $\gamma_{\rm acc}$, which is a free parameter, following a logarithmic dependence. In Table~\ref{table: mass loading parameters}, we show the parameters of the mass-loading jet model, indicating whether they are fixed or fitted parameters. 

\subsection{Jet segments beyond the mass loading region}
Given our assumption that once mass-loading stops, the total energy flux is again conserved, i.e., $\mu$ is constant. Thus we fix $\mu$ at its value at the end of the mass-loading region, and to better constrain the profile of $\sigma$ and $h$ beyond the mass-loading region, we fit a first order polynomial between $10^4$ and $10^5\,\rm r_g$, with coefficients: 
\begin{equation}\label{eq: sigma above mass loading}
    \log_{10}(\sigma) = -0.097\, x - 0.178,
\end{equation}
\begin{equation}
    \log_{10}(h) = -0.245\, x - 0.576,
\end{equation}
where $x$ is the same as above. Here we choose to interpolate the profile of $\gamma$ and $\sigma$ from the simulation data that closely follows the expected slow acceleration profile seen in semi-analytical MHD solutions of particle-dominated ($\sigma \lesssim 1$) jets \citep[see e.g.,][]{Tchekhovskoy_2009}.

Having derived the values of $\mu$, $\sigma$ and $h$, we calculate the bulk Lorentz factor for every jet segment above the $z_{\rm diss}$
\begin{equation}\label{eq: h for mass loading}
    \gamma (z\ge z_{\rm diss})=\frac{\mu}{\sigma+h +1}.
\end{equation}

\subsection{Particle acceleration and mass loaded jets}
At the location where matter is entrained into the jets, particles start to accelerate to non-thermal energy as well. Based on the definition of $h$, we solve for the energy density of the protons
\begin{equation}\label{eq: normalisation of accelerated particles}
    U_p = \frac{h\rho c^2 - \Gamma_e U_e}{\Gamma_p},
\end{equation}
where we calculate $U_e$ from equation~(\ref{eq: U internal energy density integral}) for an MJ+non-thermal power-law distribution of electrons with a fixed ratio of 10 per cent between the thermal and the non-thermal electrons, and a fixed power-law slope $p$. We finally, derive the normalisation of the non-thermal protons
\begin{equation}
    K_p = \frac{U_p}{\rm{ m_pc^2}\int \varepsilon^{-p+1}\exp(-\varepsilon/\varepsilon_{\rm max}){\rm d}\varepsilon},
\end{equation}
where
\begin{equation}
    \frac{{\rm d}n_p}{{\rm d}\varepsilon} = K_p \varepsilon^{-p} \exp (-\varepsilon/\varepsilon_{\rm max}).
\end{equation}

Following the above approach, we manage to self-consistently connect the mass-loading that leads to an increase of the specific enthalpy $h$ to the electromagnetic radiation due to the proton acceleration.

%% file: Sections/Mass_loaded_jets_results.tex
\section{results for mass-loaded jets}\label{sec: results on mass-loaded jets}
\subsection{Total energy flux evolution for mass-loaded jets}\label{sec: mu for mass-loading jets}

In Fig.~\ref{fig: mu for mass loading, shifted and renormalised}, we present the energy components for two different mass-loaded jets following the prescription of Section~\ref{sec: mass loaded jets}. We assume that both jets are Poynting flux dominated at the jet base with an initial magnetisation of $\sigma_0=40$ and accelerate to a bulk Lorentz factor of $\gamma_{\rm acc}=3$. In the left panel, we assume one electron per proton at the jet base $\eta_e=1$, and in the right, we assume a pair-dominated jet of $\eta_e = 10000$.  In both cases, we set the temperature of the thermal electrons at the jet base at $k_BT_e = 200\,\rm keV$. In the particular case of the pair-dominated jets, the specific enthalpy reaches values that are comparable to or even exceeding that of the bulk Lorentz factor and the magnetisation, especially at the loading region (see also equation~\ref{eq: h specific enthalpy versus eta}). Despite the initially pair-dominated jet base, the matter entrained into the jets is in approximately equal number of electrons and protons because we assume that the most likely composition of an accretion disc wind is a neutral gas of electrons and protons. The jet composition hence changes from pair-dominated at the regions before the loading to almost equal number of protons and pairs \citep{Angles-Castillo2021decelaration}.

In the right panel of Fig.~\ref{fig: mu for mass loading, shifted and renormalised}, we see that the increased number of pairs at the jet base leads to an increase of $h$. In the extreme case where $\eta_e \gg 1000$, the peak of the profile of $h$ may lead to an artificial and unphysical increase of $\mu$ in the loading region. In Appendix~\ref{app: constrain h} we discuss how we constrain the increase of $h$ to avoid such an artificial ``mass loss''.

\subsection{Electromagnetic spectra of steady state mass-loaded jets}

\begin{figure*}
    \centering
	\begin{minipage}{\columnwidth}
        \includegraphics[width=1.05\columnwidth]{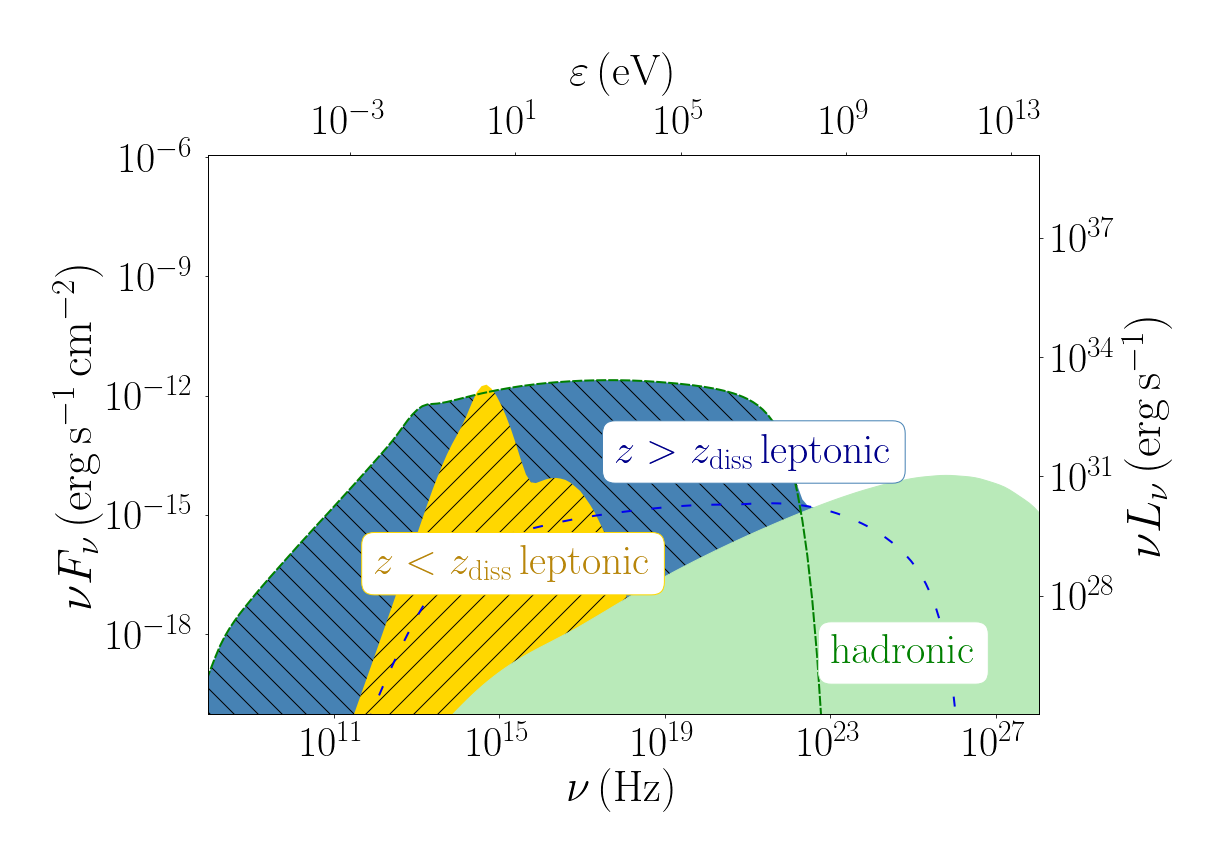}
    \end{minipage}
    \begin{minipage}{\columnwidth}
        \includegraphics[width=1.05\columnwidth]{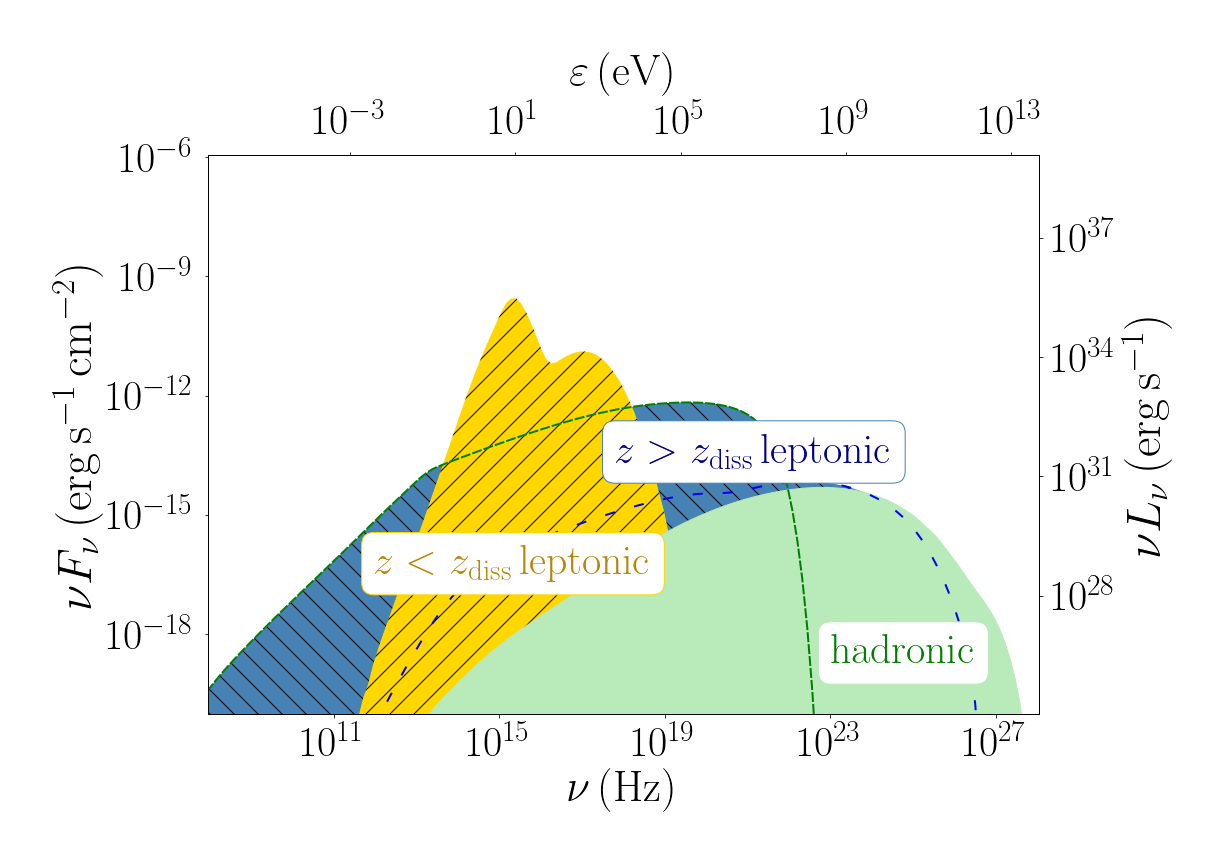}
    \end{minipage}
    \caption{\textit{Left}: The predicted spectral energy distribution of a mass-loaded jet that corresponds to the dynamical quantities of 
    Fig.~\ref{fig: jet evolution with mass loading and comparison to CLTM19} for a 10\,$\rm M_{\odot}$ \bh at 3\,kpc. We assume a jet base of 200\,keV and radius of 10\,$r_g$. We show the contribution of the jet-segments before the mass loading (yellow shaded region), and the contribution of the mass-loaded segments of both leptonic (blue-shaded) and hadronic (green-shaded). The hadronic contributes includes both the neutral pion decay and the synchrotron radiation of the secondary electrons/positrons. \textit{Right}: Similar to the left, but for a non-loaded jet with similar initial conditions. 
    } 
    \label{fig: SED for fiducial mass loading jet}
\end{figure*}

In Fig.~\ref{fig: SED for fiducial mass loading jet}, we plot in the left the predicted SED of the fiducial mass-loaded jet model based on the dynamical quantities that we show in Fig.~\ref{fig: jet evolution with mass loading and comparison to CLTM19}. We further assume a jet base of radius $10\,r_g$, an electron temperature of $200\, \rm keV$ at the jet base, an injected jet power of $10^{-3}\, L_{\rm Edd}$ and the power-law slope of the accelerated particles $p=2.2$ for both leptons and protons. Similar to above, we show the contribution of the leptonic emission of the jet segments before the dissipation/loading region, the leptonic contribution from the dissipation/loading region and beyond, and the hadronic contribution that is due to \pg. In the right subplot, we show the spectrum of a non-loaded jet with similar initial conditions. The main differences are in the jet emission from the jet base (yellow-shaded region) and the hadronic contribution. The jet-base emission is higher in the non-loaded case due to the magnetisation profile we assume here that leads to greater values for the first few jet segments up to the acceleration region (see, e.g., Fig.~\ref{fig: mu and components along the jet}).  

\begin{figure*}
    \centering
	\begin{minipage}{\columnwidth}
    \includegraphics[width=1.05\columnwidth]{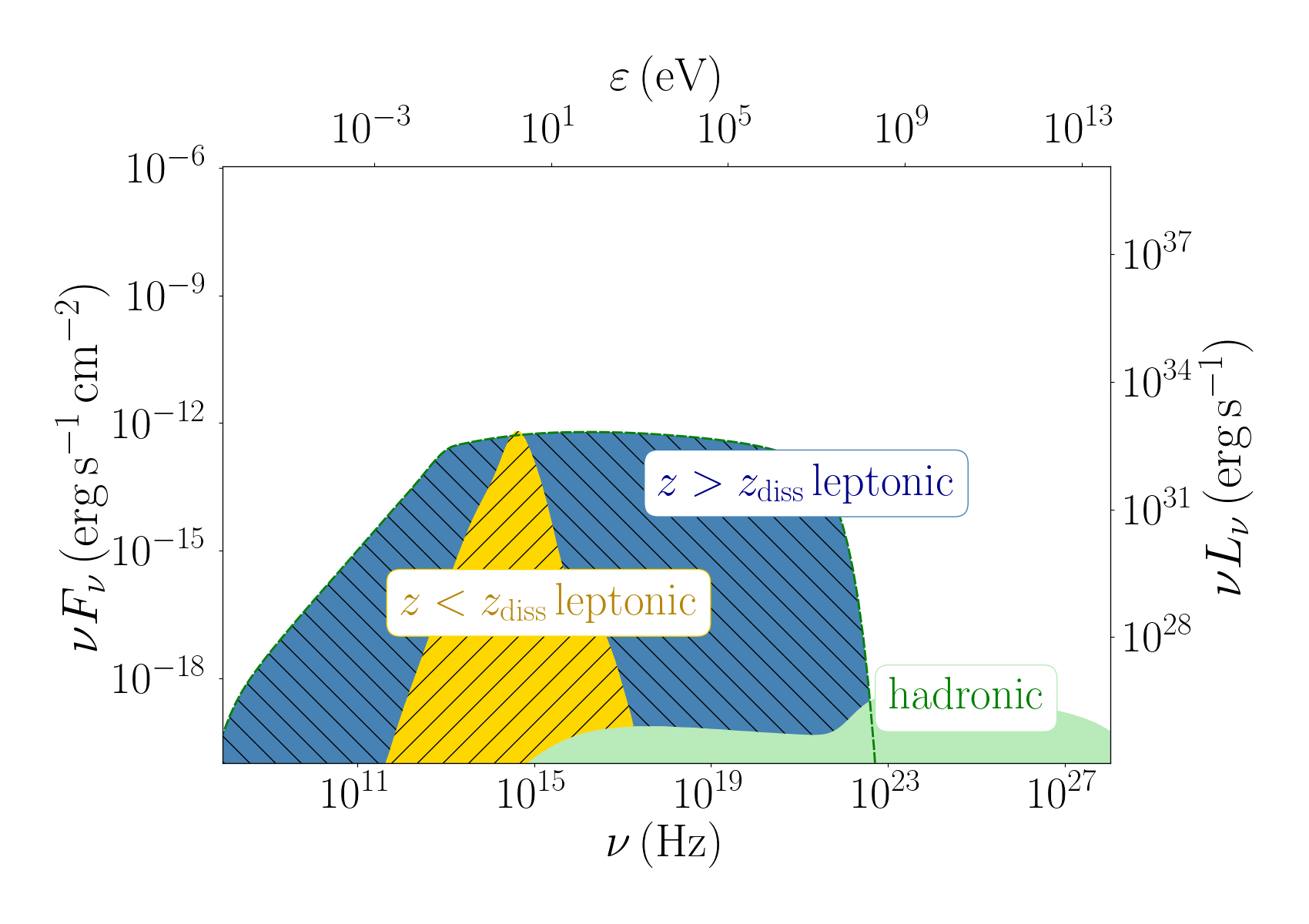}    
    \end{minipage}
    \begin{minipage}{\columnwidth}
    \includegraphics[width=1.05\columnwidth]{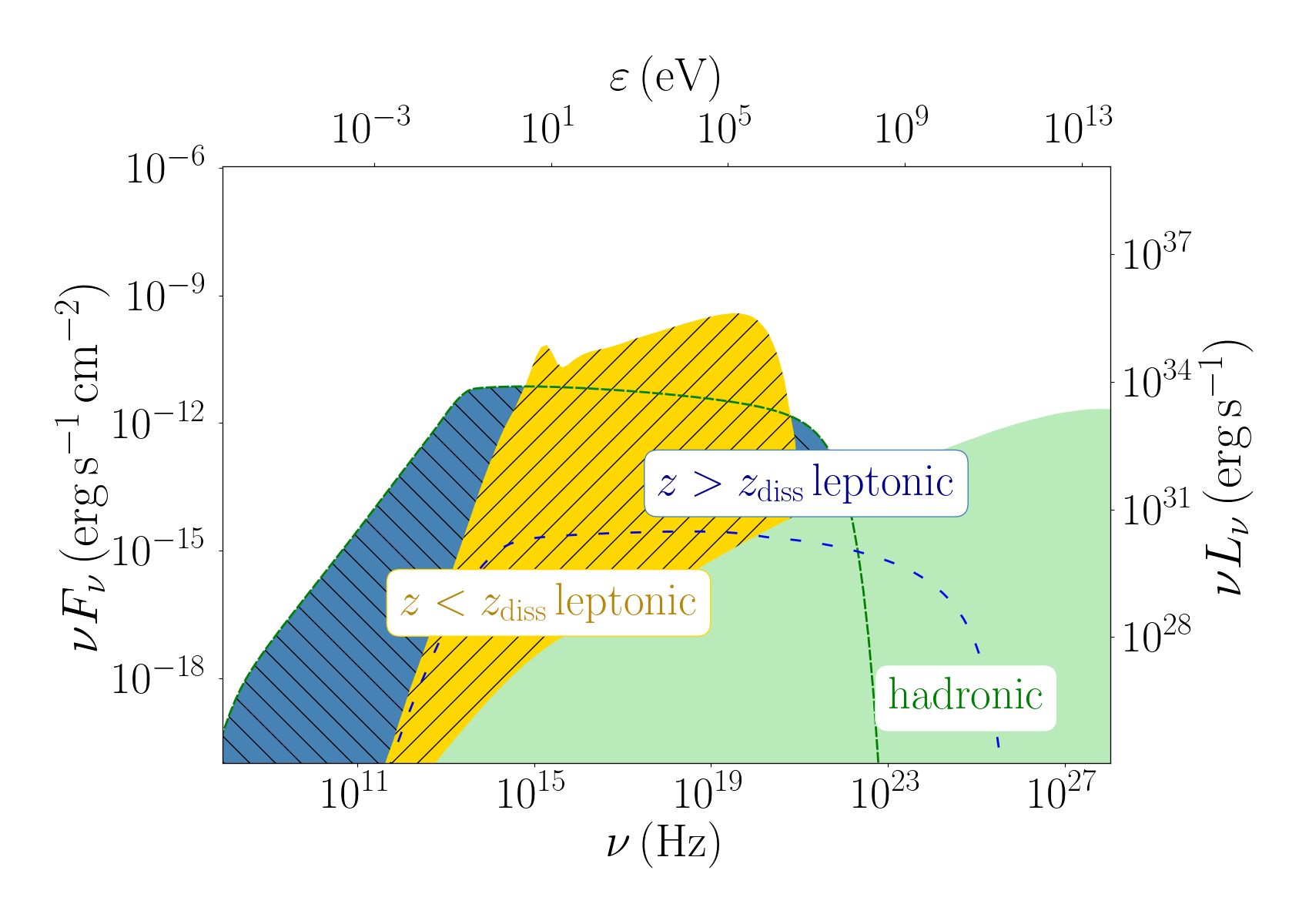}    
    \end{minipage}
    \caption{Similar to Fig.~\ref{fig: SED for fiducial mass loading jet} but for the mass-loaded jets that correspond to the dynamical quantities of Fig.~\ref{fig: mu for mass loading, shifted and renormalised}. The overall spectral distribution can significantly change under the assumption of a pair-dominated jet base ($\eta_e = 10000$) in the \textit{right} plot.
} 
    \label{fig: SEDs for mass loading jets}
\end{figure*}

In Fig.~\ref{fig: SEDs for mass loading jets} we plot the SEDs that correspond to the two models of Fig.~\ref{fig: mu for mass loading, shifted and renormalised}, where we account for mass loading at a distance $100\,r_g$ and  we assume the injected jet power to be $10^{-3}\, L_{\rm Edd}$. We show the different components of the spectrum in Appendix~\ref{app: sed components}.



\begin{figure*}
    \centering
	\begin{minipage}{\columnwidth}
    \includegraphics[width=1.1\columnwidth]{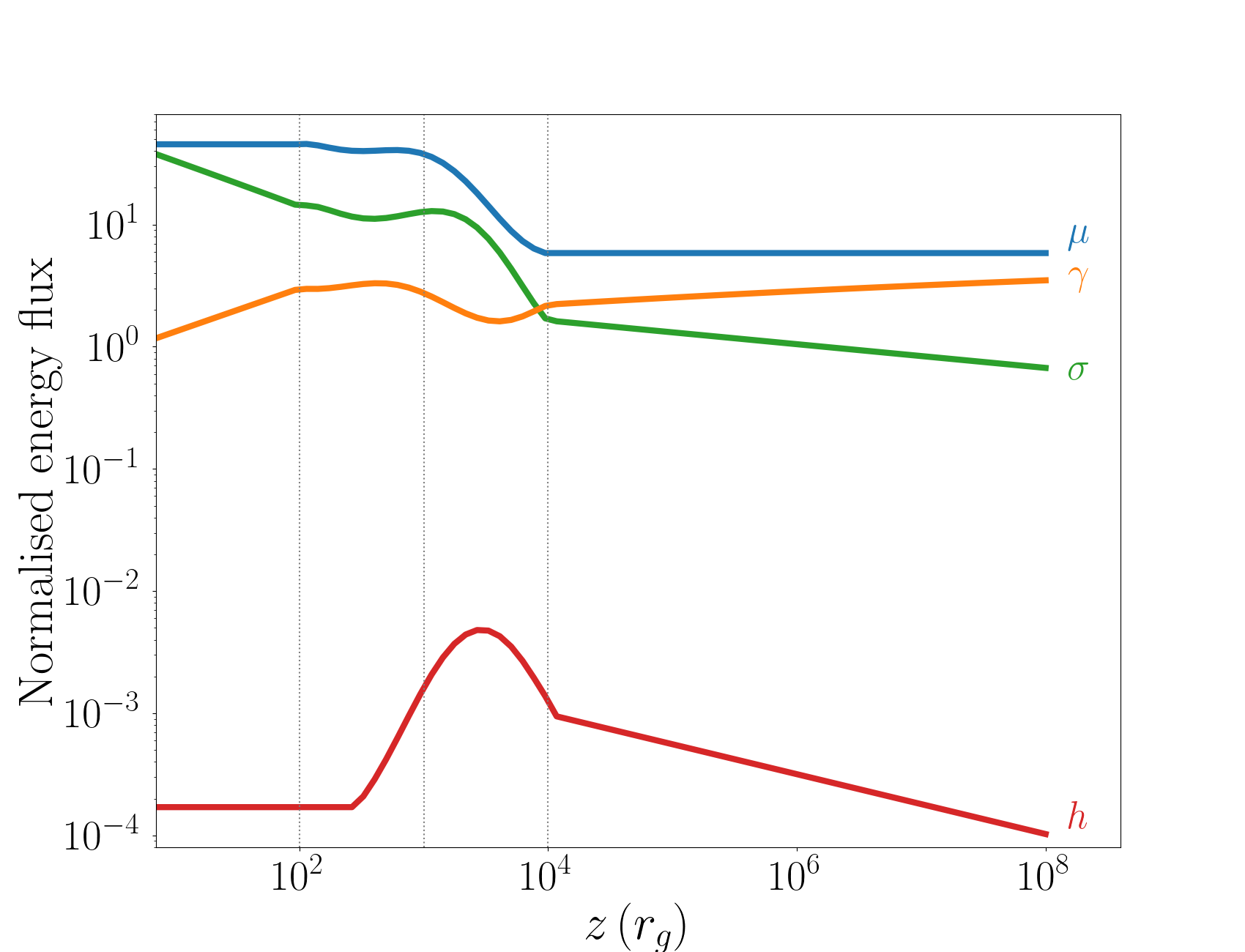}
    \end{minipage}
    \begin{minipage}{\columnwidth}
    \includegraphics[width=1.1\columnwidth]{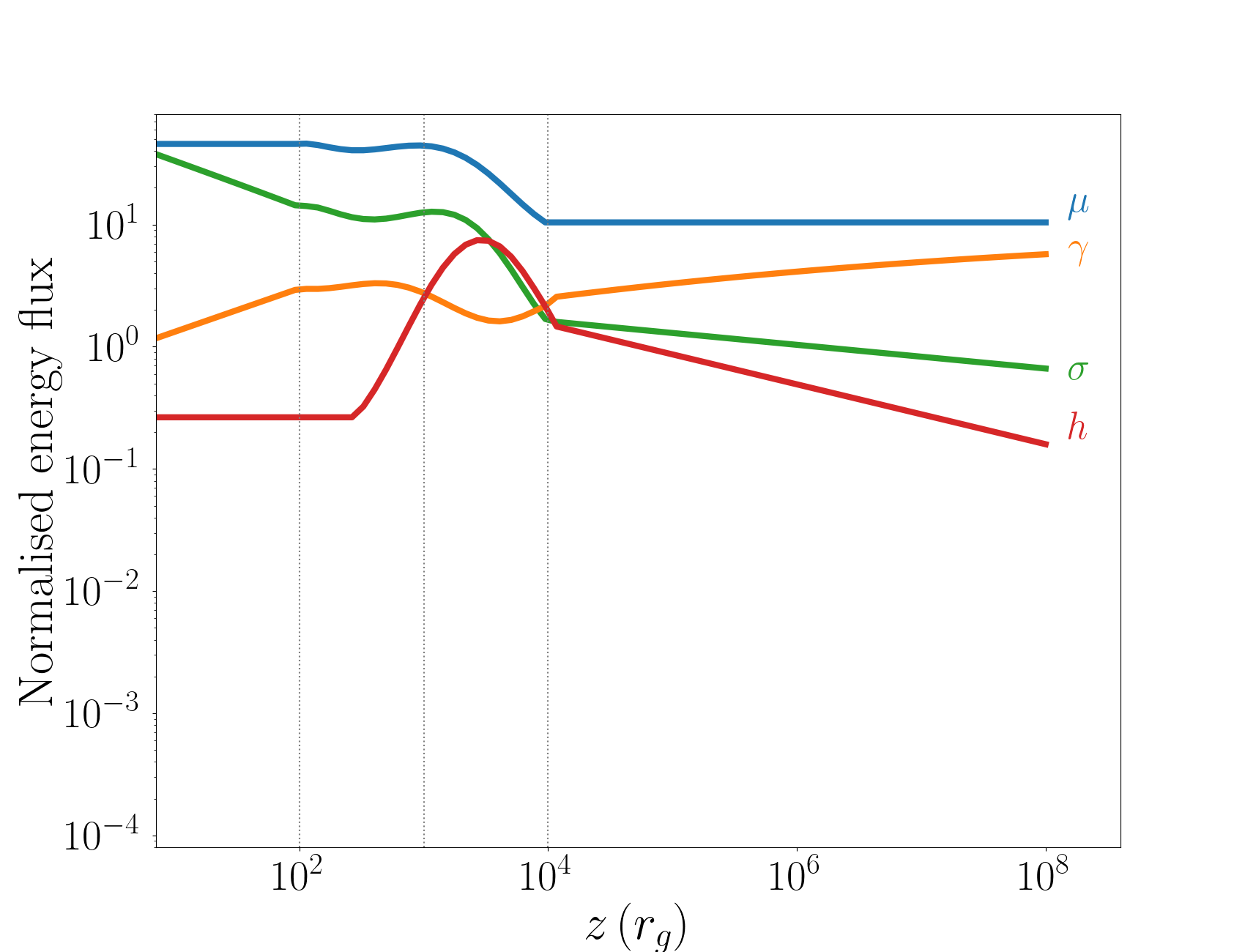}    
    \end{minipage}
    \caption{Similar to Fig.~\ref{fig: jet evolution with mass loading and comparison to CLTM19}, but for: \textit{left} an initial lepton temperature at the jet base of $k_BT_e=200\,\rm keV$ and one electron per proton ($\eta_e=1$), and in the \textit{right} for a pair-dominated jet ($\eta_e=10000$). Both scenarios are for an initial magnetisation of $\sigma_0=40$ and $\gamma_{\rm acc}=3$. The increased pair content of the right subplot leads to an increased initial specific enthalpy of the jets.
} 
    \label{fig: mu for mass loading, shifted and renormalised}
\end{figure*}

%% file: Sections/Discussion.tex
\section{Discussion}\label{sec: discussion}
\subsection{Steady state jets}
In the first part of this work, we present the analytical jet model that includes the specific enthalpy in the jet kinematics and the spatial evolution.

\subsubsection{Specific enthalpy, particle acceleration and jet evolution}

The specific enthalpy $h$ is a good estimate of whether a jet is cold or hot, with values of $h \ll 1+\sigma$ to indicate a cold flow, and values of $h\gtrsim 1+\sigma$ to indicate a hot flow. Astrophysical jets launched by black holes are overall considered cold and strongly magnetised. The majority of semi-analytical models that focus on the radiative output rather than the detailed description of the jets, neglect the specific enthalpy for simplicity \citep[][]{Markoff2005,Bosch_Ramon2006,Vila2012,zdziarski2014jet}. When particles accelerate though, and in particular in the case where these accelerated particles carry a significant fraction of the jet energy, the specific enthalpy increases. As we show in Fig.~\ref{fig: specific enthalpy}, the exact value of the specific enthalpy may get values that can easily compare to the bulk Lorentz factor (values of the order of 1 and $\sim 10$) and/or the jet magnetisation (values greater than unity for a magnetised outflow). The exact value of $h$ strongly depends on three aspects: the matter composition of the jet, the efficiency of the leptonic acceleration, and whether hadrons accelerate as well or not.  

\paragraph{Leptonic acceleration}
In the case where only leptons accelerate inside the jets, we expect the electron average Lorentz factor to increase as the acceleration efficiency increases (top subplots of Fig.~\ref{fig: specific enthalpy}) and hence the specific enthalpy to increase as well, according to equation~(\ref{eq: h specific enthalpy versus eta}).
The total specific enthalpy however
depends on the jet composition as well. When a jet is of one electron per proton ($\eta_e =1$), the values of $h$ are $\sim 0.01$ regardless of the exact average Lorentz factor of the electrons (as long as the average Lorentz factor of the leptons remains less than $\rm m_p/m_e\simeq 1836$). This is the typical scenario that current GRMHD and semi-analytical jet models consider when studying the exact jet evolution, both in space and time. As we mention above though, based on observations of both extragalactic and Galactic jets, it is very likely that jets are pair-dominated (or at least the scenario of one proton per electron is disfavoured in some cases). Such a jet content leads to an increase in the specific enthalpy compared to the case of $\eta_e=1$ (see equation~\ref{eq: h specific enthalpy versus eta}). 
The specific enthalpy hence of a jet that is pair-dominated at launching may contribute significantly to the spatial evolution of the jet, and the more relativistic (or warmer) the distribution of pairs, the larger the impact of $h$ on the jet evolution. 
A pair dominated jet in fact requires specific enthalpy that can be two to three orders of magnitude larger than the jet case of an equal number of electrons and protons (see, e.g., top plots of Fig.~\ref{fig: specific enthalpy}). Consequently, to achieve bulk flow acceleration up to the same bulk Lorentz factor, a pair-dominated jet, also requires a larger value of magnetisation at the jet base if energy flux is conserved along the jet.

\paragraph{Lepto-hadronic acceleration}
The energy content of the particles can further increase when jets accelerate both leptons and hadrons to non-thermal energies. In fact, the more efficient the particle acceleration, the larger the specific enthalpy, 
which may get values of the order of $\Gamma_p \langle \varepsilon_p \rangle$, regardless of the jet content, as long as $\eta_e \leq \rm m_p/m_e$ (Fig.~\ref{fig: specific enthalpy}).
It is hard to predict the exact value of the specific enthalpy in a jet that efficiently accelerates particles, but overall, it may get values equal to or even exceed that of the bulk Lorentz factor and/or the magnetisation, that would mean that the outflow converts to particle dominated instead. We hence suggest that the specific enthalpy should be treated with extra care and should not always be considered negligible.

\subsubsection{Specific enthalpy and spectrum}

The SED of the steady jets strongly depends not only on the hadronic acceleration (or lack of it), but also on the jet content. The most important difference is in the GeV-to-TeV spectrum. 
A pair dominated jet is characterised by the ICS and any contribution from the hadronic processes is suppressed. 
In the case of a jet with equal number of protons and pairs, and accounting for an efficient hadronic acceleration, the hadronic component dominates in the GeV/TeV bands via the neutral pion decay, which has a distinguishable shape than that of ICS in the Klein-Nishina regime.

The IR-to-X-ray spectrum of \bhs may be contaminated by different components, such as the companion star and/or the accretion disc.
In the case of a pair-dominated jet, though, the X-ray spectrum shows the multiple Compton scatterings due to the increased electron density that can potentially replicate the role of the theoretical corona \citep[][]{Markoff2005,Markoff_2015,Lucchini2021correlation,cao2021evidence}.
Such an X-ray signature can prove a useful tool to distinguish between different jet compositions, especially with the next-generation X-ray telescopes, such as for instance the Imaging X-ray Polarimetry Explorer (IXPE; \citealt[][]{Weisskopf2016IXPE}), the Advanced Telescope for High-energy Astrophysics (Athena; \citealt[][]{Nandra2013Athena}) and the Advanced X-ray Imaging Satellite (AXIS; \citealt[][]{Mushotzky2019AXIS}).

\subsection{Mass-loaded jets -- \hadjet}
The initial jet composition at the jet base significantly alters the specific enthalpy of the jet along its axis, even if we assume that at the mass-loading region the jet converts to a pair-proton outflow. We see, in particular, that a pair-proton jet base with a thermal pair distribution that peaks at some energy of the order of 500\,keV, which is a reasonable value for \bhs, resulting in insignificant specific enthalpy compared to the rest energy components, namely the magnetisation and the bulk Lorentz factor (see, e.g., the left subplot of Fig.~\ref{fig: mu for mass loading, shifted and renormalised}). If the jet base, on the other hand, is pair-dominated, then similar to our discussion above, the initial specific enthalpy at the jet base is increased and hence its effect on the jet dynamical evolution might be more important because the energy content carried by particles might be similar to the bulk kinetic energy (see, e.g, the right subplot of Fig.~\ref{fig: mu for mass loading, shifted and renormalised}).

\begin{figure}
    \includegraphics[width=1.1\columnwidth]{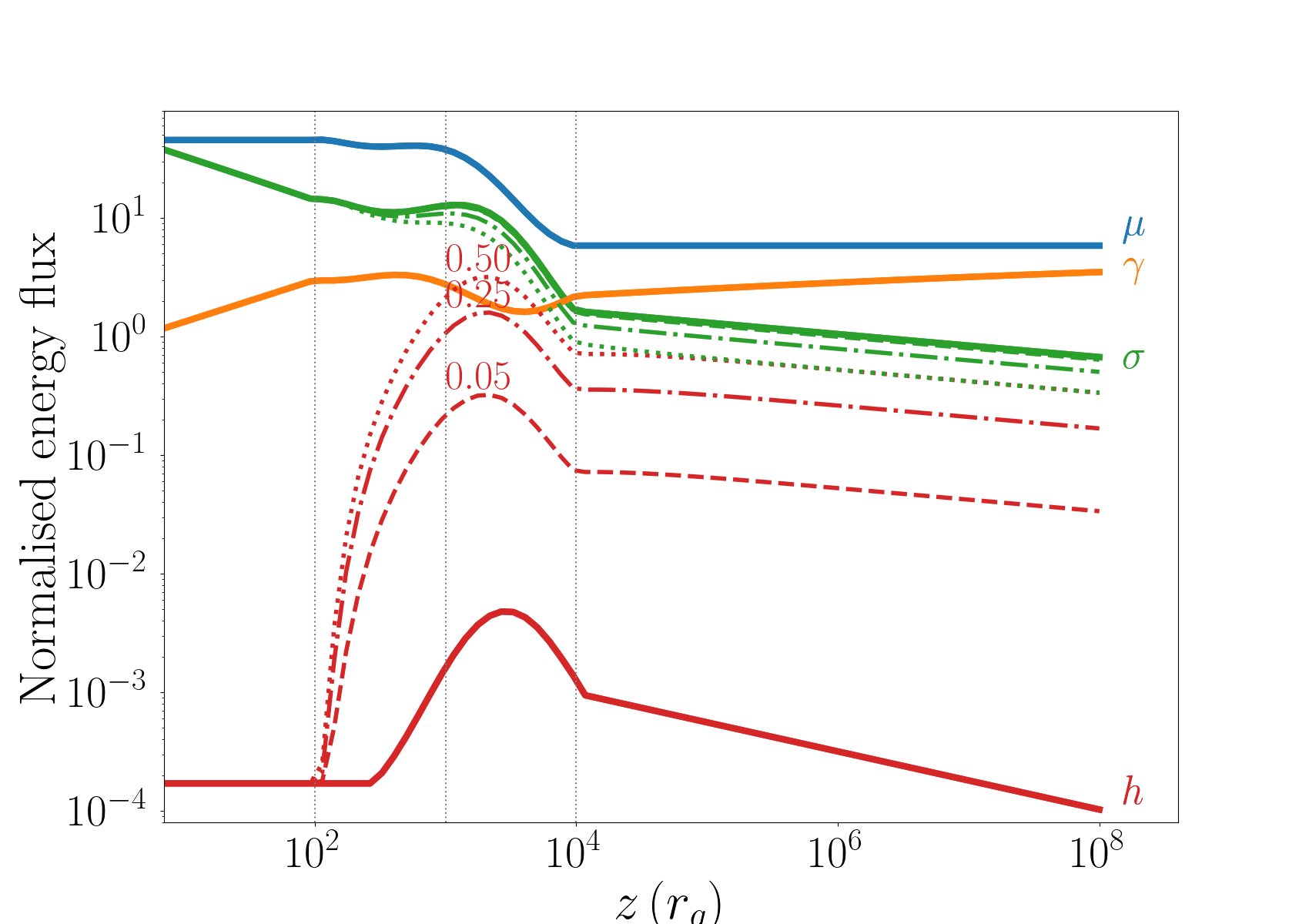}
    \caption{Similar to the left sub-plot of Fig.~\ref{fig: mu for mass loading, shifted and renormalised}, but with different $f_{\rm heat}$ parameters as shown in the plot. The $f_{\rm heat}$ parameter expresses the fraction of the magnetic energy that is allocated to the specific enthalpy to allow a further exploration of dissipation beyond our single GRMHD-based paramerisation.  
    } 
    \label{fig: mu et al for different fheat}
\end{figure}

The initial conditions at the jet base have a significant impact on the electromagnetic spectrum that is our tool to distinguish between the different scenarios. For the two different scenarios we study here, where the one shows a pair-proton jet base and the other a pair dominated jet base, there are two prominent differences in the multiwavelength SEDs. The most important one is in the GeV/TeV regime, where the larger specific enthalpy of the initially pair-dominated jet base allows for more energy to be transferred to protons. The increased energy available for non-thermal proton acceleration allows for a stronger TeV flux, which is dominated by the neutral pion decay due to \pg interactions.
Such TeV flux, depending on the distance of the \bh \citep[see, e.g.,][]{kantzas2022gx} might be significant to be detected by current TeV facilities, such as the Large High Altitude Air Shower Observatory (LHAASO) or future \gr facilities, such as the Cherenkov Telescope Array (CTA).
The fact that an initially pair dominated jet can potentially lead to a stronger TeV flux may sound counterintuitive, but in fact it is natural in our treatment due to the assumption that the mass loading is linked with energy dissipation into particle acceleration. 
The increase of the specific enthalpy depends on the initial conditions of the jet launching, and in this work we base our formalism on one specific setup of GRMHD simulations. A different setup is very likely to lead to less efficient heating of the jets, the specific enthalpy nevertheless will still increase due to energy transfer (see discussion of \cltm). To explore  the full range of possible physical scenarios with  GRMHD simulations is currently too computationally expensive.  We can however examine semi-analytically how the impact on the jet kinematics depends on the level of dissipation by replacing the heating parameter $f_{\rm heat}$ (that was used in previous work to estimate the heating of the thermal particles at the particle acceleration region; see, e.g., discussion in \citealt{Lucchini2021correlation}) with the fraction of the magnetic energy that is additionally allowed to go into energising particles. With such a parameterisation, $h$ will increase by a factor $f_{\rm heat}\sigma$ along the jet, whereas the magnetisation will be reduced as $(1-f_{\rm heat})\sigma$. We show in Fig.~\ref{fig: mu et al for different fheat} the impact of this free parameter in the energy components. To avoid a steep increase of $h$ that looks like a step-function, we use a function $\tanh^2{(z/z_{\rm diss})}$, instead. The underlying model is that of the left-hand panel of Fig.~\ref{fig: mu for mass loading, shifted and renormalised} where we assume a ``hot'' jet base (500\,keV) and one proton per electron. 

A further spectral difference between a pair-proton and a pair-dominated jet base is in the lower energy regime of the spectrum, and in particular, in the UV-to-X-ray spectrum. 
For the same initial magnetisation and injected power, 
the number density of the pairs at the pair-dominated jet base is enhanced (see, e.g., equation~\ref{eq: n0 at the jet base final}) resulting in increased Compton scatterings that lead to a significant difference in the $\sim$1--$100\,\rm keV$ range. The X-ray spectrum in particular shows a hard spectral index ($\nu F_{\nu}\propto \nu^{-\alpha+1}$, with $\alpha<1$; see right-hand plot in Fig.~\ref{fig: SEDs for mass loading jets}) that is similar to the expected output of a thermal corona \citep[][]{Sunyaev1980Comptonization,Haardt1993,Titarchuk1994Comptonization,narayan1994advection,Magdziarz1995}.

\subsection{Proton energy crisis}\label{sec: discussion on proton power}

\begin{figure}
    \includegraphics[width=1.1\columnwidth]{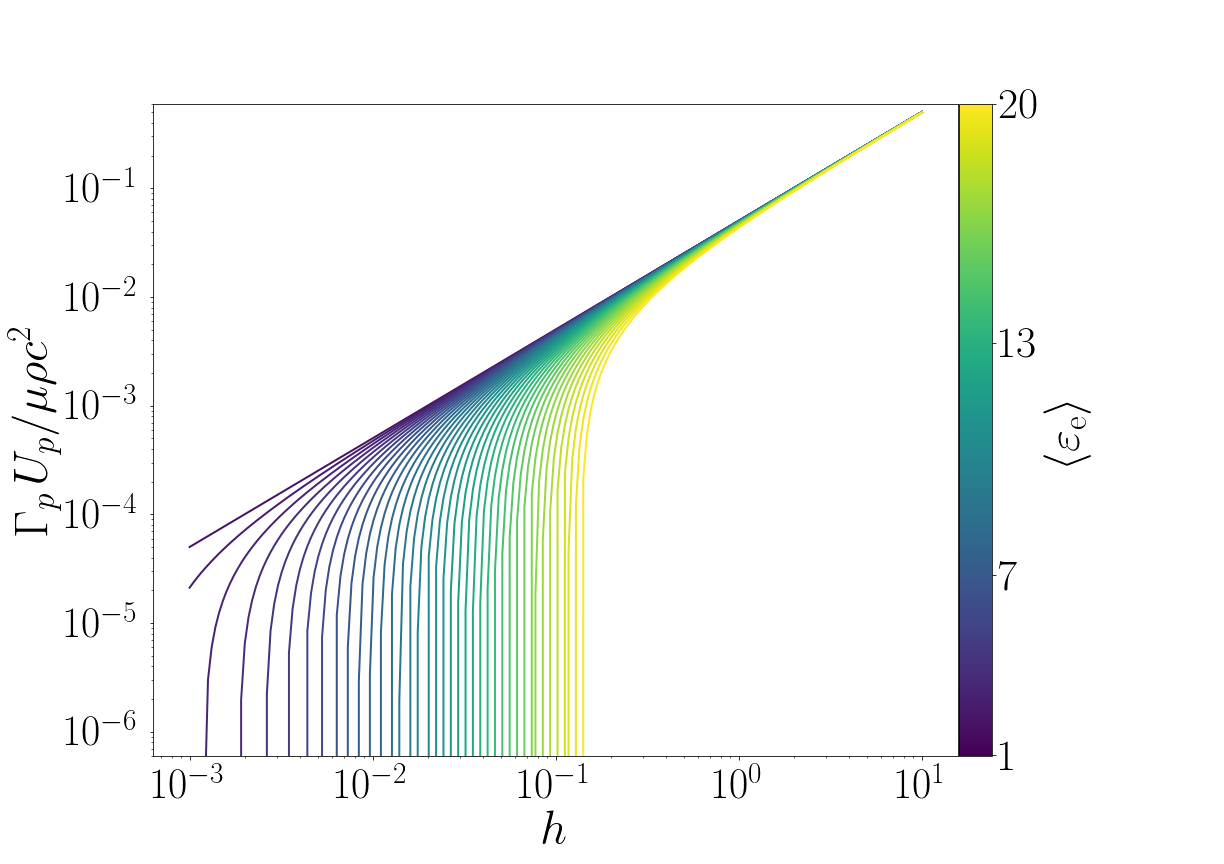}
    \caption{The specific enthalpy of the protons $\Gamma_p U_p/\rho c^2$ divided by $\mu$ shows the total energy that is allocated to protons with respect to the total available jet energy, as a function of the jet specific enthalpy $h$. We plot the proton energy density for a number of different electron energy densities that correspond to different values of $\langle \varepsilon_e \rangle$ as shown in the colormap, and we use $\eta_e=10$. 
    } 
    \label{fig: proton energy density vs h and ge}
\end{figure}

With the conserved, mass-loading jet model we develop here, we are able to constrain the total energy that is allocated to the protons and is used to accelerate them to non-thermal energies. In that way, the total energy carried by the accelerated protons never exceeds the available energy of the jets that has been a major issue in the past \citep[][]{boettcher2013leptohadronic,zdziarski2015hadronic,Liodakis2020,kantzas2022gx}. 
In Fig.~\ref{fig: proton energy density vs h and ge}, we plot the specific enthalpy of the protons $\Gamma_p U_p /\rho c^2$ divided by $\mu$ as a function of the total jet enthalpy $h$. This quantity expresses the fraction of the total energy flux of the jet that is used by the accelerated protons, and we show its dependence on $h$
for different average electron Lorentz factors, as indicated by the colormap. Regardless of the average electron energy $\langle \varepsilon_e \rangle$, the protons can hardly carry more than $\sim$10 per cent of the total energy in the jets because
higher fractions would require specific enthalpy $h$ of the order of a few or above (upper-right corner of the plot) resulting in strongly magnetised flows ($\sigma \gtrsim \gamma h$). Moreover, for particular values of $\langle \varepsilon_e \rangle$ (see the blue lines for instance that correspond to values of the order of 1 to 7), the protons can only be accelerated at $z_{\rm diss}$ and beyond if the total specific enthalpy $h$ is greater than some critical value $h>h_{\rm crit}$ where
\begin{equation}
h_{\rm crit} = \dfrac{(\langle \varepsilon_e \rangle -1)\Gamma_e}{1+\dfrac{\rm m_p/\rm m_e}{\eta_e}},
\end{equation}
hence the cutoffs for different $\langle \varepsilon_e \rangle$ at small values of $h$. In this particular figure, we use $\eta_e=10$, but as we show in Appendix~\ref{app: proton power} for smaller (larger) values of $\eta_e$ the only difference is that the cutoffs are located to smaller (larger) values of $h$.

From Fig.~\ref{fig: proton energy density vs h and ge}, we see that the energy of the accelerated protons never exceeds that of the jet because the specific enthalpy of the non-thermal protons is always less than the total normalised energy flux ($\Gamma_p U_p/\rho c^2<\mu$) and hence never violates the energy budget.

